\documentclass[
reprint,
amsmath,
amssymb,
aps,
prd,
nofootinbib,
floatfix]{revtex4-2}

\usepackage{graphicx}
\usepackage{dcolumn}
\usepackage{bm}
\usepackage{hyperref}
\usepackage{physics}
\usepackage{aas_macros}

\usepackage[capitalize]{cleveref}

\hypersetup{
colorlinks=true,
allcolors=blue
}

\DeclareRobustCommand{\ion}[2]{%
\relax\ifmmode
\ifx\testbx\f@series
{\mathbf{#1\,\mathsc{#2}}}\else
{\mathrm{#1\,\mathsc{#2}}}\fi
\else\textup{#1\,{\mdseries\textsc{#2}}}%
\fi}

\newcommand{\changeone}[1]{#1}

\newcommand{\changetwo}[1]{#1}

\begin{document}

\title{
Cross-correlating IceCube neutrinos with a large set of galaxy samples around redshift $z \sim 1$
}

\author{Aaron Ouellette}
\email{aaronjo2@illinois.edu}
\author{Gilbert Holder}
\affiliation{Department of Physics \& Illinois Center for Advanced Studies of the Universe, \\
University of Illinois Urbana-Champaign, Urbana, IL, 61801 USA}
\affiliation{Center for AstroPhysical Surveys, National Center for Supercomputing Applications, Urbana, IL 61801 USA}
\date{\today}

\begin{abstract}
    The IceCube neutrino telescope has detected a diffuse flux of high-energy astrophysical neutrinos, but the sources of this flux have largely remained elusive. Using the 10-year IceCube public dataset, we search for correlations between neutrino events and tracers of large-scale structure (LSS). We conduct a combined cross-correlation analysis using several wide-area galaxy catalogs spanning a redshift range of $z=0.1$ to $z\sim2.5$ as well as maps of the cosmic infrared background. We do not detect a definitive signal, but find tantalizing hints of a potential positive correlation between neutrinos and the tracers of LSS. We additionally construct a simple model to interpret galaxy-neutrino cross-correlations in terms of the redshift distribution of neutrino sources. We put upper limits on the clustering amplitude of neutrinos based on the measured cross-correlations with galaxies and forecast the improvements on these constraints that can be obtained using future detectors. We show that, in the future, neutrino-galaxy cross-correlations should be a powerful probe to constrain properties of neutrino source populations.

\end{abstract}

\maketitle

\section{Introduction}
High-energy astrophysical neutrinos have now been observed by the IceCube Neutrino Observatory through several detection channels \cite{Aartsen2013,Abbasi2022,2402.18026,2403.02516}. In addition, IceCube has found evidence for neutrino emission from several astrophysical sources including two active galaxies \cite{1807.08794,2211.09972} and the Galactic disk \cite{2307.04427}. Despite these detections, which astrophysical populations contribute the bulk of the diffuse neutrino flux remains an open question.

A common method used to search for neutrino emission from a class of objects is a stacking analysis, where the neutrino emission is stacked at the locations of objects in some test catalog in order to look for potential excess emission over the background. Some examples of this kind of  analysis include Refs.  \cite{2207.04946,2111.10169,2107.03149,2206.02054}, where the potential sources considered included blazars, active galactic nuclei (AGN), star-burst galaxies, and galaxy clusters. Stacking analyses allow potentially weak signals from individual sources to be amplified, but relies on the choice of a catalog that contains many of the same galaxies that are neutrino sources.

Cross-correlations with large scale structure (LSS) can also be used to search for neutrino sources \cite{Fang2020,Negro2023}, assuming only that the neutrino sources reside in galaxies or galaxy clusters, which are known to be tracers of large scale structure. For a cross-correlation analysis, there is no need to design a source catalog that is likely to match actual neutrino sources, instead the method exploits the fact that matter is highly clustered on large scales in the Universe and thus any tracers of LSS (e.g. galaxies) are spatially correlated with each other. Cross-correlation analyses are ubiquitous in cosmology in terms of both the two-point correlation function in real space or the power spectrum in Fourier or harmonic space. Ref. \cite{Fang2020} first proposed using the cross-correlation between neutrinos and galaxy samples to constrain the fraction of observed neutrinos that are correlated with LSS over a certain redshift range, \changeone{while Ref.~\cite{Negro2023} conducted a detailed analysis of the cross-correlation between neutrinos and the gamma-ray background including a quantification of the IceCube point spread function (PSF).}

In this paper we build on the analysis of Ref. \cite{Fang2020} by including several additional important components. \changeone{Similar to Ref~\cite{Negro2023},} we include modeling of the IceCube PSF and show that it has a very significant impact on the measured cross-correlation. \changeone{But, we use additional event-level smoothing in order to maximize the potential signal-to-noise and attempt to account for the spatially varying PSF.} We additionally model the details of galaxy clustering to illustrate how different galaxy samples can be sensitive to the redshift distribution of neutrino sources. We conduct an analysis of the 10-year IceCube public dataset of track-like events and provide forecasts for how well a future detector could constrain the properties of the neutrino source population. 

Motivated by the fact that both AGN activity and star formation in the Universe peak between redshifts $z\sim1-3$ \cite{Pei1995,Madau1996,Madau2014,Aird2015}, we are especially interested in looking for correlations between high-energy neutrinos and LSS around redshift $z\sim1$. Whether neutrino sources can be detected at such large cosmological distances depends on the rarity of luminous high-energy neutrino sources, so we conduct our cross-correlation analysis with tracers of LSS over a very broad redshift range from $z=0.1$ to $z\sim2.5$, including maps of the cosmic infrared background (CIB) that trace the entire star-formation history of the Universe.

This paper is structured as follows. In \cref{sec:data}, we introduce the neutrino and galaxy data sets used in our analysis. We provide an overview of the theory needed to model large-scale cross-correlations between neutrinos and galaxies in \cref{sec:theory}. We describe the details of the analysis in \cref{sec:analysis} and present the results in \cref{sec:results}. Finally, we conclude in \cref{sec:discussion}.

Throughout this paper we assume a standard flat $\Lambda$CDM universe based on the Planck 2018 results \cite{Planck2020}: $(\Omega_m, \Omega_b, h, \sigma_8, n_s) = (0.31, 0.049, 0.677, 0.81, 0.967)$.

\section{Data}\label{sec:data}
\subsection{IceCube neutrinos}
IceCube detects neutrinos through multiple different channels. Here we focus on ``track-like" events, where a muon, produced by the interaction of a muon neutrino with the Antarctic ice, produces a long track of Cherenkov radiation that can be reasonably accurately traced back to an angular position on the sky. 

The 10-year IceCube point-source data release\footnote{\url{http://doi.org/DOI:10.21234/sxvs-mt83}} \cite{2101.09836} consists of a catalog of just over a million track-like neutrino candidates detected between April 2008 and July 2018 with sky coordinates, reconstructed energies, and estimated angular uncertainties. The event selection was designed to identify tracks from high-energy muons passing through the IceCube detector and uses different selection criteria in the Northern and Southern skies. 

In the southern sky (corresponding to events that originate above the detector), while strict cuts on reconstruction quality and minimum energy are used, events in the public data release are still dominated by two foregrounds: atmospheric muons and atmospheric muon neutrinos, both of which are produced from cosmic rays interacting with the atmosphere. In the northern sky (corresponding to events that originate below the detector), atmospheric muons are blocked by the Earth, while the atmospheric neutrino foreground still remains. Due to the reduced foregrounds, the minimum energy cut is lower than in the southern sky. The reconstructed neutrino energies in the northern sky are more like lower limits rather than unbiased estimates due to energy losses, especially at high energies, from interactions inside the Earth \cite{2101.09836}.

We limit our analysis to the northern sky (defined as dec. $> -5^\circ$) in order to avoid the very large atmospheric muon foreground that is present in the southern sky.

\subsection{Galaxy samples}\label{sec:gal_data}
\begin{figure}
    \centering
    \includegraphics[width=\linewidth]{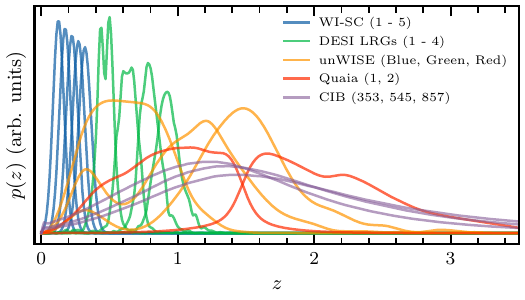}
    \caption{Redshift distributions for all of the galaxy samples used in this analysis. A summary of the properties of each galaxy sample is given in \cref{tab:galaxies}.}
    \label{fig:gal_dndzs}
\end{figure}

For our cross-correlation analysis, we look for large galaxy samples that span a variety of redshift ranges and have wide and uniform coverage in the northern sky. We provide details on the samples used below and \changeone{summarize their properties in \cref{fig:gal_dndzs} and \cref{tab:galaxies}}.

\subsubsection{WISE-SuperCOSMOS}
WISE-SuperCOSMOS\footnote{\url{http://ssa.roe.ac.uk/WISExSCOS.html}} \cite[WI-SC,][]{Bilicki2016} is a full-sky low-redshift photometric galaxy catalog. These galaxies were obtained by cross-matching data from the Wide-field Infrared Survey Explorer \cite[WISE,][]{Wright2010} with the photographic plates of SuperCOSMOS \cite{Hambly2001,Peacock2016}. WI-SC contains $\sim 20$ million sources up to redshift of $z=0.4$. The photometric redshifts have a mean error of $\sigma_z / (1+z) \approx 0.035$. We divide the catalog into 5 equal-size redshift bins between $z=0.1$ and $z=0.35$ (labeled as WI-SC 1 through 5 in \cref{tab:galaxies}). We estimate the redshift distribution of each bin by convolving the photometric redshift distributions with Gaussians given by the mean photometric redshift errors. We use the sky mask constructed by Ref. \cite{Bilicki2016} that masks the galactic disk, bulge, the Magellanic clouds, and other areas of large stellar contamination or dust extinction. As noted in Refs. \cite{Peacock2018,Xavier2019}, some stellar contamination and obscuration still remains at the few per cent level, but we treat this as negligible due to the fact that our measurements are severely limited by the quality of the neutrino data.

\subsubsection{DESI LRGs}
To probe redshifts between $z\sim0.4$ and $z\sim1$, we use luminous red galaxy (LRG) samples selected from DESI Legacy Imaging Surveys DR9\footnote{\url{https://www.legacysurvey.org/dr9/description/}} by Ref. \cite{Zhou2023}. These samples have wide sky coverage and spectroscopic redshift distributions, and are designed for cross-correlation applications with 4 tomographic bins. We specifically use the ``extended LRG" sample that has 2-3 times as many galaxies as the main sample presented in Ref. \cite{Zhou2023} in order to gain lower shot noise at the expense of a slightly less clean sample. In \cref{tab:galaxies}, these samples are labeled as DESI 1--4.

\subsubsection{unWISE galaxies}
To probe redshifts up to $z\sim2$, we use the unWISE galaxy samples presented by Refs. \cite{Krolewski2020,Krolewski2021}. These galaxies were selected from the unWISE catalog \cite{Schlafly2019} that contains over two billion objects observed by WISE. Ref. \cite{Krolewski2020} divided the unWISE catalog based on magnitude and color and rejected stars using Gaia data \cite{GaiaCollaboration2016} to create three galaxy samples with median redshifts of $\sim0.6$, $1.1$, and $1.5$. These galaxy samples are referred to as unWISE Blue, Green, and Red, respectively. The unWISE catalog is based on only two infrared bands, so these galaxy samples do not have photometric redshifts, but the overall redshift distributions were characterized through both cross-correlations with spectroscopic samples and cross-matches with COSMOS data \cite{Krolewski2020}. In our analysis we use the cross-correlation redshift distribution as this also includes information about the bias evolution of the unWISE samples. We also use the mask constructed by Ref. \cite{Krolewski2020} that is based on the Planck lensing mask and includes corrections for area lost due to sub-pixel masking around bright sources. 

\subsubsection{Quaia}
Quaia\footnote{\url{https://zenodo.org/records/10403370}} is an all-sky quasar catalog based on quasar candidate catalogs from Gaia and infrared data from unWISE \cite{Storey-Fisher2023}. The catalog contains almost 1.3 million sources brighter than $G=20.5$. Redshifts were estimated for each object by matching Gaia and unWISE photometry and Gaia-estimated redshifts to SDSS DR16Q spectroscopic redshifts \cite{Lyke2020}. Following Ref. \cite{Alonso2023}, we divide the $G < 20.5$ catalog into two redshift bins using a boundary of $z=1.47$ and estimate their redshift distributions through PDF stacking, where we assume that the redshift estimate for each event is drawn from a Gaussian with a standard deviation equal to the estimated redshift error. We call these samples Quaia 1 and Quaia 2. We additionally use the published selection function models\footnote{\url{https://zenodo.org/records/8098636}} for the two redshift bins to construct sky masks and correct for non-uniform sky coverage. 

\subsubsection{CIB maps}
In addition to galaxy catalogs, we use the large-scale CIB maps generated by Ref. \cite{Lenz2019} based on the Planck High-Frequency Instrument sky maps at 353, 545, and 857 GHz. Ref. \cite{Lenz2019} removed the Galactic dust emission from these maps using a template based on \ion{H}{I} column density. These maps trace the integrated extragalactic infrared emission and thus act as maps of unresolved galaxies over a very wide redshift range. 

We use the sky mask constructed by Ref. \cite{Lenz2019} based on the Planck 20\% Galactic plane mask and an upper limit on the \ion{H}{I} column density of $N_\text{\ion{H}{I}} = 4\times10^{20}$ cm$^{-2}$. This choice for the limit on the \ion{H}{I} column density trades larger sky coverage for more Galactic contamination. Since we only use cross-spectra in our analysis and do not use the CIB auto-spectra, this contamination should not bias our results. We also confirmed that using smaller \ion{H}{I} thresholds does not significantly change our final results.

\begin{table}
    \centering
    \setlength{\tabcolsep}{8pt}
    \begin{tabular}{lcccc}
    Galaxy sample & $\bar{n}$ [deg$^{-2}$] & $\bar{z}$ & $\delta_z$ & $b_g$ \\ \hline \\
    WI-SC 1 & 122 & 0.13 & 0.04 & 1.0 \\
    WI-SC 2 & 137 & 0.18 & 0.04 & 1.1 \\
    WI-SC 3 & 142 & 0.23 & 0.04 & 1.1 \\
    WI-SC 4 & 121 & 0.27 & 0.04 & 1.1 \\
    WI-SC 5 & 46 & 0.32 & 0.05 & 1.3 \\ \hline \\
    DESI 1 & 186 & 0.47 & 0.07 & 1.7 \\
    DESI 2 & 311 & 0.63 & 0.08 & 1.9 \\
    DESI 3 & 422 & 0.79 & 0.09 & 2.0 \\
    DESI 4 & 438 & 0.93 & 0.10 & 2.1 \\ \hline \\
    unWISE Blue & 3409 & 0.64 & 0.29 & 1.7 \\
    unWISE Green & 1846 & 1.1 & 0.47 & 2.4 \\
    unWISE Red & 109\footnote{Value from A. Krolewski (private communication), differs from that reported in Ref. \cite{Krolewski2020}.} & 1.5 & 0.47 & 3.4 \\ \hline \\
    Quaia 1 & 23 & 0.97 & 0.38 & 1.9$^\ast$ \\
    Quaia 2 & 26 & 2.1 & 0.56 & 3.4$^\ast$ \\ \hline \\
    CIB 353 & -- & 2.0 & 1.1 & 3.9$^\ast$ \\
    CIB 545 & -- & 1.8 & 1.0 & 3.5$^\ast$ \\
    CIB 857 & -- & 1.5 & 0.9 & 2.9$^\ast$ 
    \end{tabular}
    \caption{Galaxy samples used in the cross-correlation analysis. We list the mean number density per square degree ($\bar{n}$), the mean redshift ($\bar{z}$), standard deviation of the redshift distribution ($\delta_z$), and linear bias ($b_g$) obtained through fitting the galaxy auto-spectra. Note that most samples are fit using a constant bias, while samples marked with ($\ast$) use a redshift dependent bias (see \cref{sec:theory} for details) and we report the average bias ($b_\text{eff} = \int b_g(z) p_g(z)\, dz$). The values of $\bar{z}$, $\delta_z$, and $b_g$ for the CIB samples are calculated based on the model from Ref. \cite{Maniyar2018}.}
    \label{tab:galaxies}
\end{table}

\section{Theory}\label{sec:theory}
We are interested in modeling the angular cross-spectrum between galaxies with known properties and some unknown astrophysical neutrino sources. In both cases, the projected overdensity field on the sky $\delta (\vu{n})$ is related to the 3D overdensity field $\Delta(\vb{x}, t)$
\begin{equation}
    \delta(\vu{n}) = \int d\chi\, p(\chi) \Delta(\chi \vu{n}, t(\chi)),
\end{equation}
where $p(\chi) = H(z)p(z)$ is the normalized comoving distance distribution of galaxies or neutrino sources and $p(z)$ is the corresponding normalized redshift distribution\footnote{Using units with $c=1$.}.

Using the Limber approximation \cite{Limber1953,Loverde2008}(which is generally valid for scales $\ell \gtrsim 20$), we can relate an angular power spectrum $C^{XY}_\ell$ to the corresponding 3D power spectrum $P_{XY}(k, z)$:
\begin{equation}\label{eq:limber}
    C^{XY}_\ell = \int \frac{d\chi}{\chi^2} p_X(\chi)p_Y(\chi)P_{XY}\left(\frac{\ell+1/2}{\chi}, z(\chi)\right),
\end{equation}
where $X, Y \in \{g, \nu\}$ represent galaxy or neutrino source overdensity fields.

We can then further use the linear bias approximation that is valid on large scales to write $P_{XY}$ in terms of the matter power spectrum $P_m$:
\begin{equation}
    P_{XY}(k, z) = b_X b_Y P_m(k, z) + A_\text{SN},
\end{equation}
where $A_\text{SN}$ is a constant shot noise amplitude and $b_X$ and $b_Y$ are the linear bias factors for the corresponding fields. 

In the case of the neutrinos, the situation is complicated by the very large background of atmospheric neutrinos. Assuming that the atmospheric neutrinos are uniformly distributed across the sky and uncorrelated with the astrophysical ones, $b_\nu$ is completely degenerate with $f_\text{astro}$, the fraction of neutrinos that are astrophysical \cite[see appendix A in][]{Fang2020}. This is the exact same effect that an unclustered stellar contamination has on the measurement of galaxy bias \cite{Krolewski2020}. Since astrophysical and atmospheric neutrinos are completely indistinguishable on an event-by-event basis, the combination $b_\nu f_\text{astro}$ determines the amplitude of $C_\ell^{g\nu}$.

Putting everything together, we model the galaxy auto-spectra as
\begin{equation}\label{eq:cgg}
    C^{gg}_\ell = \int \frac{d\chi}{\chi^2} \left[b_g(z) p_g(z) H(z)\right]^2 P_m\left(k_\ell, z\right) + A^{gg}_\text{SN},
\end{equation}
where, in general, the galaxy bias can vary with redshift to account for galaxy evolution over cosmic time and we have defined $k_\ell = (\ell + 1/2) / \chi$. 

The neutrino auto-spectrum is modeled as
\begin{equation}
    C^{\nu\nu}_\ell = (f_\text{astro}b_\nu)^2 \int \frac{d\chi}{\chi^2} \left[p_\nu(z) H(z)\right]^2 P_m\left(k_\ell, z\right) + A^{\nu\nu}_\text{SN}.
\end{equation}
While redshift evolution of $b_\nu$ is almost certainly expected if neutrino sources have a wide redshift distribution, there is not enough information to disentangle the degeneracy between $b_\nu$ and $p_\nu(z)$, so we model it as a constant effective bias.

The galaxy-neutrino cross-spectrum is modeled as
\begin{equation}\label{eq:cgnu}
    C^{g\nu}_\ell = f_\text{astro}b_\nu \int \frac{d\chi}{\chi^2} p_\nu(z) b_g(z) p_g(z)\left[H(z)\right]^2 P_m\left(k_\ell, z\right)
\end{equation}

We ignore the shot noise contribution to the cross-spectra, because it is significant only if a large fraction of the neutrinos and the galaxies are in common between the catalogs. Since most of the neutrinos are expected to be non-astrophysical and there are vastly more galaxies in each galaxy sample than there are astrophysical neutrinos, the shot noise term should be  negligible.

We calculate all of these theory quantities using the Core Cosmology Library\footnote{\url{https://github.com/LSSTDESC/CCL}} (\verb|CCL|) \cite{Chisari2019}. The matter power spectrum $P_m(k,z)$ is calculated using the HaloFit model \cite{Takahashi2012} as implemented in \verb|CCL|. The remaining components that go into the theory prediction for the galaxy-neutrino cross-spectrum (\cref{eq:cgnu}) for a given galaxy sample are: the galaxy bias $b_g$ and redshift distribution $p_g(z)$, the neutrino source redshift distribution $p_\nu(z)$, and the neutrino clustering amplitude $b_\nu f_\text{astro}$. 

For the galaxy catalogs, the redshift distributions are determined using photometric redshift estimates or cross-correlations with spectroscopic samples as described for each galaxy sample in \cref{sec:gal_data}. We determine the galaxy bias factors by directly fitting \cref{eq:cgg} to the measured galaxy auto-spectra as described below in \cref{sec:model}. The only galaxy sample that we model with a redshift-dependent bias is Quaia, where we use the quasar bias evolution model from Ref. \cite{Laurent2017}:
\begin{equation}
    b(z) = 0.278((1+z)^2 - 6.565) + 2.393.
\end{equation}
Following Ref. \cite{Alonso2023}, we fit a constant amplitude times this functional form to the Quaia auto-spectra. Since the WI-SC and DESI samples have relatively narrow redshift distributions, we assume that redshift evolution in the bias of these sample is negligible, while the unWISE cross-correlation redshift distributions already include the effect of bias evolution, and so we fit a single constant effective bias to these samples.

The CIB maps have to be treated slightly differently since they do not consist of resolved galaxies. Ultimately the CIB model still consists of a galaxy bias and redshift distribution, but the redshift distribution is written in terms of the infrared emissivity of galaxies across redshift and the cosmic star-formation rate density. We directly follow Section 2 of Ref. \cite{Maniyar2018} to construct a linear model of the CIB anisotropies and we use the best fit parameters reported in their Table 3.

Finally, assuming a model for the neutrino source redshift distribution $p_\nu(z)$ allows us to put constraints on the properties of the neutrino source population. The amplitude of the cross-correlation determines the quantity $b_\nu f_\text{astro}$ which directly depends on how neutrino source trace the underlying matter distribution, while the full shape of the cross-spectrum depends on the details of the redshift distribution of the sources. In practice, though, as we will see below, these constraints are heavily degraded by the statistical errors in the neutrino maps.

\section{Analysis Methods}\label{sec:analysis}

\subsection{Maps}

\begin{figure}
    \centering
    \includegraphics[width=\linewidth]{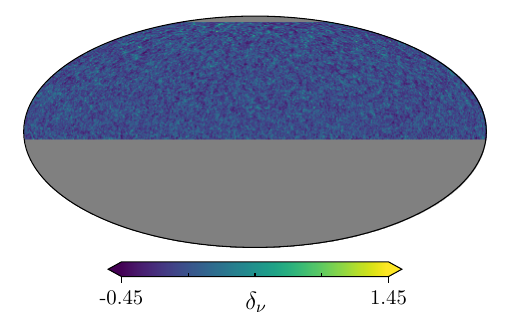}
    \caption{Neutrino over-density map for all events above 1 TeV, where each event has been smoothed using its respective angular uncertainty.}
    \label{fig:nu_map}
\end{figure}

To measure the neutrino-galaxy cross-correlations, we construct projected overdensity maps of each tracer using  \verb|HEALPix| \cite{Gorski2005} as implemented by \verb|healpy| \cite{Zonca2019} with the resolution parameter $N_\text{side} = 512$, corresponding to an angular resolution of $\sim 7'$. 

For DESI, unWISE, and the CIB, high resolution \verb|HEALPix| maps were already available which we downgraded to the target resolution by calculating the spherical harmonic coefficients $a_{\ell m}$ for each map, zeroing out all coefficients with $\ell \ge 3N_\text{side}$, and finally converting back to real space maps. This procedure ensures that no aliasing occurs.

For WI-SC and Quaia, we constructed the maps directly from the catalogs using their respective masks. With Quaia, we additionally directly follow the procedure laid out in \cite{Alonso2023} to take into account the Quaia selection function.

We turn the CIB differential intensity maps into unitless fluctuation maps similar to the galaxy overdensity maps by dividing by the CIB monopole given in Table 12 of Ref. \cite{Planck2018}.

For the neutrinos, as a first step, we consider all events with reconstructed energies $E \ge 1$ TeV. This selection removes the lowest energy events that are expected to be dominated by atmospheric neutrinos while keeping a fairly large sample ($\sim 400,000$ events) to minimize statistical errors. Later, we also present results for neutrinos binned by energy, but the analysis steps are exactly the same as described here for the single bin. 

We construct a smoothed neutrino counts map $N_\nu$ for a given event selection by treating each individual event as a 2D circular Gaussian centered at the reported angular position and using the reported angular uncertainty as the standard deviation. This procedure is similar to that used by the IceCube collaboration to construct spatial likelihoods for point source searches \cite{Braun2008,Abbasi2011}. In our analysis, constructing a smoothed counts map allows us to maximize the signal-to-noise ratio (SNR) by weighting events according to the quality of their localization \changeone{in a manner very similar to a Wiener filter. We have confirmed that our final results are consistent whether we use the smoothed map described here or the raw un-smoothed neutrino counts map, but using the smoothed map results in tighter final constraints showing that the smoothing operation results in a closer-to-optimal analysis}.

Next, we use randomized data to construct an estimate of the local mean neutrino counts $\bar{N}_\nu$. At a given energy, the effective area of the IceCube detector depends only on the source declination (due to the unique position at the South Pole); we can generate random realizations of the neutrino count maps by randomizing the R.A. values of each event. We construct the mean counts map by averaging 50 randomized smoothed counts maps and then making the resulting map independent of R.A. by averaging over declination bands.

\begin{figure}
    \centering
    \includegraphics[width=\linewidth]{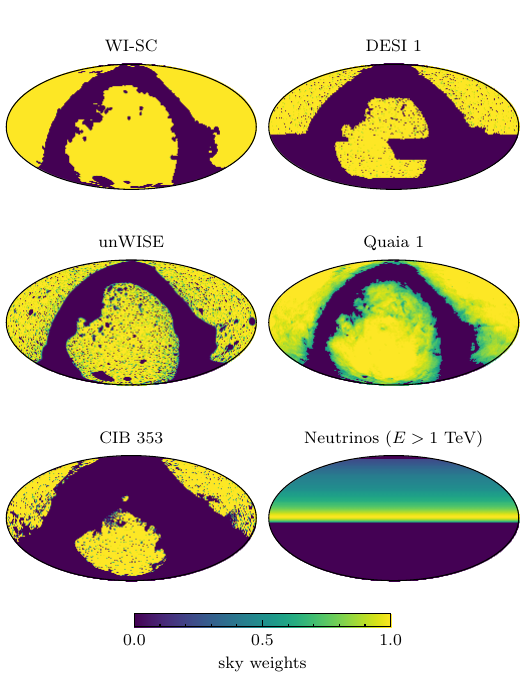}
    \caption{Sky masks or weights maps (in equatorial coordinates) for each of the tracers used in the analysis. For WI-SC and unWISE, the same mask is used for each sample in the group. DESI, Quaia, and the CIB all have masks specific to each sample in the group, so we only show a representative.}
    \label{fig:masks}
\end{figure}

The neutrino overdensity is $\delta_\nu = N_\nu / \bar{N}_\nu - 1$. We show the neutrino overdensity map in \cref{fig:nu_map}. The neutrino sky mask is based on a binary mask and inverse-variance weights. The binary mask masks the sky where dec. $\le -5^\circ$ and dec. $\ge 80^\circ$. The lower limit removes the astrophysical muon background, while the upper limit excludes the pole where the mean map becomes less accurate due to few events. We then also use a weights map that is proportional to the mean counts map in order to weight the neutrino overdensity in an inverse-variance fashion. We show the resulting weights map together with the masks used for all of the galaxy samples in \cref{fig:masks}. The effective fraction of the sky that is covered by both the neutrino mask and a galaxy sample ranges from 9\% for the CIB maps to 23\% for the WI-SC samples.

\subsection{IceCube effective beam}
While galaxy positions are generally measured at a precision of arc-seconds, IceCube's degree-scale sky localization of track-like events introduces very significant smoothing that must be taken into account by modeling the detector point-spread function (PSF, or beam window function in harmonic space). In general, the IceCube beam window function is neither constant nor isotropic, since the estimated event angular uncertainties depend on energy and declination. 

We assume that on large scales the overall effect of the combination of the detector PSF and our additional smoothing of each event using its estimated PSF on the measured power spectra can be described by an effective beam $B_\ell^\text{eff}$, such that 
\begin{equation}\label{eq:beam_auto}
    C^{\nu\nu, \text{obs}}_\ell = (B^\text{eff}_\ell)^2\, C^{\nu\nu, \text{true}}_\ell
\end{equation}
and
\begin{equation}\label{eq:beam_cross}
    C^{g\nu, \text{obs}}_\ell = B^\text{eff}_\ell\, C^{g\nu, \text{true}}_\ell.
\end{equation}
 
Under the assumption that the PSF of each individual event is a circular Gaussian and that it only varies on very large scales across the sky, we can estimate the effective beam directly from the data. The method we use is as follows. First, we assume that we can write the neutrino map as observed by IceCube (i.e., before our additional smoothing) as a sum over $N$ sets of events, indexed by $i$, that share the same PSF:
\begin{equation}
    \nu = \frac{1}{N} \sum_i^N (B^i \otimes s + n^i),
\end{equation}
where $B^i$ is the PSF of event set $i$, $\otimes$ represents the convolution operation, $s$ is some common signal map representing the clustering of astrophysical neutrinos, and $n^i$ is the non-clustered noise of event set $i$ representing shot noise and/or the atmospheric foreground. After our smoothing operation which smooths each event with the reported angular uncertainty, the neutrino map gains a second power of the PSF:
\begin{equation}
    \tilde{\nu} = \frac{1}{N} \sum_i B^i \otimes (B^i \otimes s + n^i).
\end{equation}
In harmonic space, the convolutions become products, so we can write
\begin{align}
    \nu_{\ell m} &= \frac{1}{N} \sum_i (B^i_\ell s_{\ell m} + n^i_{\ell m}) \\
    \tilde{\nu}_{\ell m} &= \frac{1}{N} \sum_i \left((B^i_\ell)^2 s_{\ell m} + B^i_\ell n^i_{\ell m}\right).
\end{align}
Then, the auto-spectrum of the map observed by IceCube is given by
\begin{equation}\label{eq:1}
    \langle \nu_{\ell m}^2 \rangle = \overline{B_\ell}^2 \langle s_{\ell m}^2 \rangle + \langle \overline{n_{\ell m}}^2 \rangle,
\end{equation}
where we have used the bar to denote the average over the $N$ sets of neutrino events. 

Similarly, the auto-spectrum of the neutrino map that has been smoothed at the event level is given by
\begin{equation}
    \langle \tilde{\nu}_{\ell m}^2 \rangle = \overline{B_\ell^2}^2 \langle s_{\ell m} \rangle + \langle \overline{B_\ell n_{\ell m}}^2 \rangle.
\end{equation}
The second term in the above equation can be simplified since noise between events are uncorrelated, or $\langle n^i_{\ell m} n^j_{\ell m} \rangle = 0$ for $i \ne j$:
\begin{align}
    \langle \overline{B_\ell n_{\ell m}}^2 \rangle &= \langle \frac{1}{N}\sum_i \left(B^i_\ell n^i_{\ell m}\right)^2 + \text{cross terms}\rangle \\
    &= \overline{B_\ell^2} \langle \overline{n^2_{\ell m}} \rangle.
\end{align}
The same trick can be used on the second term in Eq~\ref{eq:1}, so we get
\begin{align}\label{eq:nsm_auto}
    \langle \nu_{\ell m}^2 \rangle &= \overline{B_\ell}^2 \langle s_{\ell m}^2 \rangle + \langle \overline{n_{\ell m}^2} \rangle \\
    \label{eq:sm_auto}
    \langle \tilde{\nu}_{\ell m}^2 \rangle &= \overline{B_\ell^2}^2 \langle s_{\ell m} \rangle + \overline{B_\ell^2} \langle \overline{n_{\ell m}^2} \rangle.
\end{align}

Finally, if we have some other sample $g$ (for example, some galaxy sample) that also traces the same signal map $s$, the cross-spectra with the neutrino maps are
\begin{align}\label{eq:nsm_cross}
    \langle \nu_{\ell m} g_{\ell m} \rangle &= \overline{B_\ell} \langle s_{\ell m}^2 \rangle \\
    \label{eq:sm_cross}
    \langle \tilde{\nu}_{\ell m} g_{\ell m} \rangle &= \overline{B_\ell^2} \langle s_{\ell m} \rangle.
\end{align}

Comparing Eqs.~\ref{eq:nsm_auto}-\ref{eq:sm_cross}, we come up with two methods to estimate the effective beam $B_\ell^\text{eff}$. First, from Eqs.~\ref{eq:sm_auto} and \ref{eq:sm_cross} we see that the effective beam of our smoothed neutrino map can be estimated simply by averaging the squares of the individual Gaussian beams of each event:
\begin{equation}
    B_\ell^\text{eff} = \overline{B_\ell^2}.
\end{equation}
We call this the analytic estimate of the effective beam.

Secondly, by comparing Eqs.~\ref{eq:nsm_auto} and \ref{eq:sm_auto}, we can see that the effective beam can also be estimated by looking at the ratio of the auto-spectra of the two neutrino maps. Since we are very much in the noise dominated regime, $\langle s_{\ell m}^2 \rangle$ is small relative to the white noise, so
\begin{equation}\label{eq:beam_est}
    B_\ell^\text{eff} = \overline{B_\ell^2} = C_\ell^{\tilde{\nu}\tilde{\nu}} / C_\ell^{\nu\nu}.
\end{equation}
We treat this second estimate as our fiducial estimate of the effective beam, since we suspect that the auto-spectra might capture some of the information about spatial variations of the beam. 

\begin{figure}
    \centering
    \includegraphics[width=\linewidth]{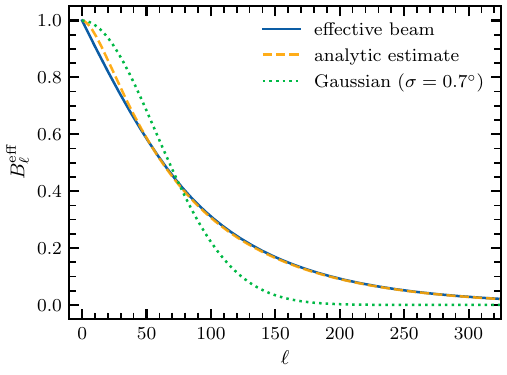}
    \caption{Effective beam (accounting for the instrument PSF and our applied smoothing) for the smoothed neutrino map consisting of all events with $E > 1$ TeV. The analytic estimate is calculated by averaging the Gaussian beams of each event. We also plot a Gaussian beam corresponding to the median angular uncertainty of all selected events to show that the overall effective beam is highly non-Gaussian. Essentially no information remains in the power spectra past $\ell \sim 350$.}
    \label{fig:beam}
\end{figure}

We plot both estimates of the effective beam in Fig.~\ref{fig:beam} and find that they agree very well, except at very large spatial scales where spatial variations in the beam become more significant. We also see that the effective beam is very non-Gaussian, even though we assumed that the PSF for each event was Gaussian. Correctly modelling this beam is important for any cross-correlation analyses, see also Fig. 7 in Appendix A of Ref.~\cite{Negro2023} where the same conclusion is reached. While the effective beam estimated by Ref.~\cite{Negro2023} is significantly different from ours due to different event selections, when attempting to match their selection, we find that our final results shift by at most $0.4\sigma$ when using their estimated beam instead of ours. As a final note, if the raw neutrino map as observed by IceCube is used (as in Refs.~\cite{Fang2020,Negro2023}) instead of our smoothed version, the effective beam should be estimated using
\begin{equation}
    B_\ell^\text{eff} = \overline{B_\ell} = C_\ell^{\tilde{\nu}\nu} / C_\ell^{\nu\nu},
\end{equation}
since there is only one power of the PSF present in the map instead of two as in our analysis.

There are several caveats to the above analysis, the rigorous treatment of which we leave for future work. First, while we have validated our estimator of the effective beam on simple Gaussian simulations, to fully validate it we would need detailed simulations of the IceCube detector which are not available publicly. There is an inherent uncertainty in this estimate of the effective beam due to the assumption that all events can be described by circular Gaussians. As can be seen from the catalog of high-energy neutrino alerts \cite{2304.01174}, the more detailed event spatial likelihoods are very often non-circular and often non-Gaussian. These effects will become more important in future analyses with better sensitivity. Additionally, in the case of a strongly spatially varying beam, there could be significant differences between the effective beams defined in terms of cross-spectra with different galaxy samples. In our analysis, all the galaxy samples have fairly similar coverage in the Northern sky and we have explicitly checked that the result of \cref{eq:beam_est} does not significantly change when computing the power spectra using the overlap of the neutrino mask and the different galaxy sample masks.

\subsection{Angular power spectra and covariance matrices}
We use the pseudo-$C_\ell$ algorithm implemented by \verb|NaMASTER|\footnote{\url{https://github.com/LSSTDESC/NaMaster}} \cite{Alonso2019} to compute all galaxy auto-power spectra and galaxy-neutrino cross-power spectra. Full details can be found in Refs. \cite{Hivon2002,Alonso2019}, here we briefly sketch the method. 

An observed field $\tilde{X}$ with incomplete sky coverage can be modelled as the product of the true underlying field $X$ with a mask $w$. Often the mask is treated simply as a binary map where $w=1$ for pixels on the sky that have been observed and $w=0$ where they have not, but in general $w$ can be designed to optimally weight different regions of the sky based on varying noise levels. The pseudo-$C_\ell$ for two fields $X$ and $Y$ is then defined as
\begin{equation}
    \tilde{C}_\ell^{XY} = \frac{1}{2\ell+1} \sum_m \tilde{X}_{\ell m}\tilde{Y}^\star_{\ell m}.
\end{equation}
Since the observed and true fields are related through a product with the mask in real space, in harmonic space they are related through a convolution. The expectation value of the pseudo-$C_\ell$ is then related to the underlying true $C_\ell$
\begin{equation}\label{eq:coupling}
    \langle \tilde{C}_\ell \rangle = \sum_{\ell'} M_{\ell \ell'} C_{\ell'},
\end{equation}
where $M_{\ell \ell'}$ is the mode-coupling matrix that can be computed directly from the power spectra of the masks. \cref{eq:coupling} is not directly invertible due to the loss of information on the masked sky, so the pseudo-$C_\ell$ is binned into bandpowers by assuming that it is piecewise constant in a number of discrete bins and a ``binned" mode-coupling matrix is computed that can be inverted to obtain an estimate of the true power spectrum. If one or both of the fields are smoothed by a PSF, then we divide the pseudo-$C_\ell$ by the corresponding beam window function before inverting the mode coupling matrix. \changeone{We do not include the effect of finite map resolution at this stage, but instead include the pixel window function in the model for the cross-correlations, though we note that this effect is completely negligible ($<1$\% change in our final results) at $N_\text{side} = 512$ due to the aggressive smoothing introduced by the IceCube effective beam.}

\verb|NaMaster| efficiently handles all of these operations and also provides a way to bin a theory power spectrum using the exact same bandpowers as a measured power spectrum to enable direct comparison.

We binned the power-spectra using linear bins of width $\Delta \ell = 50$ in the range $2 \le \ell < 3N_\text{side}$. While we calculated the cross-spectra all the way out to $\ell=1536$ to avoid power leakage, we discard all bandpowers above $\ell = 350$ for the analysis in order to avoid systematic errors from mis-characterizing the IceCube PSF at small scales and from the breakdown of the linear bias assumption at non-linear scales. \changeone{At the same time, all of the signal is concentrated at low $\ell$ due to the aggressive smoothing of the neutrino map, so we see almost no change in our final results when significantly increasing this upper scale cut.} We additionally discard the first bandpower at $\ell < 50$. \changeone{This is a relatively conservative choice driven by several considerations. First, this reduces any systematic effects from large spatial variations in the effective beam. As seen in Fig.~\ref{fig:beam}, our fiducial effective beam (using the auto-spectra of the neutrino maps) begins to diverge from the analytic estimate (assuming no spatial variation) below $\ell = 50$. This low-$\ell$ cut is also used when estimating the auto-spectra of the galaxy samples in order to avoid potential large-scale systematics in the galaxy maps, and allows us to safely use the Limber approximation (Eq.~\ref{eq:limber}) when modelling all of the power spectra. Finally, we note that in our analysis, there are negligible changes in our final results whether we include or discard this first bandpower. In future analyses with higher sensitivity, it will be necessary to more rigorously test all of these assumptions that go into choosing scale cuts.}

\begin{figure}
    \centering
    \includegraphics[width=\linewidth]{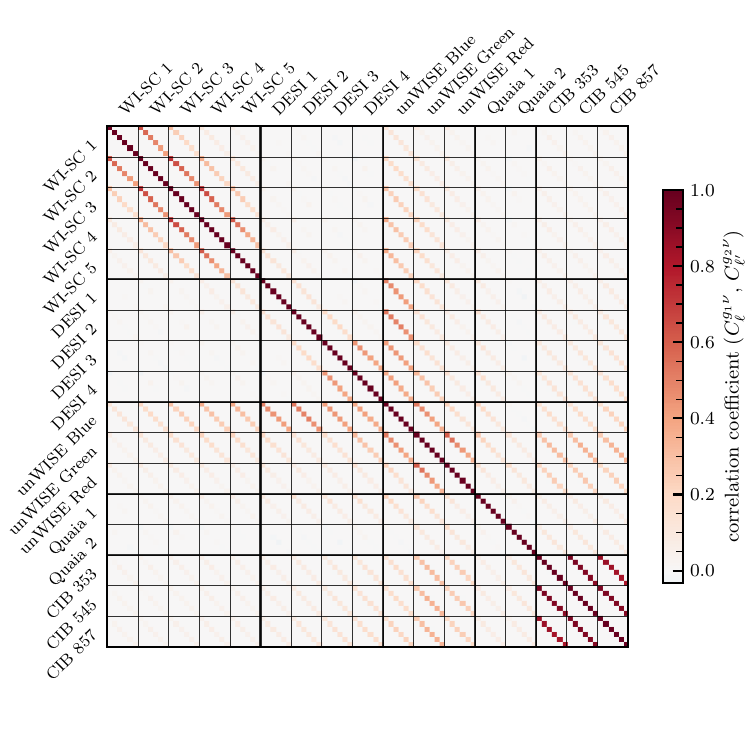}
    \caption{Correlation matrix for all of the measured cross-spectra. Cross-spectra involving galaxy samples with similar redshift distributions show significant correlations.}
    \label{fig:cov}
\end{figure}

We calculate the covariance matrix for each power spectrum analytically using the narrow kernel approximation implemented in \verb|NaMaster| \cite{Garcia-Garcia2019} which assumes that all fields are Gaussian and accounts for the mask-induced mode coupling. For the galaxy-neutrino cross-spectra we calculate the full combined covariance ${\rm Cov} \left(C^{g\nu}_\ell, C^{g'\nu}_{\ell'}\right)$ in order to account for correlations between galaxy samples $g$ and $g'$. We confirm that the diagonal elements of the covariance matrix agree very well with the variance computed from 50 realizations of randomized neutrino maps. We find that the off-diagonal elements between $\ell$-bins in the covariance from mode coupling is roughly on the order of $10^{-3}$ times the diagonal elements, while off-diagonal elements between galaxy samples can be significant. In \cref{fig:cov} we show the full correlation matrix (defined as $R_{ij} = C_{ij} / \sqrt{C_{ii}C_{jj}}$, where $C_{ij}$ are the elements of the covariance matrix) for the measured cross-spectra. Because of the substantial overlap in the redshift ranges of the various galaxy samples, there is substantial covariance between the cross-spectra that must be taken into account in any data interpretation.

We assume that our data (the power spectra) all follow multivariate Gaussians with covariances as calculated above. To fit a model to the data (either individual galaxy auto-spectra or a set of galaxy-neutrino cross-spectra), we use chi-squared minimization with
\begin{equation}
    \chi^2 = ({\bf d} - {\bf m})^T C^{-1} ({\bf d} - {\bf m}),
\end{equation}
where ${\bf d}$ is the data vector, $C$ is the corresponding covariance matrix, and ${\bf m}$ is the model prediction given some parameter values. For the combined analysis of all of the cross-correlations, we construct the data vector by concatenating all of the measured galaxy-neutrino cross-correlations between $50 < \ell < 350$.

\section{Results}\label{sec:results}
\begin{figure}
    \centering
    \includegraphics[width=\linewidth]{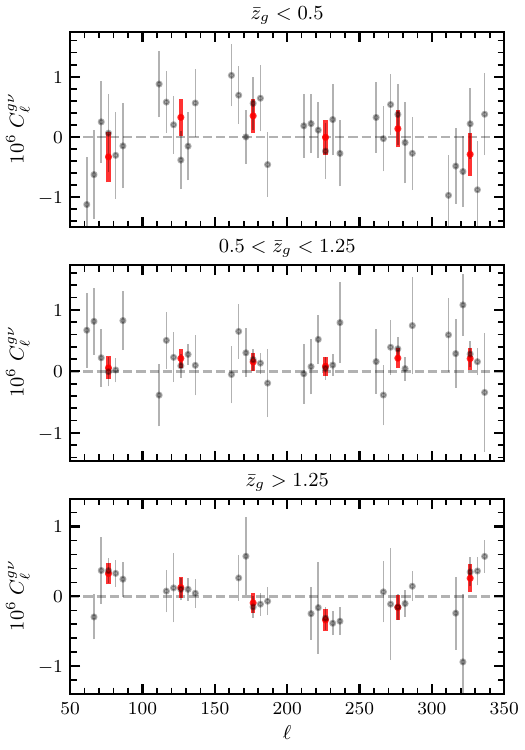}
    \caption{Measured galaxy-neutrino cross-spectra grouped by the mean redshift of the galaxy sample. In each panel, we plot the individual cross-spectra as grey points. The cross-spectra for each galaxy sample are offset in the $\ell$-direction to aid in visualization. To increase the SNR, we also plot the weighted average of each group of cross-spectra in red.}
    \label{fig:x-spectra}
\end{figure}

In \cref{fig:x-spectra}, we plot all of the measured galaxy-neutrino cross-spectra grouped by the mean redshift of the galaxy sample. In each panel, the grey points represent a cross-spectrum between the neutrinos and an individual galaxy sample. We also overlay in red the weighted average (taking into account all relevant galaxy-galaxy correlations) of all of the individual cross-spectra in each panel in order to increase the SNR.

\begin{figure}
    \centering
    \includegraphics[width=\linewidth]{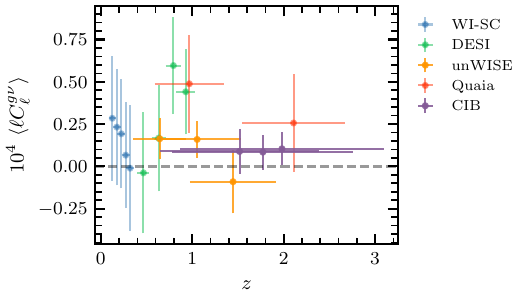}
    \caption{Amplitude of galaxy-neutrino cross-correlations as a function of the mean galaxy sample redshifts. The galaxy samples are: WI-SC (red), DESI (green), unWISE (purple), Quaia (orange), and the CIB (blue). The horizontal error bars on each point represent the standard deviation of the galaxy sample's redshift distribution.}
    \label{fig:ell-cls}
\end{figure}

Taken as a set, the measured cross-spectra are consistent with zero. We calculate a combined null chi-square value of $\chi^2_\text{null} = {\bf d}^T C^{-1} {\bf d} = 106.2$ with 102 total data points, giving a probability-to-exceed (PTE) of 0.37, indicating that we are unable to reject the null hypothesis that neutrino events are not correlated with LSS.

We do see hints of a positive excess, though, in the cross-spectra with galaxy samples in the $0.5 < z < 1.25$ redshift range, where all of the $\ell$ bins show positive correlation. To explore this further, we calculate the quantity $\langle \ell C^{g\nu}_\ell \rangle$ for each galaxy sample, where we take the average in an inverse-variance weighted fashion over the range $50 < \ell < 350$. This quantity is plotted against the mean redshift of each galaxy sample in \cref{fig:ell-cls}. \changetwo{On the scales of interest, $\ell C_\ell^{g\nu}$ is not expected to depend strongly on $\ell$ (see \cref{fig:theory} in \cref{sec:app})}, so averaging this quantity over a range in $\ell$ provides a rough model-independent measurement of the amplitude of clustering. In \cref{fig:ell-cls}, we see that several galaxy samples have a positive cross-correlation amplitude with the neutrinos at a level of $\ge 1.5\sigma$, specifically: DESI 3 (2.1$\sigma$), DESI 4 (1.8$\sigma$), unWISE Green (1.5$\sigma$), and Quaia 1 (1.7$\sigma$). These are not independent measurements, though, since the galaxy samples are correlated with each other. Using the full covariance matrix (\cref{fig:cov}), we find that combining all of the galaxy samples results in positive cross-correlation at a redshift of $z\sim1$ at roughly a $2\sigma$ level.

\subsection{Model fit}\label{sec:model}
To interpret the measured cross-correlations and attempt to constrain the neutrino source populations, we fit models to the galaxy auto-spectra and the galaxy-neutrino cross-spectra.

For each of the galaxy samples, we use \verb|NaMASTER| to calculate the auto-spectrum $C^{gg}_\ell$ and its covariance using the same procedure as described above, except that no effective beam is used. We then fit a simple two parameter model based on \cref{eq:cgg} where both the galaxy bias and shot noise amplitudes are free parameters. For all galaxy samples other than Quaia, we assume a constant bias. This should be a valid approximation for the WI-SC and DESI samples that have narrow redshift distributions, while the unWISE redshift distribution, directly measured through cross-correlations, includes the redshift dependence of the bias \cite{Krolewski2020}. For Quaia, we use the quasar bias evolution model from Ref. \cite{Laurent2017} and simply fit the overall amplitude; this was shown to be a  good fit to the Quaia sample \cite{Alonso2023}. We obtain good fits for all of the galaxy auto-spectra (PTE $\ge 0.1$) and list the resulting effective bias values in \cref{tab:galaxies}. These fits generally agree with previous values found in the literature \cite{Xavier2019,Krolewski2020,Alonso2023} with the caveat that we are using a very simplified model that does not include magnification bias, non-Gaussianities, or other higher-order effects. To model cross-correlations with the CIB, we use the best-fit model from Ref. \cite{Maniyar2018} which is valid on linear scales. 

To model the neutrino cross-correlations we use \cref{eq:cgnu}, treating the combination $b_\nu f_\text{astro}$ as a free parameter. Our cross-correlation measurements have very low SNR, so we do not attempt to fit a model for the neutrino redshift distribution $p_\nu(z)$. Instead, we construct three potential models for the neutrino redshift distribution and compute the best fit amplitude $b_\nu f_\text{astro}$ for each template.

\begin{figure}
    \centering
    \includegraphics[width=\linewidth]{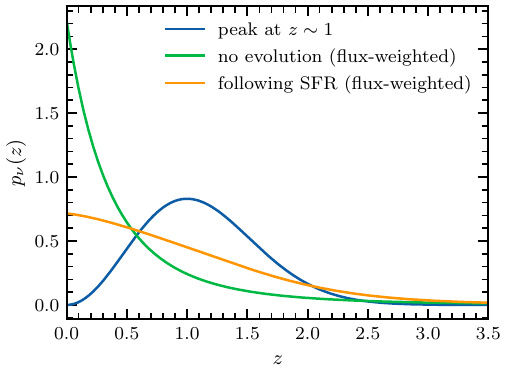}
    \caption{Model neutrino redshift distributions used to fit the measured cross-correlations. The first model is described by \cref{eq:dndz} and represents a neutrino source distribution that peaks at $z\sim1$. The second model includes flux-weighting and assumes a constant co-moving number density of neutrino sources. The last model also includes flux-weighting, but assumes the number density of neutrino sources is proportional to the cosmic star-formation rate density.}
    \label{fig:nu_dndz}
\end{figure}

The first model for the neutrino redshift distribution is inspired by the potential peak at $z\sim1$ seen in \cref{fig:ell-cls} and is based on the parameterization used to model the redshift distribution of galaxies observed by the Vera Rubin Observatory \cite{0912.0201}:
\begin{equation}\label{eq:dndz}
    p_\nu(z) \propto z^2 \exp \left(-z^2\right).
\end{equation}
While this model is potentially able to fit the peak seen in \cref{fig:ell-cls}, a more physically motivated model should include flux-weighting. Unlike samples of galaxies where a galaxy is in the catalog if it is above some detection limit, regardless of how bright it is, a sample of detected neutrinos will be more likely to come from objects with higher fluxes. To model this, we use \verb|FIRESONG|\footnote{\url{https://github.com/icecube/FIRESONG}} \cite{Tung2021} to construct potential realizations of flux-weighted neutrino redshift distributions. These distributions should describe the proportion of detected neutrinos rather than neutrino sources at each redshift. We construct two model distributions using \verb|FIRESONG|. The first assumes that the co-moving number density of neutrino sources is constant across redshift, while the second assumes that the number density of neutrino sources is proportional to the cosmic star formation rate density as parameterized by Madau and Dickinson \cite[Eq. 15]{Madau2014}. Both source distributions are weighted by the observed source fluxes, including the neutrino K-correction, where we have assumed that all sources have a power-law spectrum with index $\gamma = 2.37$ \cite{Abbasi2022} and identical neutrino luminosities. All three neutrino redshift distributions are shown in \cref{fig:nu_dndz}.

Once $p_\nu(z)$ is specified, the model for the galaxy-neutrino cross-spectra becomes fully linear in $b_\nu f_\text{astro}$. For each $p_\nu(z)$ model, we compute a template $\vb{t}$ that represents the model prediction for all of the galaxy-neutrino cross-correlations between $50 < \ell < 350$ evaluated at $b_\nu f_\text{astro} = 1$. We then analytically solve for the best fit value and $1\sigma$ error of $b_\nu f_\text{astro}$:
\begin{align}\label{eq:fit}
    b_\nu f_\text{astro} &= \frac{\vb{d}^T C^{-1} \vb{t}}{\vb{t}^T C^{-1} \vb{t}}, \\
    \label{eq:fit_err}
    \sigma(b_\nu f_\text{astro}) &= \left(\vb{t}^T C^{-1} \vb{t}\right)^{-1/2},
\end{align}
where $\vb{d}$ is the data vector containing all of the measured galaxy-neutrino cross-spectra between $50 < \ell < 350$ and $C$ is the full covariance matrix. \changeone{To confirm that this assumption of Gaussianity is appropriate, we have run a Kolmogorov-Smirnov test on the $b_\nu f_\text{astro}$ values resulting from 25 random neutrino map simulations, and for all $p_\nu(z)$ models, find $p$-values $> 0.4$. We found similarly high $p$-values for the energy-binned neutrino maps that are presented in the following section. Thus, when computing error bars or upper limits on $b_\nu f_\text{astro}$, we simply assume Gaussian distributions in all cases.} We have also ignored any uncertainty in the galaxy or CIB models since the data is completely dominated by noise from the neutrinos.

The results from fitting these three models to the data are summarized in \cref{tab:results}. While all three models are preferred by the data over the null model, the significance is very low and the differences between the $\chi^2$ values for the different models are not large enough to claim a preference for any given model. Again, we see that the peak in the cross-correlations at $z\sim1$ is approximately $2\sigma$ significant.

\begin{table}
    \centering
    \setlength{\tabcolsep}{8pt}
    \renewcommand{\arraystretch}{1.2}
    \begin{tabular}{p{2.5cm} | c c c c}
        Model & $b_\nu f_\text{astro}$ & $\chi^2$ / dof & $\Delta \chi^2$ \\ \hline
        Null model & 0 & 106.2 / 102 &  \\
        Peak at $z\sim1$ & $0.26\pm0.13$ & 101.8 / 101 & -4.4 \\
        No evolution\newline (flux-weighted) & $0.05\pm0.05$ & 105.2 / 101 & -0.9 \\
        Following SFR\newline (flux-weighted) & $0.14\pm0.10$ & 104.1 / 101 & -2.1
    \end{tabular}
    \caption{Results from fitting the amplitude $b_\nu f_\text{astro}$ to the galaxy-neutrino cross-spectra, assuming the three models for $p_\nu(z)$ shown in \cref{fig:nu_dndz}. The last column lists the improvement in the $\chi^2$ value relative to the null model.}
    \label{tab:results}
\end{table}

We have no doubt that higher statistical significance for a positive $b_\nu f_\text{astro}$ could be achieved by carefully tailoring the neutrino redshift distribution $p_\nu(z)$, but without a strong physics motivation this risks a descent into ``p-hacking.'' 

\subsection{Energy binning}\label{sec:bins}
The quantity $f_\text{astro}$ for track-like events is expected to depend very strongly on energy, due to the softer energy spectrum of atmospheric neutrinos compared to the astrophysical ones. \changeone{Thus, to increase our sensitivity,} we divide the neutrino events into three equal-size logarithmic energy \changeone{proxy} bins between $10^{3}$ GeV and $10^{6}$ GeV. These bins have roughly 390,000, 9,700, and 218 events respectively. \changeone{We again emphasize that this energy proxy does not directly correspond to the true neutrino energy (see Ref~\cite{2101.09836}), \changetwo{but higher energy proxies do generally correlate with higher true neutrino energies.}} Based on Ref. \cite{Abbasi2022}, \changeone{which measures the energy spectrum of a somewhat similar event selection to ours (track-like events in the northern sky)}, we expect that $f_\text{astro}$ for each bin to be on the order of 0.006, 0.06, and 0.45 respectively. 

For each energy bin, we construct overdensity maps, estimate the effective beam, and compute cross-spectra with the galaxy samples in the exact same manner as described for the entire neutrino sample.

\begin{figure*}
    \centering
    \includegraphics[width=\linewidth]{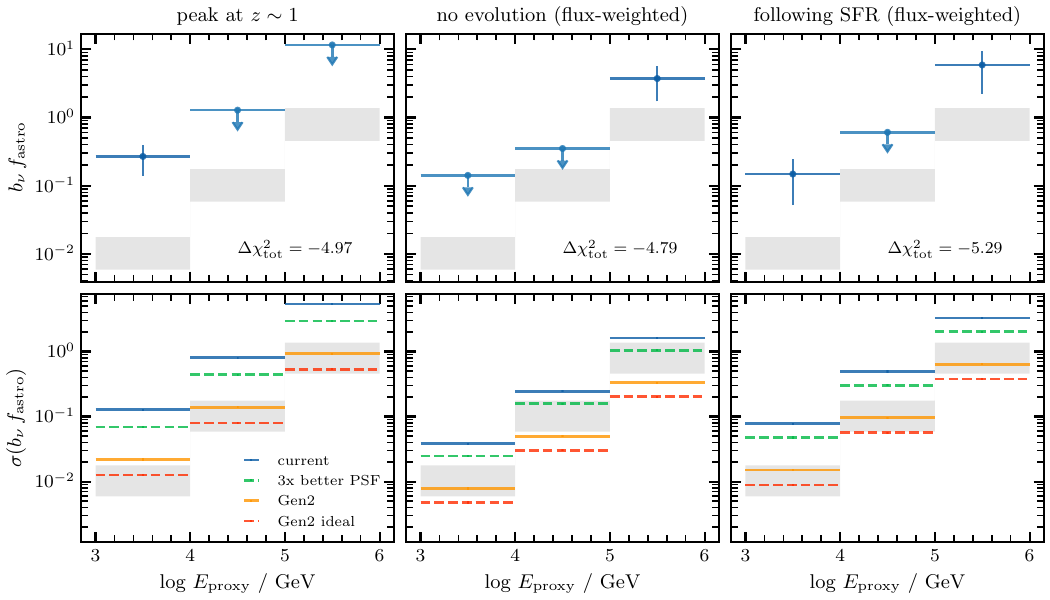}
    \caption{\textbf{Top row:} Energy-binned constraints on the galaxy-neutrino correlation amplitude for three different scenarios with fixed neutrino source redshift distributions. We plot the mean inferred values of $b_\nu f_\text{astro}$ with $1\sigma$ error bars for measurements that are more than $1\sigma$ significant, otherwise we plot 95\% confidence-level upper limits. Note that the significance of these measurements can seem exaggerated due to the logarithmic axis scaling. For each model we also report the total improvement in the $\chi^2$ value compared to the null model, assuming that the energy bins are all independent. \textbf{Bottom row:} Forecasts for future measurements of galaxy-neutrino cross-correlations. We plot the sizes of $1\sigma$ error bars obtained in this work together with 3 potential future scenarios that are described in the text. In all panels, the grey shaded regions indicate the estimated astrophysical fraction $f_\text{astro}$ based on the energy spectrum measurement from Ref. \cite{Abbasi2022} with $1 \le b_\nu \le 3$ (roughly plausible values for the linear bias of different source populations).}
    \label{fig:constraints_forecasts}
\end{figure*}

We again use the three assumed models for the neutrino redshift distribution and plot the resulting constraints on $b_\nu f_\text{astro}$ for the three energy bins in the top row of \cref{fig:constraints_forecasts} and compare them to the expected values of $f_\text{astro}$ from energy spectrum measurements. We find that in the simple ``peak at $z\sim1$" model, the $2\sigma$ significant correlation between neutrinos and LSS is driven entirely by the lowest energy bin, potentially in slight conflict with expectations based on the energy spectrum measurement of $f_\text{astro}$. On the other hand, in the flux-weighted models we get potential detections coming from the high energy events. In the ``no evolution" (``following SFR") model, the lowest and highest energy bins are significant at $1\sigma$ and $1.9\sigma$ ($1.5\sigma$ and $1.6\sigma$) respectively. All of these significance values are low and the $\Delta \chi^2$ values for each of the models are not different enough to prefer one of them over the other. At this point, it is unclear whether we have found a real correlation between neutrinos and LSS or simply a noise fluctuation. 

\subsection{Forecasts}
To get a sense of what can be achieved with cross-correlations between neutrinos and LSS, we forecast the SNR achievable in an ideal cross-correlation analysis using data from a future detector. For these forecasts, we assume that all shot noise is Poissonian, or $A_\text{SN} = 1 / \bar{n}$, where $\bar{n}$ is the average number density over the sky of objects being considered. We calculate the covariance between power spectra assuming that all fields are Gaussian and ignoring mask-induced mode coupling:
\begin{equation}
    {\rm Cov} \left(C^{g\nu}_\ell, C^{g'\nu}_{\ell'}\right) = \delta_{\ell\ell'} \frac{C^{g\nu}_\ell C^{g'\nu}_\ell + C^{gg'}_\ell C^{\nu\nu}_\ell}{(2\ell+1) f_\text{sky} B_\ell^2},
\end{equation}
where $f_\text{sky}$ is the fraction of sky covered by both the galaxy and neutrino masks and $B_\ell$ is the effective beam for the neutrino sample being considered.

For simplicity we only consider the galaxy catalogs, excluding the CIB maps, when calculating forecasts, since they provide almost all of the constraining power in this analysis, and use the effective bias values reported in \cref{tab:galaxies}. For the neutrinos, we use the same three redshift distribution models described in \cref{sec:model} and assume a linear bias of $b_\nu = 1$. The fraction of neutrinos that are of astrophysical origin is heavily dependent on the specific selection criteria used to build the neutrino sample. Since we are using track-like events in the northern sky, we assume that $f_\text{astro}$ should be reasonably similar to the measurements by Ref. \cite{Abbasi2022} of the diffuse neutrino flux using track-like events in the northern sky.

We consider three possible scenarios for future neutrino detectors. The first scenario assumes the same number of detected events as the current 10-year dataset, but with 3 times better angular uncertainties. The second scenario, ``Gen2", assumes 10 times as many events with 3 times better angular uncertainties (roughly in line with what is expected from IceCube-Gen2 \cite{2008.04323}). Finally, the third scenario, ``Gen2 ideal", assumes 10 times as many events with essentially no angular uncertainties. We then calculate the expected errors on the measurements of $b_\nu f_\text{astro}$, $\sigma(b_\nu f_\text{astro})$, assuming ideal measurements of the cross-correlations out to $\ell_\text{max} = 750$. As a check, we confirm that a forecast of the expected constraining power of current data matches the constraints we got in the analysis of actual data. 

We plot the resulting forecasts of $\sigma(b_\nu f_\text{astro})$  compared with the errors obtained in the current analysis in the bottom row of \cref{fig:constraints_forecasts}. As expected, in the noise-dominated regime, we find that $\sigma(b_\nu f_\text{astro})$ is independent of $b_\nu f_\text{astro}$ and scales linearly with $1/\sqrt{N_\nu}$, where $N_\nu$ is the total number of neutrinos in the sample. We find that a Gen2-like detector should provide measurements of the galaxy-neutrino cross-spectra with roughly 5 to 6 times better SNR than the current measurements, which should be sufficient to probe interesting values of $b_\nu f_\text{astro}$. In addition, future data would provide significantly better power to discriminate between different models of the neutrino redshift distribution.

\section{Discussion}\label{sec:discussion}
We have done a detailed analysis of the cross-correlation between IceCube neutrinos and large-scale structure around redshift $z=1$. This is the first time such an analysis has been carried out using the 10-year public IceCube data and a large number of galaxy samples. We do not find a significant signal, but we do see tantalizing hints of a positive correlation at roughly the $2\sigma$ level. The main limiting factor in this analysis is the very high level of noise present in the neutrino maps, but improved modelling of the IceCube PSF (especially its uncertainty) will also be necessary to determine if this is a physical result or a noise fluctuation.

We have parameterized the amplitude of neutrino clustering in terms of the neutrino source linear bias $b_\nu$ and the astrophysical fraction of neutrinos $f_\text{astro}$, which are completely degenerate in this cross-correlation analysis. The degeneracy between $b_\nu$ and $f_\text{astro}$ can be broken by including information about the neutrino energy spectrum, which allows direct measurements of $f_\text{astro}$. In theory it should be possible to do a simultaneous fit to the energy spectrum and the cross-spectra with LSS for a given neutrino event selection in order to get constraints on both $f_\text{astro}$ and the linear bias of neutrino sources $b_\nu$.

In our analysis we are not able to meaningfully constrain the redshift distribution of astrophysical neutrinos. Physical models of neutrino production in astrophysical sources could help put theoretical priors on the neutrino source redshift distribution. This, in turn, will allow better constraints on $b_\nu$ as well as on the shape of the neutrino redshift distribution $p_\nu(z)$. Ultimately, this will allow us to pin down the astrophysical populations that build up the diffuse astrophysical neutrino flux. This approach is also very similar to that taken by Ref. \cite{Muzio2023}, where the authors drew the connection between neutrino sky anisotropies and a possible cutoff in the energy spectrum of ultrahigh energy neutrinos.

While the current analysis of neutrino-galaxy cross-correlations is severely limited by the quality of the neutrino data, there are some improvements that could be done on the galaxy side as well. Given that the flux-weighted neutrino distributions are heavily skewed to low-redshift (\cref{fig:nu_dndz}), detailed analysis using low-redshift galaxy samples, especially the DESI Bright Galaxy Sample \cite{Hahn2023}, will be important to discriminate between these models. At low redshifts, though, more careful modelling than what we have done will be necessary in order to account for the more non-linear and non-Gaussian galaxy fields. While we have used a few low-redshift samples in this work, the majority of our galaxy samples are at significantly higher redshifts where the simple linear bias model should be adequate on the scales we have considered. In the future it will also be necessary to do a full combined fit for the galaxy-neutrino cross-correlations, taking into account the uncertainties in the galaxy data.

We have also presented forecasts for future measurements of $b_\nu f_\text{astro}$. In the future, with data from the IceCube-Gen2 detector, cross-correlations with LSS have very significant potential to constrain the properties of neutrino sources at low to medium redshift. Together with on-going searches for neutrino point-sources, cross-correlations with LSS should be able to provide complementary constraints on the properties of neutrino source populations.

\begin{acknowledgments}
We thank Alex Krolewski for sharing the unWISE galaxy maps and providing comments on our manuscript.
We also thank Abhishek Maniyar and Ziang Yan for guidance with the CIB model. We thank the anonymous referee for providing comments that helped improve this paper. This work was partially supported by the Center for AstroPhysical Surveys (CAPS) at the National Center for Supercomputing Applications (NCSA), University of Illinois Urbana-Champaign. A.O. acknowledges support from the Illinois Center for Advanced Studies of the Universe (ICASU)/CAPS/NCSA Graduate Fellowship. This work made use of the Illinois Campus Cluster, a computing resource that is operated by the Illinois Campus Cluster Program (ICCP) in conjunction with the National Center for Supercomputing Applications (NCSA) and which is supported by funds from the University of Illinois at Urbana-Champaign.

This research has made use of NASA's Astrophysics Data System and \verb|adstex| (\url{https://github.com/yymao/adstex}). We also acknowledge the open-source software that was invaluable in this research, including: Python, \verb|NumPy| \cite{Harris2020}, \verb|SciPy| \cite{Virtanen2020}, and \verb|matplotlib| \cite{Hunter2007}.
\end{acknowledgments}

\appendix
\section{Theory curves for cross-spectra}\label{sec:app}
\begin{figure*}
    \centering
    \includegraphics[width=\linewidth]{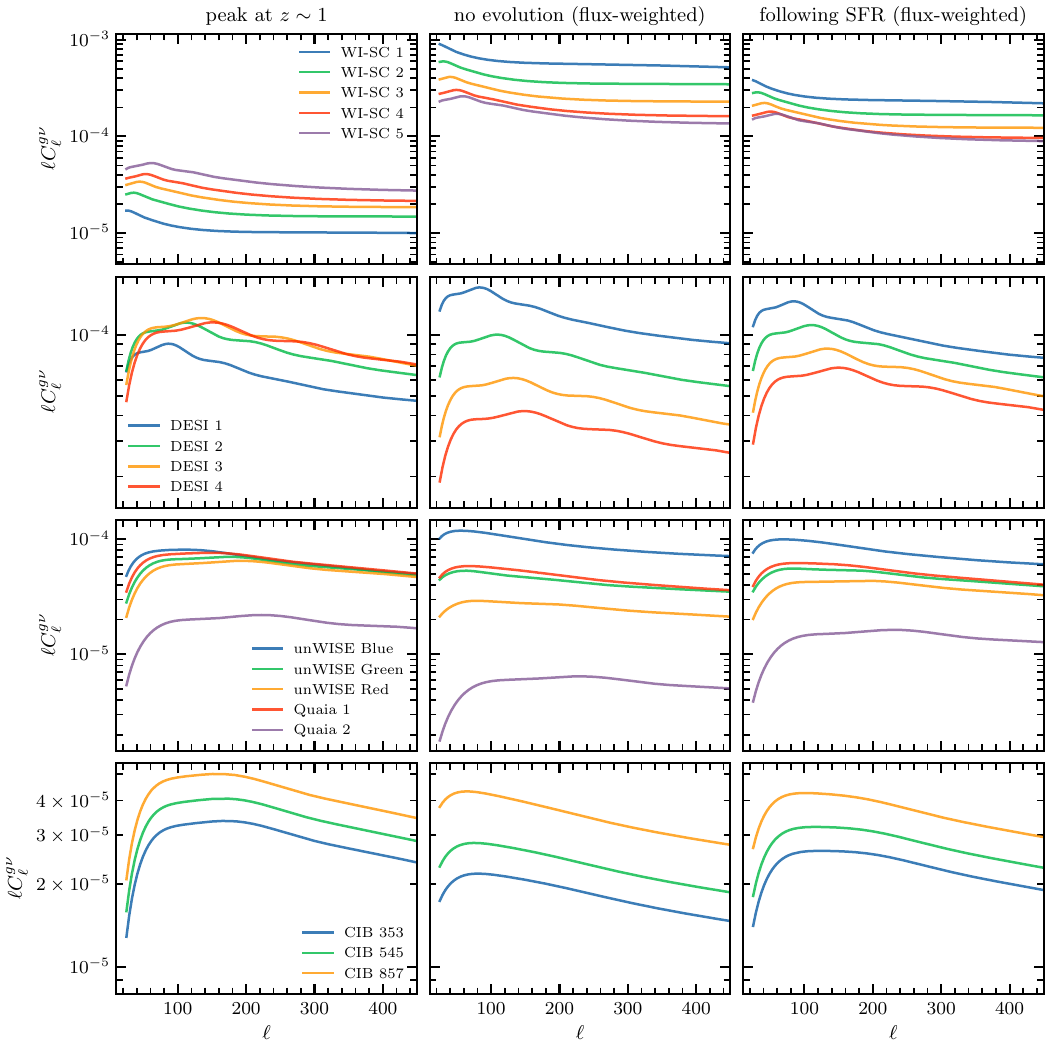}
    \caption{Theory curves (un-binned) for the neutrino-galaxy cross-spectra $C_\ell^{g\nu}$. Each column corresponds to a different model for the neutrino redshift distribution (as given in Section~\ref{sec:model}), while each row corresponds to a different set of galaxy samples as given in the legends. These curves are all computed assuming $b_\nu f_\text{astro} = 1$ and linear bias for the galaxies as given in Table~\ref{tab:galaxies}. Scanning across the columns, it is easy to see how changing the model for the neutrino redshift distribution significantly impacts the relative amplitudes of the cross-spectra.}
    \label{fig:theory}
\end{figure*}

In Fig.~\ref{fig:theory} we show all of the theory curves for the neutrino-galaxy cross-spectra computed for each of the model neutrino redshift distributions that we have tested. These curves are calculated for $b_\nu f_\text{astro} = 1$ so that they can be used as model templates in Eqs.~\ref{eq:fit} and \ref{eq:fit_err}.

\bibliography{references}

\begin{thebibliography}{57}%
\makeatletter
\providecommand \@ifxundefined [1]{%
 \@ifx{#1\undefined}
}%
\providecommand \@ifnum [1]{%
 \ifnum #1\expandafter \@firstoftwo
 \else \expandafter \@secondoftwo
 \fi
}%
\providecommand \@ifx [1]{%
 \ifx #1\expandafter \@firstoftwo
 \else \expandafter \@secondoftwo
 \fi
}%
\providecommand \natexlab [1]{#1}%
\providecommand \enquote  [1]{``#1''}%
\providecommand \bibnamefont  [1]{#1}%
\providecommand \bibfnamefont [1]{#1}%
\providecommand \citenamefont [1]{#1}%
\providecommand \href@noop [0]{\@secondoftwo}%
\providecommand \href [0]{\begingroup \@sanitize@url \@href}%
\providecommand \@href[1]{\@@startlink{#1}\@@href}%
\providecommand \@@href[1]{\endgroup#1\@@endlink}%
\providecommand \@sanitize@url [0]{\catcode `\\12\catcode `\$12\catcode `\&12\catcode `\#12\catcode `\^12\catcode `\_12\catcode `\%12\relax}%
\providecommand \@@startlink[1]{}%
\providecommand \@@endlink[0]{}%
\providecommand \url  [0]{\begingroup\@sanitize@url \@url }%
\providecommand \@url [1]{\endgroup\@href {#1}{\urlprefix }}%
\providecommand \urlprefix  [0]{URL }%
\providecommand \Eprint [0]{\href }%
\providecommand \doibase [0]{https://doi.org/}%
\providecommand \selectlanguage [0]{\@gobble}%
\providecommand \bibinfo  [0]{\@secondoftwo}%
\providecommand \bibfield  [0]{\@secondoftwo}%
\providecommand \translation [1]{[#1]}%
\providecommand \BibitemOpen [0]{}%
\providecommand \bibitemStop [0]{}%
\providecommand \bibitemNoStop [0]{.\EOS\space}%
\providecommand \EOS [0]{\spacefactor3000\relax}%
\providecommand \BibitemShut  [1]{\csname bibitem#1\endcsname}%
\let\auto@bib@innerbib\@empty
\bibitem [{\citenamefont {{Aartsen}}\ \emph {et~al.}(2013)\citenamefont {{Aartsen}}, \citenamefont {{Abbasi}}, \citenamefont {{Abdou}}, \citenamefont {{Ackermann}}, \citenamefont {{Adams}}, \citenamefont {{Aguilar}}, \citenamefont {{Ahlers}}, \citenamefont {{Altmann}}, \citenamefont {{Auffenberg}}, \citenamefont {{Bai}},\ and\ \citenamefont {et~al.}}]{Aartsen2013}%
  \BibitemOpen
  \bibfield  {author} {\bibinfo {author} {\bibfnamefont {M.~G.}\ \bibnamefont {{Aartsen}}}, \bibinfo {author} {\bibfnamefont {R.}~\bibnamefont {{Abbasi}}}, \bibinfo {author} {\bibfnamefont {Y.}~\bibnamefont {{Abdou}}}, \bibinfo {author} {\bibfnamefont {M.}~\bibnamefont {{Ackermann}}}, \bibinfo {author} {\bibfnamefont {J.}~\bibnamefont {{Adams}}}, \bibinfo {author} {\bibfnamefont {J.~A.}\ \bibnamefont {{Aguilar}}}, \bibinfo {author} {\bibfnamefont {M.}~\bibnamefont {{Ahlers}}}, \bibinfo {author} {\bibfnamefont {D.}~\bibnamefont {{Altmann}}}, \bibinfo {author} {\bibfnamefont {J.}~\bibnamefont {{Auffenberg}}}, \bibinfo {author} {\bibfnamefont {X.}~\bibnamefont {{Bai}}},\ and\ \bibinfo {author} {\bibnamefont {et~al.}},\ }\bibfield  {title} {\bibinfo {title} {{First Observation of PeV-Energy Neutrinos with IceCube}},\ }\href {https://doi.org/10.1103/PhysRevLett.111.021103} {\bibfield  {journal} {\bibinfo  {journal} {\prl}\ }\textbf {\bibinfo {volume} {111}},\ \bibinfo {eid} {021103} (\bibinfo {year} {2013})},\
  \Eprint {https://arxiv.org/abs/1304.5356} {arXiv:1304.5356 [astro-ph.HE]} \BibitemShut {NoStop}%
\bibitem [{\citenamefont {{Abbasi}}\ \emph {et~al.}(2022{\natexlab{a}})\citenamefont {{Abbasi}}, \citenamefont {{Ackermann}}, \citenamefont {{Adams}}, \citenamefont {{Aguilar}}, \citenamefont {{Ahlers}}, \citenamefont {{Ahrens}}, \citenamefont {{Alameddine}}, \citenamefont {{Alispach}}, \citenamefont {{Alves}}, \citenamefont {{Amin}},\ and\ \citenamefont {et~al.}}]{Abbasi2022}%
  \BibitemOpen
  \bibfield  {author} {\bibinfo {author} {\bibfnamefont {R.}~\bibnamefont {{Abbasi}}}, \bibinfo {author} {\bibfnamefont {M.}~\bibnamefont {{Ackermann}}}, \bibinfo {author} {\bibfnamefont {J.}~\bibnamefont {{Adams}}}, \bibinfo {author} {\bibfnamefont {J.~A.}\ \bibnamefont {{Aguilar}}}, \bibinfo {author} {\bibfnamefont {M.}~\bibnamefont {{Ahlers}}}, \bibinfo {author} {\bibfnamefont {M.}~\bibnamefont {{Ahrens}}}, \bibinfo {author} {\bibfnamefont {J.~M.}\ \bibnamefont {{Alameddine}}}, \bibinfo {author} {\bibfnamefont {C.}~\bibnamefont {{Alispach}}}, \bibinfo {author} {\bibfnamefont {J.}~\bibnamefont {{Alves}}, \bibfnamefont {A.~A.}}, \bibinfo {author} {\bibfnamefont {N.~M.}\ \bibnamefont {{Amin}}},\ and\ \bibinfo {author} {\bibnamefont {et~al.}},\ }\bibfield  {title} {\bibinfo {title} {{Improved Characterization of the Astrophysical Muon-neutrino Flux with 9.5 Years of IceCube Data}},\ }\href {https://doi.org/10.3847/1538-4357/ac4d29} {\bibfield  {journal} {\bibinfo  {journal} {\apj}\ }\textbf {\bibinfo {volume}
  {928}},\ \bibinfo {eid} {50} (\bibinfo {year} {2022}{\natexlab{a}})},\ \Eprint {https://arxiv.org/abs/2111.10299} {arXiv:2111.10299 [astro-ph.HE]} \BibitemShut {NoStop}%
\bibitem [{\citenamefont {{Abbasi}}\ \emph {et~al.}(2024{\natexlab{a}})\citenamefont {{Abbasi}}, \citenamefont {{Ackermann}}, \citenamefont {{Adams}}, \citenamefont {{Agarwalla}}, \citenamefont {{Aguilar}}, \citenamefont {{Ahlers}}, \citenamefont {{Alameddine}}, \citenamefont {{Amin}}, \citenamefont {{Andeen}}, \citenamefont {{Anton}},\ and\ \citenamefont {et~al.}}]{2402.18026}%
  \BibitemOpen
  \bibfield  {author} {\bibinfo {author} {\bibfnamefont {R.}~\bibnamefont {{Abbasi}}}, \bibinfo {author} {\bibfnamefont {M.}~\bibnamefont {{Ackermann}}}, \bibinfo {author} {\bibfnamefont {J.}~\bibnamefont {{Adams}}}, \bibinfo {author} {\bibfnamefont {S.~K.}\ \bibnamefont {{Agarwalla}}}, \bibinfo {author} {\bibfnamefont {J.~A.}\ \bibnamefont {{Aguilar}}}, \bibinfo {author} {\bibfnamefont {M.}~\bibnamefont {{Ahlers}}}, \bibinfo {author} {\bibfnamefont {J.~M.}\ \bibnamefont {{Alameddine}}}, \bibinfo {author} {\bibfnamefont {N.~M.}\ \bibnamefont {{Amin}}}, \bibinfo {author} {\bibfnamefont {K.}~\bibnamefont {{Andeen}}}, \bibinfo {author} {\bibfnamefont {G.}~\bibnamefont {{Anton}}},\ and\ \bibinfo {author} {\bibnamefont {et~al.}},\ }\bibfield  {title} {\bibinfo {title} {{Characterization of the astrophysical diffuse neutrino flux using starting track events in IceCube}},\ }\href {https://doi.org/10.1103/PhysRevD.110.022001} {\bibfield  {journal} {\bibinfo  {journal} {\prd}\ }\textbf {\bibinfo {volume} {110}},\
  \bibinfo {eid} {022001} (\bibinfo {year} {2024}{\natexlab{a}})},\ \Eprint {https://arxiv.org/abs/2402.18026} {arXiv:2402.18026 [astro-ph.HE]} \BibitemShut {NoStop}%
\bibitem [{\citenamefont {{Abbasi}}\ \emph {et~al.}(2024{\natexlab{b}})\citenamefont {{Abbasi}}, \citenamefont {{Ackermann}}, \citenamefont {{Adams}}, \citenamefont {{Agarwalla}}, \citenamefont {{Aguilar}}, \citenamefont {{Ahlers}}, \citenamefont {{Alameddine}}, \citenamefont {{Amin}}, \citenamefont {{Andeen}}, \citenamefont {{Anton}},\ and\ \citenamefont {et~al.}}]{2403.02516}%
  \BibitemOpen
  \bibfield  {author} {\bibinfo {author} {\bibfnamefont {R.}~\bibnamefont {{Abbasi}}}, \bibinfo {author} {\bibfnamefont {M.}~\bibnamefont {{Ackermann}}}, \bibinfo {author} {\bibfnamefont {J.}~\bibnamefont {{Adams}}}, \bibinfo {author} {\bibfnamefont {S.~K.}\ \bibnamefont {{Agarwalla}}}, \bibinfo {author} {\bibfnamefont {J.~A.}\ \bibnamefont {{Aguilar}}}, \bibinfo {author} {\bibfnamefont {M.}~\bibnamefont {{Ahlers}}}, \bibinfo {author} {\bibfnamefont {J.~M.}\ \bibnamefont {{Alameddine}}}, \bibinfo {author} {\bibfnamefont {N.~M.}\ \bibnamefont {{Amin}}}, \bibinfo {author} {\bibfnamefont {K.}~\bibnamefont {{Andeen}}}, \bibinfo {author} {\bibfnamefont {G.}~\bibnamefont {{Anton}}},\ and\ \bibinfo {author} {\bibnamefont {et~al.}},\ }\bibfield  {title} {\bibinfo {title} {{Observation of Seven Astrophysical Tau Neutrino Candidates with IceCube}},\ }\href {https://doi.org/10.1103/PhysRevLett.132.151001} {\bibfield  {journal} {\bibinfo  {journal} {\prl}\ }\textbf {\bibinfo {volume} {132}},\ \bibinfo {eid} {151001}
  (\bibinfo {year} {2024}{\natexlab{b}})},\ \Eprint {https://arxiv.org/abs/2403.02516} {arXiv:2403.02516 [astro-ph.HE]} \BibitemShut {NoStop}%
\bibitem [{\citenamefont {{IceCube Collaboration}}\ \emph {et~al.}(2018)\citenamefont {{IceCube Collaboration}}, \citenamefont {{Aartsen}}, \citenamefont {{Ackermann}}, \citenamefont {{Adams}}, \citenamefont {{Aguilar}}, \citenamefont {{Ahlers}}, \citenamefont {{Ahrens}}, \citenamefont {{Samarai}}, \citenamefont {{Altmann}}, \citenamefont {{Andeen}},\ and\ \citenamefont {et~al.}}]{1807.08794}%
  \BibitemOpen
  \bibfield  {author} {\bibinfo {author} {\bibnamefont {{IceCube Collaboration}}}, \bibinfo {author} {\bibfnamefont {M.~G.}\ \bibnamefont {{Aartsen}}}, \bibinfo {author} {\bibfnamefont {M.}~\bibnamefont {{Ackermann}}}, \bibinfo {author} {\bibfnamefont {J.}~\bibnamefont {{Adams}}}, \bibinfo {author} {\bibfnamefont {J.~A.}\ \bibnamefont {{Aguilar}}}, \bibinfo {author} {\bibfnamefont {M.}~\bibnamefont {{Ahlers}}}, \bibinfo {author} {\bibfnamefont {M.}~\bibnamefont {{Ahrens}}}, \bibinfo {author} {\bibfnamefont {I.~A.}\ \bibnamefont {{Samarai}}}, \bibinfo {author} {\bibfnamefont {D.}~\bibnamefont {{Altmann}}}, \bibinfo {author} {\bibfnamefont {K.}~\bibnamefont {{Andeen}}},\ and\ \bibinfo {author} {\bibnamefont {et~al.}},\ }\bibfield  {title} {\bibinfo {title} {{Neutrino emission from the direction of the blazar TXS 0506+056 prior to the IceCube-170922A alert}},\ }\href {https://doi.org/10.1126/science.aat2890} {\bibfield  {journal} {\bibinfo  {journal} {Science}\ }\textbf {\bibinfo {volume} {361}},\ \bibinfo
  {pages} {147} (\bibinfo {year} {2018})},\ \Eprint {https://arxiv.org/abs/1807.08794} {arXiv:1807.08794 [astro-ph.HE]} \BibitemShut {NoStop}%
\bibitem [{\citenamefont {{IceCube Collaboration}}\ \emph {et~al.}(2022)\citenamefont {{IceCube Collaboration}}, \citenamefont {{Abbasi}}, \citenamefont {{Ackermann}}, \citenamefont {{Adams}}, \citenamefont {{Aguilar}}, \citenamefont {{Ahlers}}, \citenamefont {{Ahrens}}, \citenamefont {{Alameddine}}, \citenamefont {{Alispach}}, \citenamefont {{Alves}},\ and\ \citenamefont {et~al.}}]{2211.09972}%
  \BibitemOpen
  \bibfield  {author} {\bibinfo {author} {\bibnamefont {{IceCube Collaboration}}}, \bibinfo {author} {\bibfnamefont {R.}~\bibnamefont {{Abbasi}}}, \bibinfo {author} {\bibfnamefont {M.}~\bibnamefont {{Ackermann}}}, \bibinfo {author} {\bibfnamefont {J.}~\bibnamefont {{Adams}}}, \bibinfo {author} {\bibfnamefont {J.~A.}\ \bibnamefont {{Aguilar}}}, \bibinfo {author} {\bibfnamefont {M.}~\bibnamefont {{Ahlers}}}, \bibinfo {author} {\bibfnamefont {M.}~\bibnamefont {{Ahrens}}}, \bibinfo {author} {\bibfnamefont {J.~M.}\ \bibnamefont {{Alameddine}}}, \bibinfo {author} {\bibfnamefont {C.}~\bibnamefont {{Alispach}}}, \bibinfo {author} {\bibfnamefont {J.}~\bibnamefont {{Alves}}, \bibfnamefont {A.~A.}},\ and\ \bibinfo {author} {\bibnamefont {et~al.}},\ }\bibfield  {title} {\bibinfo {title} {{Evidence for neutrino emission from the nearby active galaxy NGC 1068}},\ }\href {https://doi.org/10.1126/science.abg3395} {\bibfield  {journal} {\bibinfo  {journal} {Science}\ }\textbf {\bibinfo {volume} {378}},\ \bibinfo {pages} {538}
  (\bibinfo {year} {2022})},\ \Eprint {https://arxiv.org/abs/2211.09972} {arXiv:2211.09972 [astro-ph.HE]} \BibitemShut {NoStop}%
\bibitem [{\citenamefont {{Icecube Collaboration}}\ \emph {et~al.}(2023)\citenamefont {{Icecube Collaboration}}, \citenamefont {{Abbasi}}, \citenamefont {{Ackermann}}, \citenamefont {{Adams}}, \citenamefont {{Aguilar}}, \citenamefont {{Ahlers}}, \citenamefont {{Ahrens}}, \citenamefont {{Alameddine}}, \citenamefont {{Alves}}, \citenamefont {{Amin}},\ and\ \citenamefont {et~al.}}]{2307.04427}%
  \BibitemOpen
  \bibfield  {author} {\bibinfo {author} {\bibnamefont {{Icecube Collaboration}}}, \bibinfo {author} {\bibfnamefont {R.}~\bibnamefont {{Abbasi}}}, \bibinfo {author} {\bibfnamefont {M.}~\bibnamefont {{Ackermann}}}, \bibinfo {author} {\bibfnamefont {J.}~\bibnamefont {{Adams}}}, \bibinfo {author} {\bibfnamefont {J.~A.}\ \bibnamefont {{Aguilar}}}, \bibinfo {author} {\bibfnamefont {M.}~\bibnamefont {{Ahlers}}}, \bibinfo {author} {\bibfnamefont {M.}~\bibnamefont {{Ahrens}}}, \bibinfo {author} {\bibfnamefont {J.~M.}\ \bibnamefont {{Alameddine}}}, \bibinfo {author} {\bibfnamefont {A.~A.}\ \bibnamefont {{Alves}}}, \bibinfo {author} {\bibfnamefont {N.~M.}\ \bibnamefont {{Amin}}},\ and\ \bibinfo {author} {\bibnamefont {et~al.}},\ }\bibfield  {title} {\bibinfo {title} {{Observation of high-energy neutrinos from the Galactic plane}},\ }\href {https://doi.org/10.1126/science.adc9818} {\bibfield  {journal} {\bibinfo  {journal} {Science}\ }\textbf {\bibinfo {volume} {380}},\ \bibinfo {pages} {1338} (\bibinfo {year}
  {2023})},\ \Eprint {https://arxiv.org/abs/2307.04427} {arXiv:2307.04427 [astro-ph.HE]} \BibitemShut {NoStop}%
\bibitem [{\citenamefont {{Abbasi}}\ \emph {et~al.}(2022{\natexlab{b}})\citenamefont {{Abbasi}}, \citenamefont {{Ackermann}}, \citenamefont {{Adams}}, \citenamefont {{Aguilar}}, \citenamefont {{Ahlers}}, \citenamefont {{Ahrens}}, \citenamefont {{Alameddine}}, \citenamefont {{Alves}}, \citenamefont {{Amin}}, \citenamefont {{Andeen}},\ and\ \citenamefont {et~al.}}]{2207.04946}%
  \BibitemOpen
  \bibfield  {author} {\bibinfo {author} {\bibfnamefont {R.}~\bibnamefont {{Abbasi}}}, \bibinfo {author} {\bibfnamefont {M.}~\bibnamefont {{Ackermann}}}, \bibinfo {author} {\bibfnamefont {J.}~\bibnamefont {{Adams}}}, \bibinfo {author} {\bibfnamefont {J.~A.}\ \bibnamefont {{Aguilar}}}, \bibinfo {author} {\bibfnamefont {M.}~\bibnamefont {{Ahlers}}}, \bibinfo {author} {\bibfnamefont {M.}~\bibnamefont {{Ahrens}}}, \bibinfo {author} {\bibfnamefont {J.~M.}\ \bibnamefont {{Alameddine}}}, \bibinfo {author} {\bibfnamefont {A.~A.}\ \bibnamefont {{Alves}}}, \bibinfo {author} {\bibfnamefont {N.~M.}\ \bibnamefont {{Amin}}}, \bibinfo {author} {\bibfnamefont {K.}~\bibnamefont {{Andeen}}},\ and\ \bibinfo {author} {\bibnamefont {et~al.}},\ }\bibfield  {title} {\bibinfo {title} {{Search for Astrophysical Neutrinos from 1FLE Blazars with IceCube}},\ }\href {https://doi.org/10.3847/1538-4357/ac8de4} {\bibfield  {journal} {\bibinfo  {journal} {\apj}\ }\textbf {\bibinfo {volume} {938}},\ \bibinfo {eid} {38} (\bibinfo {year}
  {2022}{\natexlab{b}})},\ \Eprint {https://arxiv.org/abs/2207.04946} {arXiv:2207.04946 [astro-ph.HE]} \BibitemShut {NoStop}%
\bibitem [{\citenamefont {{Abbasi}}\ \emph {et~al.}(2022{\natexlab{c}})\citenamefont {{Abbasi}}, \citenamefont {{Ackermann}}, \citenamefont {{Adams}}, \citenamefont {{Aguilar}}, \citenamefont {{Ahlers}}, \citenamefont {{Ahrens}}, \citenamefont {{Alameddine}}, \citenamefont {{Alispach}}, \citenamefont {{Alves}}, \citenamefont {{Amin}},\ and\ \citenamefont {et~al.}}]{2111.10169}%
  \BibitemOpen
  \bibfield  {author} {\bibinfo {author} {\bibfnamefont {R.}~\bibnamefont {{Abbasi}}}, \bibinfo {author} {\bibfnamefont {M.}~\bibnamefont {{Ackermann}}}, \bibinfo {author} {\bibfnamefont {J.}~\bibnamefont {{Adams}}}, \bibinfo {author} {\bibfnamefont {J.~A.}\ \bibnamefont {{Aguilar}}}, \bibinfo {author} {\bibfnamefont {M.}~\bibnamefont {{Ahlers}}}, \bibinfo {author} {\bibfnamefont {M.}~\bibnamefont {{Ahrens}}}, \bibinfo {author} {\bibfnamefont {J.~M.}\ \bibnamefont {{Alameddine}}}, \bibinfo {author} {\bibfnamefont {C.}~\bibnamefont {{Alispach}}}, \bibinfo {author} {\bibfnamefont {A.~A.}\ \bibnamefont {{Alves}}}, \bibinfo {author} {\bibfnamefont {N.~M.}\ \bibnamefont {{Amin}}},\ and\ \bibinfo {author} {\bibnamefont {et~al.}},\ }\bibfield  {title} {\bibinfo {title} {{Search for neutrino emission from cores of active galactic nuclei}},\ }\href {https://doi.org/10.1103/PhysRevD.106.022005} {\bibfield  {journal} {\bibinfo  {journal} {\prd}\ }\textbf {\bibinfo {volume} {106}},\ \bibinfo {eid} {022005} (\bibinfo
  {year} {2022}{\natexlab{c}})},\ \Eprint {https://arxiv.org/abs/2111.10169} {arXiv:2111.10169 [astro-ph.HE]} \BibitemShut {NoStop}%
\bibitem [{\citenamefont {{Abbasi}}\ \emph {et~al.}(2022{\natexlab{d}})\citenamefont {{Abbasi}}, \citenamefont {{Ackermann}}, \citenamefont {{Adams}}, \citenamefont {{Aguilar}}, \citenamefont {{Ahlers}}, \citenamefont {{Ahrens}}, \citenamefont {{Alispach}}, \citenamefont {{Alves}}, \citenamefont {{Amin}}, \citenamefont {{An}},\ and\ \citenamefont {et~al.}}]{2107.03149}%
  \BibitemOpen
  \bibfield  {author} {\bibinfo {author} {\bibfnamefont {R.}~\bibnamefont {{Abbasi}}}, \bibinfo {author} {\bibfnamefont {M.}~\bibnamefont {{Ackermann}}}, \bibinfo {author} {\bibfnamefont {J.}~\bibnamefont {{Adams}}}, \bibinfo {author} {\bibfnamefont {J.~A.}\ \bibnamefont {{Aguilar}}}, \bibinfo {author} {\bibfnamefont {M.}~\bibnamefont {{Ahlers}}}, \bibinfo {author} {\bibfnamefont {M.}~\bibnamefont {{Ahrens}}}, \bibinfo {author} {\bibfnamefont {C.}~\bibnamefont {{Alispach}}}, \bibinfo {author} {\bibfnamefont {A.~A.}\ \bibnamefont {{Alves}}}, \bibinfo {author} {\bibfnamefont {N.~M.}\ \bibnamefont {{Amin}}}, \bibinfo {author} {\bibfnamefont {R.}~\bibnamefont {{An}}},\ and\ \bibinfo {author} {\bibnamefont {et~al.}},\ }\bibfield  {title} {\bibinfo {title} {{Search for High-energy Neutrinos from Ultraluminous Infrared Galaxies with IceCube}},\ }\href {https://doi.org/10.3847/1538-4357/ac3cb6} {\bibfield  {journal} {\bibinfo  {journal} {\apj}\ }\textbf {\bibinfo {volume} {926}},\ \bibinfo {eid} {59} (\bibinfo {year}
  {2022}{\natexlab{d}})},\ \Eprint {https://arxiv.org/abs/2107.03149} {arXiv:2107.03149 [astro-ph.HE]} \BibitemShut {NoStop}%
\bibitem [{\citenamefont {{Abbasi}}\ \emph {et~al.}(2022{\natexlab{e}})\citenamefont {{Abbasi}}, \citenamefont {{Ackermann}}, \citenamefont {{Adams}}, \citenamefont {{Aguilar}}, \citenamefont {{Ahlers}}, \citenamefont {{Ahrens}}, \citenamefont {{Alameddine}}, \citenamefont {{Alves}}, \citenamefont {{Amin}}, \citenamefont {{Andeen}},\ and\ \citenamefont {et~al.}}]{2206.02054}%
  \BibitemOpen
  \bibfield  {author} {\bibinfo {author} {\bibfnamefont {R.}~\bibnamefont {{Abbasi}}}, \bibinfo {author} {\bibfnamefont {M.}~\bibnamefont {{Ackermann}}}, \bibinfo {author} {\bibfnamefont {J.}~\bibnamefont {{Adams}}}, \bibinfo {author} {\bibfnamefont {J.~A.}\ \bibnamefont {{Aguilar}}}, \bibinfo {author} {\bibfnamefont {M.}~\bibnamefont {{Ahlers}}}, \bibinfo {author} {\bibfnamefont {M.}~\bibnamefont {{Ahrens}}}, \bibinfo {author} {\bibfnamefont {J.~M.}\ \bibnamefont {{Alameddine}}}, \bibinfo {author} {\bibfnamefont {A.~A.}\ \bibnamefont {{Alves}}}, \bibinfo {author} {\bibfnamefont {N.~M.}\ \bibnamefont {{Amin}}}, \bibinfo {author} {\bibfnamefont {K.}~\bibnamefont {{Andeen}}},\ and\ \bibinfo {author} {\bibnamefont {et~al.}},\ }\bibfield  {title} {\bibinfo {title} {{Searching for High-energy Neutrino Emission from Galaxy Clusters with IceCube}},\ }\href {https://doi.org/10.3847/2041-8213/ac966b} {\bibfield  {journal} {\bibinfo  {journal} {\apjl}\ }\textbf {\bibinfo {volume} {938}},\ \bibinfo {eid} {L11} (\bibinfo
  {year} {2022}{\natexlab{e}})},\ \Eprint {https://arxiv.org/abs/2206.02054} {arXiv:2206.02054 [astro-ph.HE]} \BibitemShut {NoStop}%
\bibitem [{\citenamefont {{Fang}}\ \emph {et~al.}(2020)\citenamefont {{Fang}}, \citenamefont {{Banerjee}}, \citenamefont {{Charles}},\ and\ \citenamefont {{Omori}}}]{Fang2020}%
  \BibitemOpen
  \bibfield  {author} {\bibinfo {author} {\bibfnamefont {K.}~\bibnamefont {{Fang}}}, \bibinfo {author} {\bibfnamefont {A.}~\bibnamefont {{Banerjee}}}, \bibinfo {author} {\bibfnamefont {E.}~\bibnamefont {{Charles}}},\ and\ \bibinfo {author} {\bibfnamefont {Y.}~\bibnamefont {{Omori}}},\ }\bibfield  {title} {\bibinfo {title} {{A Cross-correlation Study of High-energy Neutrinos and Tracers of Large-scale Structure}},\ }\href {https://doi.org/10.3847/1538-4357/ab8561} {\bibfield  {journal} {\bibinfo  {journal} {\apj}\ }\textbf {\bibinfo {volume} {894}},\ \bibinfo {eid} {112} (\bibinfo {year} {2020})},\ \Eprint {https://arxiv.org/abs/2002.06234} {arXiv:2002.06234 [astro-ph.HE]} \BibitemShut {NoStop}%
\bibitem [{\citenamefont {{Negro}}\ \emph {et~al.}(2023)\citenamefont {{Negro}}, \citenamefont {{Crnogor{\v{c}}evi{\'c}}}, \citenamefont {{Burns}}, \citenamefont {{Charles}}, \citenamefont {{Marcotulli}},\ and\ \citenamefont {{Caputo}}}]{Negro2023}%
  \BibitemOpen
  \bibfield  {author} {\bibinfo {author} {\bibfnamefont {M.}~\bibnamefont {{Negro}}}, \bibinfo {author} {\bibfnamefont {M.}~\bibnamefont {{Crnogor{\v{c}}evi{\'c}}}}, \bibinfo {author} {\bibfnamefont {E.}~\bibnamefont {{Burns}}}, \bibinfo {author} {\bibfnamefont {E.}~\bibnamefont {{Charles}}}, \bibinfo {author} {\bibfnamefont {L.}~\bibnamefont {{Marcotulli}}},\ and\ \bibinfo {author} {\bibfnamefont {R.}~\bibnamefont {{Caputo}}},\ }\bibfield  {title} {\bibinfo {title} {{A Cross-correlation Study between IceCube Neutrino Events and the FERMI Unresolved Gamma-Ray Sky}},\ }\href {https://doi.org/10.3847/1538-4357/acd172} {\bibfield  {journal} {\bibinfo  {journal} {\apj}\ }\textbf {\bibinfo {volume} {951}},\ \bibinfo {eid} {83} (\bibinfo {year} {2023})},\ \Eprint {https://arxiv.org/abs/2304.10934} {arXiv:2304.10934 [astro-ph.HE]} \BibitemShut {NoStop}%
\bibitem [{\citenamefont {{Pei}}(1995)}]{Pei1995}%
  \BibitemOpen
  \bibfield  {author} {\bibinfo {author} {\bibfnamefont {Y.~C.}\ \bibnamefont {{Pei}}},\ }\bibfield  {title} {\bibinfo {title} {{The Luminosity Function of Quasars}},\ }\href {https://doi.org/10.1086/175105} {\bibfield  {journal} {\bibinfo  {journal} {\apj}\ }\textbf {\bibinfo {volume} {438}},\ \bibinfo {pages} {623} (\bibinfo {year} {1995})}\BibitemShut {NoStop}%
\bibitem [{\citenamefont {{Madau}}\ \emph {et~al.}(1996)\citenamefont {{Madau}}, \citenamefont {{Ferguson}}, \citenamefont {{Dickinson}}, \citenamefont {{Giavalisco}}, \citenamefont {{Steidel}},\ and\ \citenamefont {{Fruchter}}}]{Madau1996}%
  \BibitemOpen
  \bibfield  {author} {\bibinfo {author} {\bibfnamefont {P.}~\bibnamefont {{Madau}}}, \bibinfo {author} {\bibfnamefont {H.~C.}\ \bibnamefont {{Ferguson}}}, \bibinfo {author} {\bibfnamefont {M.~E.}\ \bibnamefont {{Dickinson}}}, \bibinfo {author} {\bibfnamefont {M.}~\bibnamefont {{Giavalisco}}}, \bibinfo {author} {\bibfnamefont {C.~C.}\ \bibnamefont {{Steidel}}},\ and\ \bibinfo {author} {\bibfnamefont {A.}~\bibnamefont {{Fruchter}}},\ }\bibfield  {title} {\bibinfo {title} {{High-redshift galaxies in the Hubble Deep Field: colour selection and star formation history to z\raisebox{-0.5ex}\textasciitilde4}},\ }\href {https://doi.org/10.1093/mnras/283.4.1388} {\bibfield  {journal} {\bibinfo  {journal} {\mnras}\ }\textbf {\bibinfo {volume} {283}},\ \bibinfo {pages} {1388} (\bibinfo {year} {1996})},\ \Eprint {https://arxiv.org/abs/astro-ph/9607172} {arXiv:astro-ph/9607172 [astro-ph]} \BibitemShut {NoStop}%
\bibitem [{\citenamefont {{Madau}}\ and\ \citenamefont {{Dickinson}}(2014)}]{Madau2014}%
  \BibitemOpen
  \bibfield  {author} {\bibinfo {author} {\bibfnamefont {P.}~\bibnamefont {{Madau}}}\ and\ \bibinfo {author} {\bibfnamefont {M.}~\bibnamefont {{Dickinson}}},\ }\bibfield  {title} {\bibinfo {title} {{Cosmic Star-Formation History}},\ }\href {https://doi.org/10.1146/annurev-astro-081811-125615} {\bibfield  {journal} {\bibinfo  {journal} {\araa}\ }\textbf {\bibinfo {volume} {52}},\ \bibinfo {pages} {415} (\bibinfo {year} {2014})},\ \Eprint {https://arxiv.org/abs/1403.0007} {arXiv:1403.0007 [astro-ph.CO]} \BibitemShut {NoStop}%
\bibitem [{\citenamefont {{Aird}}\ \emph {et~al.}(2015)\citenamefont {{Aird}}, \citenamefont {{Coil}}, \citenamefont {{Georgakakis}}, \citenamefont {{Nandra}}, \citenamefont {{Barro}},\ and\ \citenamefont {{P{\'e}rez-Gonz{\'a}lez}}}]{Aird2015}%
  \BibitemOpen
  \bibfield  {author} {\bibinfo {author} {\bibfnamefont {J.}~\bibnamefont {{Aird}}}, \bibinfo {author} {\bibfnamefont {A.~L.}\ \bibnamefont {{Coil}}}, \bibinfo {author} {\bibfnamefont {A.}~\bibnamefont {{Georgakakis}}}, \bibinfo {author} {\bibfnamefont {K.}~\bibnamefont {{Nandra}}}, \bibinfo {author} {\bibfnamefont {G.}~\bibnamefont {{Barro}}},\ and\ \bibinfo {author} {\bibfnamefont {P.~G.}\ \bibnamefont {{P{\'e}rez-Gonz{\'a}lez}}},\ }\bibfield  {title} {\bibinfo {title} {{The evolution of the X-ray luminosity functions of unabsorbed and absorbed AGNs out to z{\ensuremath{\sim}} 5}},\ }\href {https://doi.org/10.1093/mnras/stv1062} {\bibfield  {journal} {\bibinfo  {journal} {\mnras}\ }\textbf {\bibinfo {volume} {451}},\ \bibinfo {pages} {1892} (\bibinfo {year} {2015})},\ \Eprint {https://arxiv.org/abs/1503.01120} {arXiv:1503.01120 [astro-ph.HE]} \BibitemShut {NoStop}%
\bibitem [{\citenamefont {{Planck Collaboration}}\ \emph {et~al.}(2020{\natexlab{a}})\citenamefont {{Planck Collaboration}}, \citenamefont {{Aghanim}}, \citenamefont {{Akrami}}, \citenamefont {{Ashdown}}, \citenamefont {{Aumont}}, \citenamefont {{Baccigalupi}}, \citenamefont {{Ballardini}}, \citenamefont {{Banday}}, \citenamefont {{Barreiro}}, \citenamefont {{Bartolo}}, \citenamefont {{Basak}}, \citenamefont {{Battye}}, \citenamefont {{Benabed}}, \citenamefont {{Bernard}}, \citenamefont {{Bersanelli}}, \citenamefont {{Bielewicz}}, \citenamefont {{Bock}}, \citenamefont {{Bond}}, \citenamefont {{Borrill}}, \citenamefont {{Bouchet}}, \citenamefont {{Boulanger}}, \citenamefont {{Bucher}}, \citenamefont {{Burigana}}, \citenamefont {{Butler}}, \citenamefont {{Calabrese}}, \citenamefont {{Cardoso}}, \citenamefont {{Carron}}, \citenamefont {{Challinor}}, \citenamefont {{Chiang}}, \citenamefont {{Chluba}}, \citenamefont {{Colombo}}, \citenamefont {{Combet}}, \citenamefont {{Contreras}}, \citenamefont {{Crill}},
  \citenamefont {{Cuttaia}}, \citenamefont {{de Bernardis}}, \citenamefont {{de Zotti}}, \citenamefont {{Delabrouille}}, \citenamefont {{Delouis}}, \citenamefont {{Di Valentino}}, \citenamefont {{Diego}}, \citenamefont {{Dor{\'e}}}, \citenamefont {{Douspis}}, \citenamefont {{Ducout}}, \citenamefont {{Dupac}}, \citenamefont {{Dusini}}, \citenamefont {{Efstathiou}}, \citenamefont {{Elsner}}, \citenamefont {{En{\ss}lin}}, \citenamefont {{Eriksen}}, \citenamefont {{Fantaye}}, \citenamefont {{Farhang}}, \citenamefont {{Fergusson}}, \citenamefont {{Fernandez-Cobos}}, \citenamefont {{Finelli}}, \citenamefont {{Forastieri}}, \citenamefont {{Frailis}}, \citenamefont {{Fraisse}}, \citenamefont {{Franceschi}}, \citenamefont {{Frolov}}, \citenamefont {{Galeotta}}, \citenamefont {{Galli}}, \citenamefont {{Ganga}}, \citenamefont {{G{\'e}nova-Santos}}, \citenamefont {{Gerbino}}, \citenamefont {{Ghosh}}, \citenamefont {{Gonz{\'a}lez-Nuevo}}, \citenamefont {{G{\'o}rski}}, \citenamefont {{Gratton}}, \citenamefont {{Gruppuso}},
  \citenamefont {{Gudmundsson}}, \citenamefont {{Hamann}}, \citenamefont {{Handley}}, \citenamefont {{Hansen}}, \citenamefont {{Herranz}}, \citenamefont {{Hildebrandt}}, \citenamefont {{Hivon}}, \citenamefont {{Huang}}, \citenamefont {{Jaffe}}, \citenamefont {{Jones}}, \citenamefont {{Karakci}}, \citenamefont {{Keih{\"a}nen}}, \citenamefont {{Keskitalo}}, \citenamefont {{Kiiveri}}, \citenamefont {{Kim}}, \citenamefont {{Kisner}}, \citenamefont {{Knox}}, \citenamefont {{Krachmalnicoff}}, \citenamefont {{Kunz}}, \citenamefont {{Kurki-Suonio}}, \citenamefont {{Lagache}}, \citenamefont {{Lamarre}}, \citenamefont {{Lasenby}}, \citenamefont {{Lattanzi}}, \citenamefont {{Lawrence}}, \citenamefont {{Le Jeune}}, \citenamefont {{Lemos}}, \citenamefont {{Lesgourgues}}, \citenamefont {{Levrier}}, \citenamefont {{Lewis}}, \citenamefont {{Liguori}}, \citenamefont {{Lilje}}, \citenamefont {{Lilley}}, \citenamefont {{Lindholm}}, \citenamefont {{L{\'o}pez-Caniego}}, \citenamefont {{Lubin}}, \citenamefont {{Ma}}, \citenamefont
  {{Mac{\'\i}as-P{\'e}rez}}, \citenamefont {{Maggio}}, \citenamefont {{Maino}}, \citenamefont {{Mandolesi}}, \citenamefont {{Mangilli}}, \citenamefont {{Marcos-Caballero}}, \citenamefont {{Maris}}, \citenamefont {{Martin}}, \citenamefont {{Martinelli}}, \citenamefont {{Mart{\'\i}nez-Gonz{\'a}lez}}, \citenamefont {{Matarrese}}, \citenamefont {{Mauri}}, \citenamefont {{McEwen}}, \citenamefont {{Meinhold}}, \citenamefont {{Melchiorri}}, \citenamefont {{Mennella}}, \citenamefont {{Migliaccio}}, \citenamefont {{Millea}}, \citenamefont {{Mitra}}, \citenamefont {{Miville-Desch{\^e}nes}}, \citenamefont {{Molinari}}, \citenamefont {{Montier}}, \citenamefont {{Morgante}}, \citenamefont {{Moss}}, \citenamefont {{Natoli}}, \citenamefont {{N{\o}rgaard-Nielsen}}, \citenamefont {{Pagano}}, \citenamefont {{Paoletti}}, \citenamefont {{Partridge}}, \citenamefont {{Patanchon}}, \citenamefont {{Peiris}}, \citenamefont {{Perrotta}}, \citenamefont {{Pettorino}}, \citenamefont {{Piacentini}}, \citenamefont {{Polastri}},
  \citenamefont {{Polenta}}, \citenamefont {{Puget}}, \citenamefont {{Rachen}}, \citenamefont {{Reinecke}}, \citenamefont {{Remazeilles}}, \citenamefont {{Renzi}}, \citenamefont {{Rocha}}, \citenamefont {{Rosset}}, \citenamefont {{Roudier}}, \citenamefont {{Rubi{\~n}o-Mart{\'\i}n}}, \citenamefont {{Ruiz-Granados}}, \citenamefont {{Salvati}}, \citenamefont {{Sandri}}, \citenamefont {{Savelainen}}, \citenamefont {{Scott}}, \citenamefont {{Shellard}}, \citenamefont {{Sirignano}}, \citenamefont {{Sirri}}, \citenamefont {{Spencer}}, \citenamefont {{Sunyaev}}, \citenamefont {{Suur-Uski}}, \citenamefont {{Tauber}}, \citenamefont {{Tavagnacco}}, \citenamefont {{Tenti}}, \citenamefont {{Toffolatti}}, \citenamefont {{Tomasi}}, \citenamefont {{Trombetti}}, \citenamefont {{Valenziano}}, \citenamefont {{Valiviita}}, \citenamefont {{Van Tent}}, \citenamefont {{Vibert}}, \citenamefont {{Vielva}}, \citenamefont {{Villa}}, \citenamefont {{Vittorio}}, \citenamefont {{Wandelt}}, \citenamefont {{Wehus}}, \citenamefont {{White}},
  \citenamefont {{White}}, \citenamefont {{Zacchei}},\ and\ \citenamefont {{Zonca}}}]{Planck2020}%
  \BibitemOpen
  \bibfield  {author} {\bibinfo {author} {\bibnamefont {{Planck Collaboration}}}, \bibinfo {author} {\bibfnamefont {N.}~\bibnamefont {{Aghanim}}}, \bibinfo {author} {\bibfnamefont {Y.}~\bibnamefont {{Akrami}}}, \bibinfo {author} {\bibfnamefont {M.}~\bibnamefont {{Ashdown}}}, \bibinfo {author} {\bibfnamefont {J.}~\bibnamefont {{Aumont}}}, \bibinfo {author} {\bibfnamefont {C.}~\bibnamefont {{Baccigalupi}}}, \bibinfo {author} {\bibfnamefont {M.}~\bibnamefont {{Ballardini}}}, \bibinfo {author} {\bibfnamefont {A.~J.}\ \bibnamefont {{Banday}}}, \bibinfo {author} {\bibfnamefont {R.~B.}\ \bibnamefont {{Barreiro}}}, \bibinfo {author} {\bibfnamefont {N.}~\bibnamefont {{Bartolo}}}, \bibinfo {author} {\bibfnamefont {S.}~\bibnamefont {{Basak}}}, \bibinfo {author} {\bibfnamefont {R.}~\bibnamefont {{Battye}}}, \bibinfo {author} {\bibfnamefont {K.}~\bibnamefont {{Benabed}}}, \bibinfo {author} {\bibfnamefont {J.~P.}\ \bibnamefont {{Bernard}}}, \bibinfo {author} {\bibfnamefont {M.}~\bibnamefont {{Bersanelli}}}, \bibinfo
  {author} {\bibfnamefont {P.}~\bibnamefont {{Bielewicz}}}, \bibinfo {author} {\bibfnamefont {J.~J.}\ \bibnamefont {{Bock}}}, \bibinfo {author} {\bibfnamefont {J.~R.}\ \bibnamefont {{Bond}}}, \bibinfo {author} {\bibfnamefont {J.}~\bibnamefont {{Borrill}}}, \bibinfo {author} {\bibfnamefont {F.~R.}\ \bibnamefont {{Bouchet}}}, \bibinfo {author} {\bibfnamefont {F.}~\bibnamefont {{Boulanger}}}, \bibinfo {author} {\bibfnamefont {M.}~\bibnamefont {{Bucher}}}, \bibinfo {author} {\bibfnamefont {C.}~\bibnamefont {{Burigana}}}, \bibinfo {author} {\bibfnamefont {R.~C.}\ \bibnamefont {{Butler}}}, \bibinfo {author} {\bibfnamefont {E.}~\bibnamefont {{Calabrese}}}, \bibinfo {author} {\bibfnamefont {J.~F.}\ \bibnamefont {{Cardoso}}}, \bibinfo {author} {\bibfnamefont {J.}~\bibnamefont {{Carron}}}, \bibinfo {author} {\bibfnamefont {A.}~\bibnamefont {{Challinor}}}, \bibinfo {author} {\bibfnamefont {H.~C.}\ \bibnamefont {{Chiang}}}, \bibinfo {author} {\bibfnamefont {J.}~\bibnamefont {{Chluba}}}, \bibinfo {author} {\bibfnamefont
  {L.~P.~L.}\ \bibnamefont {{Colombo}}}, \bibinfo {author} {\bibfnamefont {C.}~\bibnamefont {{Combet}}}, \bibinfo {author} {\bibfnamefont {D.}~\bibnamefont {{Contreras}}}, \bibinfo {author} {\bibfnamefont {B.~P.}\ \bibnamefont {{Crill}}}, \bibinfo {author} {\bibfnamefont {F.}~\bibnamefont {{Cuttaia}}}, \bibinfo {author} {\bibfnamefont {P.}~\bibnamefont {{de Bernardis}}}, \bibinfo {author} {\bibfnamefont {G.}~\bibnamefont {{de Zotti}}}, \bibinfo {author} {\bibfnamefont {J.}~\bibnamefont {{Delabrouille}}}, \bibinfo {author} {\bibfnamefont {J.~M.}\ \bibnamefont {{Delouis}}}, \bibinfo {author} {\bibfnamefont {E.}~\bibnamefont {{Di Valentino}}}, \bibinfo {author} {\bibfnamefont {J.~M.}\ \bibnamefont {{Diego}}}, \bibinfo {author} {\bibfnamefont {O.}~\bibnamefont {{Dor{\'e}}}}, \bibinfo {author} {\bibfnamefont {M.}~\bibnamefont {{Douspis}}}, \bibinfo {author} {\bibfnamefont {A.}~\bibnamefont {{Ducout}}}, \bibinfo {author} {\bibfnamefont {X.}~\bibnamefont {{Dupac}}}, \bibinfo {author} {\bibfnamefont {S.}~\bibnamefont
  {{Dusini}}}, \bibinfo {author} {\bibfnamefont {G.}~\bibnamefont {{Efstathiou}}}, \bibinfo {author} {\bibfnamefont {F.}~\bibnamefont {{Elsner}}}, \bibinfo {author} {\bibfnamefont {T.~A.}\ \bibnamefont {{En{\ss}lin}}}, \bibinfo {author} {\bibfnamefont {H.~K.}\ \bibnamefont {{Eriksen}}}, \bibinfo {author} {\bibfnamefont {Y.}~\bibnamefont {{Fantaye}}}, \bibinfo {author} {\bibfnamefont {M.}~\bibnamefont {{Farhang}}}, \bibinfo {author} {\bibfnamefont {J.}~\bibnamefont {{Fergusson}}}, \bibinfo {author} {\bibfnamefont {R.}~\bibnamefont {{Fernandez-Cobos}}}, \bibinfo {author} {\bibfnamefont {F.}~\bibnamefont {{Finelli}}}, \bibinfo {author} {\bibfnamefont {F.}~\bibnamefont {{Forastieri}}}, \bibinfo {author} {\bibfnamefont {M.}~\bibnamefont {{Frailis}}}, \bibinfo {author} {\bibfnamefont {A.~A.}\ \bibnamefont {{Fraisse}}}, \bibinfo {author} {\bibfnamefont {E.}~\bibnamefont {{Franceschi}}}, \bibinfo {author} {\bibfnamefont {A.}~\bibnamefont {{Frolov}}}, \bibinfo {author} {\bibfnamefont {S.}~\bibnamefont {{Galeotta}}},
  \bibinfo {author} {\bibfnamefont {S.}~\bibnamefont {{Galli}}}, \bibinfo {author} {\bibfnamefont {K.}~\bibnamefont {{Ganga}}}, \bibinfo {author} {\bibfnamefont {R.~T.}\ \bibnamefont {{G{\'e}nova-Santos}}}, \bibinfo {author} {\bibfnamefont {M.}~\bibnamefont {{Gerbino}}}, \bibinfo {author} {\bibfnamefont {T.}~\bibnamefont {{Ghosh}}}, \bibinfo {author} {\bibfnamefont {J.}~\bibnamefont {{Gonz{\'a}lez-Nuevo}}}, \bibinfo {author} {\bibfnamefont {K.~M.}\ \bibnamefont {{G{\'o}rski}}}, \bibinfo {author} {\bibfnamefont {S.}~\bibnamefont {{Gratton}}}, \bibinfo {author} {\bibfnamefont {A.}~\bibnamefont {{Gruppuso}}}, \bibinfo {author} {\bibfnamefont {J.~E.}\ \bibnamefont {{Gudmundsson}}}, \bibinfo {author} {\bibfnamefont {J.}~\bibnamefont {{Hamann}}}, \bibinfo {author} {\bibfnamefont {W.}~\bibnamefont {{Handley}}}, \bibinfo {author} {\bibfnamefont {F.~K.}\ \bibnamefont {{Hansen}}}, \bibinfo {author} {\bibfnamefont {D.}~\bibnamefont {{Herranz}}}, \bibinfo {author} {\bibfnamefont {S.~R.}\ \bibnamefont {{Hildebrandt}}},
  \bibinfo {author} {\bibfnamefont {E.}~\bibnamefont {{Hivon}}}, \bibinfo {author} {\bibfnamefont {Z.}~\bibnamefont {{Huang}}}, \bibinfo {author} {\bibfnamefont {A.~H.}\ \bibnamefont {{Jaffe}}}, \bibinfo {author} {\bibfnamefont {W.~C.}\ \bibnamefont {{Jones}}}, \bibinfo {author} {\bibfnamefont {A.}~\bibnamefont {{Karakci}}}, \bibinfo {author} {\bibfnamefont {E.}~\bibnamefont {{Keih{\"a}nen}}}, \bibinfo {author} {\bibfnamefont {R.}~\bibnamefont {{Keskitalo}}}, \bibinfo {author} {\bibfnamefont {K.}~\bibnamefont {{Kiiveri}}}, \bibinfo {author} {\bibfnamefont {J.}~\bibnamefont {{Kim}}}, \bibinfo {author} {\bibfnamefont {T.~S.}\ \bibnamefont {{Kisner}}}, \bibinfo {author} {\bibfnamefont {L.}~\bibnamefont {{Knox}}}, \bibinfo {author} {\bibfnamefont {N.}~\bibnamefont {{Krachmalnicoff}}}, \bibinfo {author} {\bibfnamefont {M.}~\bibnamefont {{Kunz}}}, \bibinfo {author} {\bibfnamefont {H.}~\bibnamefont {{Kurki-Suonio}}}, \bibinfo {author} {\bibfnamefont {G.}~\bibnamefont {{Lagache}}}, \bibinfo {author} {\bibfnamefont
  {J.~M.}\ \bibnamefont {{Lamarre}}}, \bibinfo {author} {\bibfnamefont {A.}~\bibnamefont {{Lasenby}}}, \bibinfo {author} {\bibfnamefont {M.}~\bibnamefont {{Lattanzi}}}, \bibinfo {author} {\bibfnamefont {C.~R.}\ \bibnamefont {{Lawrence}}}, \bibinfo {author} {\bibfnamefont {M.}~\bibnamefont {{Le Jeune}}}, \bibinfo {author} {\bibfnamefont {P.}~\bibnamefont {{Lemos}}}, \bibinfo {author} {\bibfnamefont {J.}~\bibnamefont {{Lesgourgues}}}, \bibinfo {author} {\bibfnamefont {F.}~\bibnamefont {{Levrier}}}, \bibinfo {author} {\bibfnamefont {A.}~\bibnamefont {{Lewis}}}, \bibinfo {author} {\bibfnamefont {M.}~\bibnamefont {{Liguori}}}, \bibinfo {author} {\bibfnamefont {P.~B.}\ \bibnamefont {{Lilje}}}, \bibinfo {author} {\bibfnamefont {M.}~\bibnamefont {{Lilley}}}, \bibinfo {author} {\bibfnamefont {V.}~\bibnamefont {{Lindholm}}}, \bibinfo {author} {\bibfnamefont {M.}~\bibnamefont {{L{\'o}pez-Caniego}}}, \bibinfo {author} {\bibfnamefont {P.~M.}\ \bibnamefont {{Lubin}}}, \bibinfo {author} {\bibfnamefont {Y.~Z.}\ \bibnamefont
  {{Ma}}}, \bibinfo {author} {\bibfnamefont {J.~F.}\ \bibnamefont {{Mac{\'\i}as-P{\'e}rez}}}, \bibinfo {author} {\bibfnamefont {G.}~\bibnamefont {{Maggio}}}, \bibinfo {author} {\bibfnamefont {D.}~\bibnamefont {{Maino}}}, \bibinfo {author} {\bibfnamefont {N.}~\bibnamefont {{Mandolesi}}}, \bibinfo {author} {\bibfnamefont {A.}~\bibnamefont {{Mangilli}}}, \bibinfo {author} {\bibfnamefont {A.}~\bibnamefont {{Marcos-Caballero}}}, \bibinfo {author} {\bibfnamefont {M.}~\bibnamefont {{Maris}}}, \bibinfo {author} {\bibfnamefont {P.~G.}\ \bibnamefont {{Martin}}}, \bibinfo {author} {\bibfnamefont {M.}~\bibnamefont {{Martinelli}}}, \bibinfo {author} {\bibfnamefont {E.}~\bibnamefont {{Mart{\'\i}nez-Gonz{\'a}lez}}}, \bibinfo {author} {\bibfnamefont {S.}~\bibnamefont {{Matarrese}}}, \bibinfo {author} {\bibfnamefont {N.}~\bibnamefont {{Mauri}}}, \bibinfo {author} {\bibfnamefont {J.~D.}\ \bibnamefont {{McEwen}}}, \bibinfo {author} {\bibfnamefont {P.~R.}\ \bibnamefont {{Meinhold}}}, \bibinfo {author} {\bibfnamefont
  {A.}~\bibnamefont {{Melchiorri}}}, \bibinfo {author} {\bibfnamefont {A.}~\bibnamefont {{Mennella}}}, \bibinfo {author} {\bibfnamefont {M.}~\bibnamefont {{Migliaccio}}}, \bibinfo {author} {\bibfnamefont {M.}~\bibnamefont {{Millea}}}, \bibinfo {author} {\bibfnamefont {S.}~\bibnamefont {{Mitra}}}, \bibinfo {author} {\bibfnamefont {M.~A.}\ \bibnamefont {{Miville-Desch{\^e}nes}}}, \bibinfo {author} {\bibfnamefont {D.}~\bibnamefont {{Molinari}}}, \bibinfo {author} {\bibfnamefont {L.}~\bibnamefont {{Montier}}}, \bibinfo {author} {\bibfnamefont {G.}~\bibnamefont {{Morgante}}}, \bibinfo {author} {\bibfnamefont {A.}~\bibnamefont {{Moss}}}, \bibinfo {author} {\bibfnamefont {P.}~\bibnamefont {{Natoli}}}, \bibinfo {author} {\bibfnamefont {H.~U.}\ \bibnamefont {{N{\o}rgaard-Nielsen}}}, \bibinfo {author} {\bibfnamefont {L.}~\bibnamefont {{Pagano}}}, \bibinfo {author} {\bibfnamefont {D.}~\bibnamefont {{Paoletti}}}, \bibinfo {author} {\bibfnamefont {B.}~\bibnamefont {{Partridge}}}, \bibinfo {author} {\bibfnamefont
  {G.}~\bibnamefont {{Patanchon}}}, \bibinfo {author} {\bibfnamefont {H.~V.}\ \bibnamefont {{Peiris}}}, \bibinfo {author} {\bibfnamefont {F.}~\bibnamefont {{Perrotta}}}, \bibinfo {author} {\bibfnamefont {V.}~\bibnamefont {{Pettorino}}}, \bibinfo {author} {\bibfnamefont {F.}~\bibnamefont {{Piacentini}}}, \bibinfo {author} {\bibfnamefont {L.}~\bibnamefont {{Polastri}}}, \bibinfo {author} {\bibfnamefont {G.}~\bibnamefont {{Polenta}}}, \bibinfo {author} {\bibfnamefont {J.~L.}\ \bibnamefont {{Puget}}}, \bibinfo {author} {\bibfnamefont {J.~P.}\ \bibnamefont {{Rachen}}}, \bibinfo {author} {\bibfnamefont {M.}~\bibnamefont {{Reinecke}}}, \bibinfo {author} {\bibfnamefont {M.}~\bibnamefont {{Remazeilles}}}, \bibinfo {author} {\bibfnamefont {A.}~\bibnamefont {{Renzi}}}, \bibinfo {author} {\bibfnamefont {G.}~\bibnamefont {{Rocha}}}, \bibinfo {author} {\bibfnamefont {C.}~\bibnamefont {{Rosset}}}, \bibinfo {author} {\bibfnamefont {G.}~\bibnamefont {{Roudier}}}, \bibinfo {author} {\bibfnamefont {J.~A.}\ \bibnamefont
  {{Rubi{\~n}o-Mart{\'\i}n}}}, \bibinfo {author} {\bibfnamefont {B.}~\bibnamefont {{Ruiz-Granados}}}, \bibinfo {author} {\bibfnamefont {L.}~\bibnamefont {{Salvati}}}, \bibinfo {author} {\bibfnamefont {M.}~\bibnamefont {{Sandri}}}, \bibinfo {author} {\bibfnamefont {M.}~\bibnamefont {{Savelainen}}}, \bibinfo {author} {\bibfnamefont {D.}~\bibnamefont {{Scott}}}, \bibinfo {author} {\bibfnamefont {E.~P.~S.}\ \bibnamefont {{Shellard}}}, \bibinfo {author} {\bibfnamefont {C.}~\bibnamefont {{Sirignano}}}, \bibinfo {author} {\bibfnamefont {G.}~\bibnamefont {{Sirri}}}, \bibinfo {author} {\bibfnamefont {L.~D.}\ \bibnamefont {{Spencer}}}, \bibinfo {author} {\bibfnamefont {R.}~\bibnamefont {{Sunyaev}}}, \bibinfo {author} {\bibfnamefont {A.~S.}\ \bibnamefont {{Suur-Uski}}}, \bibinfo {author} {\bibfnamefont {J.~A.}\ \bibnamefont {{Tauber}}}, \bibinfo {author} {\bibfnamefont {D.}~\bibnamefont {{Tavagnacco}}}, \bibinfo {author} {\bibfnamefont {M.}~\bibnamefont {{Tenti}}}, \bibinfo {author} {\bibfnamefont {L.}~\bibnamefont
  {{Toffolatti}}}, \bibinfo {author} {\bibfnamefont {M.}~\bibnamefont {{Tomasi}}}, \bibinfo {author} {\bibfnamefont {T.}~\bibnamefont {{Trombetti}}}, \bibinfo {author} {\bibfnamefont {L.}~\bibnamefont {{Valenziano}}}, \bibinfo {author} {\bibfnamefont {J.}~\bibnamefont {{Valiviita}}}, \bibinfo {author} {\bibfnamefont {B.}~\bibnamefont {{Van Tent}}}, \bibinfo {author} {\bibfnamefont {L.}~\bibnamefont {{Vibert}}}, \bibinfo {author} {\bibfnamefont {P.}~\bibnamefont {{Vielva}}}, \bibinfo {author} {\bibfnamefont {F.}~\bibnamefont {{Villa}}}, \bibinfo {author} {\bibfnamefont {N.}~\bibnamefont {{Vittorio}}}, \bibinfo {author} {\bibfnamefont {B.~D.}\ \bibnamefont {{Wandelt}}}, \bibinfo {author} {\bibfnamefont {I.~K.}\ \bibnamefont {{Wehus}}}, \bibinfo {author} {\bibfnamefont {M.}~\bibnamefont {{White}}}, \bibinfo {author} {\bibfnamefont {S.~D.~M.}\ \bibnamefont {{White}}}, \bibinfo {author} {\bibfnamefont {A.}~\bibnamefont {{Zacchei}}},\ and\ \bibinfo {author} {\bibfnamefont {A.}~\bibnamefont {{Zonca}}},\ }\bibfield
  {title} {\bibinfo {title} {{Planck 2018 results. VI. Cosmological parameters}},\ }\href {https://doi.org/10.1051/0004-6361/201833910} {\bibfield  {journal} {\bibinfo  {journal} {\aap}\ }\textbf {\bibinfo {volume} {641}},\ \bibinfo {eid} {A6} (\bibinfo {year} {2020}{\natexlab{a}})},\ \Eprint {https://arxiv.org/abs/1807.06209} {arXiv:1807.06209 [astro-ph.CO]} \BibitemShut {NoStop}%
\bibitem [{\citenamefont {{IceCube Collaboration}}\ \emph {et~al.}(2021)\citenamefont {{IceCube Collaboration}}, \citenamefont {{Abbasi}}, \citenamefont {{Ackermann}}, \citenamefont {{Adams}}, \citenamefont {{Aguilar}}, \citenamefont {{Ahlers}}, \citenamefont {{Ahrens}}, \citenamefont {{Alispach}}, \citenamefont {{Amin}}, \citenamefont {{Andeen}},\ and\ \citenamefont {et~al.}}]{2101.09836}%
  \BibitemOpen
  \bibfield  {author} {\bibinfo {author} {\bibnamefont {{IceCube Collaboration}}}, \bibinfo {author} {\bibfnamefont {R.}~\bibnamefont {{Abbasi}}}, \bibinfo {author} {\bibfnamefont {M.}~\bibnamefont {{Ackermann}}}, \bibinfo {author} {\bibfnamefont {J.}~\bibnamefont {{Adams}}}, \bibinfo {author} {\bibfnamefont {J.~A.}\ \bibnamefont {{Aguilar}}}, \bibinfo {author} {\bibfnamefont {M.}~\bibnamefont {{Ahlers}}}, \bibinfo {author} {\bibfnamefont {M.}~\bibnamefont {{Ahrens}}}, \bibinfo {author} {\bibfnamefont {C.}~\bibnamefont {{Alispach}}}, \bibinfo {author} {\bibfnamefont {N.~M.}\ \bibnamefont {{Amin}}}, \bibinfo {author} {\bibfnamefont {K.}~\bibnamefont {{Andeen}}},\ and\ \bibinfo {author} {\bibnamefont {et~al.}},\ }\bibfield  {title} {\bibinfo {title} {{IceCube Data for Neutrino Point-Source Searches Years 2008-2018}},\ }\href {https://doi.org/10.48550/arXiv.2101.09836} {\bibfield  {journal} {\bibinfo  {journal} {arXiv e-prints}\ ,\ \bibinfo {eid} {arXiv:2101.09836}} (\bibinfo {year} {2021})},\ \Eprint
  {https://arxiv.org/abs/2101.09836} {arXiv:2101.09836 [astro-ph.HE]} \BibitemShut {NoStop}%
\bibitem [{\citenamefont {{Bilicki}}\ \emph {et~al.}(2016)\citenamefont {{Bilicki}}, \citenamefont {{Peacock}}, \citenamefont {{Jarrett}}, \citenamefont {{Cluver}}, \citenamefont {{Maddox}}, \citenamefont {{Brown}}, \citenamefont {{Taylor}}, \citenamefont {{Hambly}}, \citenamefont {{Solarz}}, \citenamefont {{Holwerda}}, \citenamefont {{Baldry}}, \citenamefont {{Loveday}}, \citenamefont {{Moffett}}, \citenamefont {{Hopkins}}, \citenamefont {{Driver}}, \citenamefont {{Alpaslan}},\ and\ \citenamefont {{Bland-Hawthorn}}}]{Bilicki2016}%
  \BibitemOpen
  \bibfield  {author} {\bibinfo {author} {\bibfnamefont {M.}~\bibnamefont {{Bilicki}}}, \bibinfo {author} {\bibfnamefont {J.~A.}\ \bibnamefont {{Peacock}}}, \bibinfo {author} {\bibfnamefont {T.~H.}\ \bibnamefont {{Jarrett}}}, \bibinfo {author} {\bibfnamefont {M.~E.}\ \bibnamefont {{Cluver}}}, \bibinfo {author} {\bibfnamefont {N.}~\bibnamefont {{Maddox}}}, \bibinfo {author} {\bibfnamefont {M.~J.~I.}\ \bibnamefont {{Brown}}}, \bibinfo {author} {\bibfnamefont {E.~N.}\ \bibnamefont {{Taylor}}}, \bibinfo {author} {\bibfnamefont {N.~C.}\ \bibnamefont {{Hambly}}}, \bibinfo {author} {\bibfnamefont {A.}~\bibnamefont {{Solarz}}}, \bibinfo {author} {\bibfnamefont {B.~W.}\ \bibnamefont {{Holwerda}}}, \bibinfo {author} {\bibfnamefont {I.}~\bibnamefont {{Baldry}}}, \bibinfo {author} {\bibfnamefont {J.}~\bibnamefont {{Loveday}}}, \bibinfo {author} {\bibfnamefont {A.}~\bibnamefont {{Moffett}}}, \bibinfo {author} {\bibfnamefont {A.~M.}\ \bibnamefont {{Hopkins}}}, \bibinfo {author} {\bibfnamefont {S.~P.}\ \bibnamefont
  {{Driver}}}, \bibinfo {author} {\bibfnamefont {M.}~\bibnamefont {{Alpaslan}}},\ and\ \bibinfo {author} {\bibfnamefont {J.}~\bibnamefont {{Bland-Hawthorn}}},\ }\bibfield  {title} {\bibinfo {title} {{WISE {\texttimes} SuperCOSMOS Photometric Redshift Catalog: 20 Million Galaxies over 3/pi Steradians}},\ }\href {https://doi.org/10.3847/0067-0049/225/1/5} {\bibfield  {journal} {\bibinfo  {journal} {\apjs}\ }\textbf {\bibinfo {volume} {225}},\ \bibinfo {eid} {5} (\bibinfo {year} {2016})},\ \Eprint {https://arxiv.org/abs/1607.01182} {arXiv:1607.01182 [astro-ph.CO]} \BibitemShut {NoStop}%
\bibitem [{\citenamefont {{Wright}}\ \emph {et~al.}(2010)\citenamefont {{Wright}}, \citenamefont {{Eisenhardt}}, \citenamefont {{Mainzer}}, \citenamefont {{Ressler}}, \citenamefont {{Cutri}}, \citenamefont {{Jarrett}}, \citenamefont {{Kirkpatrick}}, \citenamefont {{Padgett}}, \citenamefont {{McMillan}}, \citenamefont {{Skrutskie}}, \citenamefont {{Stanford}}, \citenamefont {{Cohen}}, \citenamefont {{Walker}}, \citenamefont {{Mather}}, \citenamefont {{Leisawitz}}, \citenamefont {{Gautier}}, \citenamefont {{McLean}}, \citenamefont {{Benford}}, \citenamefont {{Lonsdale}}, \citenamefont {{Blain}}, \citenamefont {{Mendez}}, \citenamefont {{Irace}}, \citenamefont {{Duval}}, \citenamefont {{Liu}}, \citenamefont {{Royer}}, \citenamefont {{Heinrichsen}}, \citenamefont {{Howard}}, \citenamefont {{Shannon}}, \citenamefont {{Kendall}}, \citenamefont {{Walsh}}, \citenamefont {{Larsen}}, \citenamefont {{Cardon}}, \citenamefont {{Schick}}, \citenamefont {{Schwalm}}, \citenamefont {{Abid}}, \citenamefont {{Fabinsky}},
  \citenamefont {{Naes}},\ and\ \citenamefont {{Tsai}}}]{Wright2010}%
  \BibitemOpen
  \bibfield  {author} {\bibinfo {author} {\bibfnamefont {E.~L.}\ \bibnamefont {{Wright}}}, \bibinfo {author} {\bibfnamefont {P.~R.~M.}\ \bibnamefont {{Eisenhardt}}}, \bibinfo {author} {\bibfnamefont {A.~K.}\ \bibnamefont {{Mainzer}}}, \bibinfo {author} {\bibfnamefont {M.~E.}\ \bibnamefont {{Ressler}}}, \bibinfo {author} {\bibfnamefont {R.~M.}\ \bibnamefont {{Cutri}}}, \bibinfo {author} {\bibfnamefont {T.}~\bibnamefont {{Jarrett}}}, \bibinfo {author} {\bibfnamefont {J.~D.}\ \bibnamefont {{Kirkpatrick}}}, \bibinfo {author} {\bibfnamefont {D.}~\bibnamefont {{Padgett}}}, \bibinfo {author} {\bibfnamefont {R.~S.}\ \bibnamefont {{McMillan}}}, \bibinfo {author} {\bibfnamefont {M.}~\bibnamefont {{Skrutskie}}}, \bibinfo {author} {\bibfnamefont {S.~A.}\ \bibnamefont {{Stanford}}}, \bibinfo {author} {\bibfnamefont {M.}~\bibnamefont {{Cohen}}}, \bibinfo {author} {\bibfnamefont {R.~G.}\ \bibnamefont {{Walker}}}, \bibinfo {author} {\bibfnamefont {J.~C.}\ \bibnamefont {{Mather}}}, \bibinfo {author} {\bibfnamefont
  {D.}~\bibnamefont {{Leisawitz}}}, \bibinfo {author} {\bibfnamefont {I.}~\bibnamefont {{Gautier}}, \bibfnamefont {Thomas~N.}}, \bibinfo {author} {\bibfnamefont {I.}~\bibnamefont {{McLean}}}, \bibinfo {author} {\bibfnamefont {D.}~\bibnamefont {{Benford}}}, \bibinfo {author} {\bibfnamefont {C.~J.}\ \bibnamefont {{Lonsdale}}}, \bibinfo {author} {\bibfnamefont {A.}~\bibnamefont {{Blain}}}, \bibinfo {author} {\bibfnamefont {B.}~\bibnamefont {{Mendez}}}, \bibinfo {author} {\bibfnamefont {W.~R.}\ \bibnamefont {{Irace}}}, \bibinfo {author} {\bibfnamefont {V.}~\bibnamefont {{Duval}}}, \bibinfo {author} {\bibfnamefont {F.}~\bibnamefont {{Liu}}}, \bibinfo {author} {\bibfnamefont {D.}~\bibnamefont {{Royer}}}, \bibinfo {author} {\bibfnamefont {I.}~\bibnamefont {{Heinrichsen}}}, \bibinfo {author} {\bibfnamefont {J.}~\bibnamefont {{Howard}}}, \bibinfo {author} {\bibfnamefont {M.}~\bibnamefont {{Shannon}}}, \bibinfo {author} {\bibfnamefont {M.}~\bibnamefont {{Kendall}}}, \bibinfo {author} {\bibfnamefont {A.~L.}\
  \bibnamefont {{Walsh}}}, \bibinfo {author} {\bibfnamefont {M.}~\bibnamefont {{Larsen}}}, \bibinfo {author} {\bibfnamefont {J.~G.}\ \bibnamefont {{Cardon}}}, \bibinfo {author} {\bibfnamefont {S.}~\bibnamefont {{Schick}}}, \bibinfo {author} {\bibfnamefont {M.}~\bibnamefont {{Schwalm}}}, \bibinfo {author} {\bibfnamefont {M.}~\bibnamefont {{Abid}}}, \bibinfo {author} {\bibfnamefont {B.}~\bibnamefont {{Fabinsky}}}, \bibinfo {author} {\bibfnamefont {L.}~\bibnamefont {{Naes}}},\ and\ \bibinfo {author} {\bibfnamefont {C.-W.}\ \bibnamefont {{Tsai}}},\ }\bibfield  {title} {\bibinfo {title} {{The Wide-field Infrared Survey Explorer (WISE): Mission Description and Initial On-orbit Performance}},\ }\href {https://doi.org/10.1088/0004-6256/140/6/1868} {\bibfield  {journal} {\bibinfo  {journal} {\aj}\ }\textbf {\bibinfo {volume} {140}},\ \bibinfo {pages} {1868} (\bibinfo {year} {2010})},\ \Eprint {https://arxiv.org/abs/1008.0031} {arXiv:1008.0031 [astro-ph.IM]} \BibitemShut {NoStop}%
\bibitem [{\citenamefont {{Hambly}}\ \emph {et~al.}(2001)\citenamefont {{Hambly}}, \citenamefont {{MacGillivray}}, \citenamefont {{Read}}, \citenamefont {{Tritton}}, \citenamefont {{Thomson}}, \citenamefont {{Kelly}}, \citenamefont {{Morgan}}, \citenamefont {{Smith}}, \citenamefont {{Driver}}, \citenamefont {{Williamson}}, \citenamefont {{Parker}}, \citenamefont {{Hawkins}}, \citenamefont {{Williams}},\ and\ \citenamefont {{Lawrence}}}]{Hambly2001}%
  \BibitemOpen
  \bibfield  {author} {\bibinfo {author} {\bibfnamefont {N.~C.}\ \bibnamefont {{Hambly}}}, \bibinfo {author} {\bibfnamefont {H.~T.}\ \bibnamefont {{MacGillivray}}}, \bibinfo {author} {\bibfnamefont {M.~A.}\ \bibnamefont {{Read}}}, \bibinfo {author} {\bibfnamefont {S.~B.}\ \bibnamefont {{Tritton}}}, \bibinfo {author} {\bibfnamefont {E.~B.}\ \bibnamefont {{Thomson}}}, \bibinfo {author} {\bibfnamefont {B.~D.}\ \bibnamefont {{Kelly}}}, \bibinfo {author} {\bibfnamefont {D.~H.}\ \bibnamefont {{Morgan}}}, \bibinfo {author} {\bibfnamefont {R.~E.}\ \bibnamefont {{Smith}}}, \bibinfo {author} {\bibfnamefont {S.~P.}\ \bibnamefont {{Driver}}}, \bibinfo {author} {\bibfnamefont {J.}~\bibnamefont {{Williamson}}}, \bibinfo {author} {\bibfnamefont {Q.~A.}\ \bibnamefont {{Parker}}}, \bibinfo {author} {\bibfnamefont {M.~R.~S.}\ \bibnamefont {{Hawkins}}}, \bibinfo {author} {\bibfnamefont {P.~M.}\ \bibnamefont {{Williams}}},\ and\ \bibinfo {author} {\bibfnamefont {A.}~\bibnamefont {{Lawrence}}},\ }\bibfield  {title} {\bibinfo
  {title} {{The SuperCOSMOS Sky Survey - I. Introduction and description}},\ }\href {https://doi.org/10.1111/j.1365-2966.2001.04660.x} {\bibfield  {journal} {\bibinfo  {journal} {\mnras}\ }\textbf {\bibinfo {volume} {326}},\ \bibinfo {pages} {1279} (\bibinfo {year} {2001})},\ \Eprint {https://arxiv.org/abs/astro-ph/0108286} {arXiv:astro-ph/0108286 [astro-ph]} \BibitemShut {NoStop}%
\bibitem [{\citenamefont {{Peacock}}\ \emph {et~al.}(2016)\citenamefont {{Peacock}}, \citenamefont {{Hambly}}, \citenamefont {{Bilicki}}, \citenamefont {{MacGillivray}}, \citenamefont {{Miller}}, \citenamefont {{Read}},\ and\ \citenamefont {{Tritton}}}]{Peacock2016}%
  \BibitemOpen
  \bibfield  {author} {\bibinfo {author} {\bibfnamefont {J.~A.}\ \bibnamefont {{Peacock}}}, \bibinfo {author} {\bibfnamefont {N.~C.}\ \bibnamefont {{Hambly}}}, \bibinfo {author} {\bibfnamefont {M.}~\bibnamefont {{Bilicki}}}, \bibinfo {author} {\bibfnamefont {H.~T.}\ \bibnamefont {{MacGillivray}}}, \bibinfo {author} {\bibfnamefont {L.}~\bibnamefont {{Miller}}}, \bibinfo {author} {\bibfnamefont {M.~A.}\ \bibnamefont {{Read}}},\ and\ \bibinfo {author} {\bibfnamefont {S.~B.}\ \bibnamefont {{Tritton}}},\ }\bibfield  {title} {\bibinfo {title} {{The SuperCOSMOS all-sky galaxy catalogue}},\ }\href {https://doi.org/10.1093/mnras/stw1818} {\bibfield  {journal} {\bibinfo  {journal} {\mnras}\ }\textbf {\bibinfo {volume} {462}},\ \bibinfo {pages} {2085} (\bibinfo {year} {2016})},\ \Eprint {https://arxiv.org/abs/1607.01189} {arXiv:1607.01189 [astro-ph.CO]} \BibitemShut {NoStop}%
\bibitem [{\citenamefont {{Peacock}}\ and\ \citenamefont {{Bilicki}}(2018)}]{Peacock2018}%
  \BibitemOpen
  \bibfield  {author} {\bibinfo {author} {\bibfnamefont {J.~A.}\ \bibnamefont {{Peacock}}}\ and\ \bibinfo {author} {\bibfnamefont {M.}~\bibnamefont {{Bilicki}}},\ }\bibfield  {title} {\bibinfo {title} {{Wide-area tomography of CMB lensing and the growth of cosmological density fluctuations}},\ }\href {https://doi.org/10.1093/mnras/sty2314} {\bibfield  {journal} {\bibinfo  {journal} {\mnras}\ }\textbf {\bibinfo {volume} {481}},\ \bibinfo {pages} {1133} (\bibinfo {year} {2018})},\ \Eprint {https://arxiv.org/abs/1805.11525} {arXiv:1805.11525 [astro-ph.CO]} \BibitemShut {NoStop}%
\bibitem [{\citenamefont {{Xavier}}\ \emph {et~al.}(2019)\citenamefont {{Xavier}}, \citenamefont {{Costa-Duarte}}, \citenamefont {{Balaguera-Antol{\'\i}nez}},\ and\ \citenamefont {{Bilicki}}}]{Xavier2019}%
  \BibitemOpen
  \bibfield  {author} {\bibinfo {author} {\bibfnamefont {H.~S.}\ \bibnamefont {{Xavier}}}, \bibinfo {author} {\bibfnamefont {M.~V.}\ \bibnamefont {{Costa-Duarte}}}, \bibinfo {author} {\bibfnamefont {A.}~\bibnamefont {{Balaguera-Antol{\'\i}nez}}},\ and\ \bibinfo {author} {\bibfnamefont {M.}~\bibnamefont {{Bilicki}}},\ }\bibfield  {title} {\bibinfo {title} {{All-sky angular power spectra from cleaned WISE{\texttimes}SuperCOSMOS galaxy number counts}},\ }\href {https://doi.org/10.1088/1475-7516/2019/08/037} {\bibfield  {journal} {\bibinfo  {journal} {\jcap}\ }\textbf {\bibinfo {volume} {2019}},\ \bibinfo {eid} {037} (\bibinfo {year} {2019})},\ \Eprint {https://arxiv.org/abs/1812.08182} {arXiv:1812.08182 [astro-ph.CO]} \BibitemShut {NoStop}%
\bibitem [{\citenamefont {{Zhou}}\ \emph {et~al.}(2023)\citenamefont {{Zhou}}, \citenamefont {{Ferraro}}, \citenamefont {{White}}, \citenamefont {{DeRose}}, \citenamefont {{Sailer}}, \citenamefont {{Aguilar}}, \citenamefont {{Ahlen}}, \citenamefont {{Bailey}}, \citenamefont {{Brooks}}, \citenamefont {{Claybaugh}}, \citenamefont {{Dawson}}, \citenamefont {{de la Macorra}}, \citenamefont {{Dey}}, \citenamefont {{Doel}}, \citenamefont {{Font-Ribera}}, \citenamefont {{Forero-Romero}}, \citenamefont {{Gontcho A Gontcho}}, \citenamefont {{Guy}}, \citenamefont {{Kremin}}, \citenamefont {{Lambert}}, \citenamefont {{Le Guillou}}, \citenamefont {{Levi}}, \citenamefont {{Magneville}}, \citenamefont {{Manera}}, \citenamefont {{Meisner}}, \citenamefont {{Miquel}}, \citenamefont {{Moustakas}}, \citenamefont {{Myers}}, \citenamefont {{Newman}}, \citenamefont {{Nie}}, \citenamefont {{Percival}}, \citenamefont {{Rezaie}}, \citenamefont {{Rossi}}, \citenamefont {{Sanchez}}, \citenamefont {{Schlegel}}, \citenamefont
  {{Schubnell}}, \citenamefont {{Seo}}, \citenamefont {{Tarl{\'e}}},\ and\ \citenamefont {{Zhou}}}]{Zhou2023}%
  \BibitemOpen
  \bibfield  {author} {\bibinfo {author} {\bibfnamefont {R.}~\bibnamefont {{Zhou}}}, \bibinfo {author} {\bibfnamefont {S.}~\bibnamefont {{Ferraro}}}, \bibinfo {author} {\bibfnamefont {M.}~\bibnamefont {{White}}}, \bibinfo {author} {\bibfnamefont {J.}~\bibnamefont {{DeRose}}}, \bibinfo {author} {\bibfnamefont {N.}~\bibnamefont {{Sailer}}}, \bibinfo {author} {\bibfnamefont {J.}~\bibnamefont {{Aguilar}}}, \bibinfo {author} {\bibfnamefont {S.}~\bibnamefont {{Ahlen}}}, \bibinfo {author} {\bibfnamefont {S.}~\bibnamefont {{Bailey}}}, \bibinfo {author} {\bibfnamefont {D.}~\bibnamefont {{Brooks}}}, \bibinfo {author} {\bibfnamefont {T.}~\bibnamefont {{Claybaugh}}}, \bibinfo {author} {\bibfnamefont {K.}~\bibnamefont {{Dawson}}}, \bibinfo {author} {\bibfnamefont {A.}~\bibnamefont {{de la Macorra}}}, \bibinfo {author} {\bibfnamefont {B.}~\bibnamefont {{Dey}}}, \bibinfo {author} {\bibfnamefont {P.}~\bibnamefont {{Doel}}}, \bibinfo {author} {\bibfnamefont {A.}~\bibnamefont {{Font-Ribera}}}, \bibinfo {author} {\bibfnamefont
  {J.~E.}\ \bibnamefont {{Forero-Romero}}}, \bibinfo {author} {\bibfnamefont {S.}~\bibnamefont {{Gontcho A Gontcho}}}, \bibinfo {author} {\bibfnamefont {J.}~\bibnamefont {{Guy}}}, \bibinfo {author} {\bibfnamefont {A.}~\bibnamefont {{Kremin}}}, \bibinfo {author} {\bibfnamefont {A.}~\bibnamefont {{Lambert}}}, \bibinfo {author} {\bibfnamefont {L.}~\bibnamefont {{Le Guillou}}}, \bibinfo {author} {\bibfnamefont {M.}~\bibnamefont {{Levi}}}, \bibinfo {author} {\bibfnamefont {C.}~\bibnamefont {{Magneville}}}, \bibinfo {author} {\bibfnamefont {M.}~\bibnamefont {{Manera}}}, \bibinfo {author} {\bibfnamefont {A.}~\bibnamefont {{Meisner}}}, \bibinfo {author} {\bibfnamefont {R.}~\bibnamefont {{Miquel}}}, \bibinfo {author} {\bibfnamefont {J.}~\bibnamefont {{Moustakas}}}, \bibinfo {author} {\bibfnamefont {A.~D.}\ \bibnamefont {{Myers}}}, \bibinfo {author} {\bibfnamefont {J.~A.}\ \bibnamefont {{Newman}}}, \bibinfo {author} {\bibfnamefont {J.}~\bibnamefont {{Nie}}}, \bibinfo {author} {\bibfnamefont {W.}~\bibnamefont
  {{Percival}}}, \bibinfo {author} {\bibfnamefont {M.}~\bibnamefont {{Rezaie}}}, \bibinfo {author} {\bibfnamefont {G.}~\bibnamefont {{Rossi}}}, \bibinfo {author} {\bibfnamefont {E.}~\bibnamefont {{Sanchez}}}, \bibinfo {author} {\bibfnamefont {D.}~\bibnamefont {{Schlegel}}}, \bibinfo {author} {\bibfnamefont {M.}~\bibnamefont {{Schubnell}}}, \bibinfo {author} {\bibfnamefont {H.-J.}\ \bibnamefont {{Seo}}}, \bibinfo {author} {\bibfnamefont {G.}~\bibnamefont {{Tarl{\'e}}}},\ and\ \bibinfo {author} {\bibfnamefont {Z.}~\bibnamefont {{Zhou}}},\ }\bibfield  {title} {\bibinfo {title} {{DESI luminous red galaxy samples for cross-correlations}},\ }\href {https://doi.org/10.1088/1475-7516/2023/11/097} {\bibfield  {journal} {\bibinfo  {journal} {\jcap}\ }\textbf {\bibinfo {volume} {2023}},\ \bibinfo {eid} {097} (\bibinfo {year} {2023})},\ \Eprint {https://arxiv.org/abs/2309.06443} {arXiv:2309.06443 [astro-ph.CO]} \BibitemShut {NoStop}%
\bibitem [{\citenamefont {{Krolewski}}\ \emph {et~al.}(2020)\citenamefont {{Krolewski}}, \citenamefont {{Ferraro}}, \citenamefont {{Schlafly}},\ and\ \citenamefont {{White}}}]{Krolewski2020}%
  \BibitemOpen
  \bibfield  {author} {\bibinfo {author} {\bibfnamefont {A.}~\bibnamefont {{Krolewski}}}, \bibinfo {author} {\bibfnamefont {S.}~\bibnamefont {{Ferraro}}}, \bibinfo {author} {\bibfnamefont {E.~F.}\ \bibnamefont {{Schlafly}}},\ and\ \bibinfo {author} {\bibfnamefont {M.}~\bibnamefont {{White}}},\ }\bibfield  {title} {\bibinfo {title} {{unWISE tomography of Planck CMB lensing}},\ }\href {https://doi.org/10.1088/1475-7516/2020/05/047} {\bibfield  {journal} {\bibinfo  {journal} {\jcap}\ }\textbf {\bibinfo {volume} {2020}},\ \bibinfo {eid} {047} (\bibinfo {year} {2020})},\ \Eprint {https://arxiv.org/abs/1909.07412} {arXiv:1909.07412 [astro-ph.CO]} \BibitemShut {NoStop}%
\bibitem [{\citenamefont {{Krolewski}}\ \emph {et~al.}(2021)\citenamefont {{Krolewski}}, \citenamefont {{Ferraro}},\ and\ \citenamefont {{White}}}]{Krolewski2021}%
  \BibitemOpen
  \bibfield  {author} {\bibinfo {author} {\bibfnamefont {A.}~\bibnamefont {{Krolewski}}}, \bibinfo {author} {\bibfnamefont {S.}~\bibnamefont {{Ferraro}}},\ and\ \bibinfo {author} {\bibfnamefont {M.}~\bibnamefont {{White}}},\ }\bibfield  {title} {\bibinfo {title} {{Cosmological constraints from unWISE and Planck CMB lensing tomography}},\ }\href {https://doi.org/10.1088/1475-7516/2021/12/028} {\bibfield  {journal} {\bibinfo  {journal} {\jcap}\ }\textbf {\bibinfo {volume} {2021}},\ \bibinfo {eid} {028} (\bibinfo {year} {2021})},\ \Eprint {https://arxiv.org/abs/2105.03421} {arXiv:2105.03421 [astro-ph.CO]} \BibitemShut {NoStop}%
\bibitem [{\citenamefont {{Schlafly}}\ \emph {et~al.}(2019)\citenamefont {{Schlafly}}, \citenamefont {{Meisner}},\ and\ \citenamefont {{Green}}}]{Schlafly2019}%
  \BibitemOpen
  \bibfield  {author} {\bibinfo {author} {\bibfnamefont {E.~F.}\ \bibnamefont {{Schlafly}}}, \bibinfo {author} {\bibfnamefont {A.~M.}\ \bibnamefont {{Meisner}}},\ and\ \bibinfo {author} {\bibfnamefont {G.~M.}\ \bibnamefont {{Green}}},\ }\bibfield  {title} {\bibinfo {title} {{The unWISE Catalog: Two Billion Infrared Sources from Five Years of WISE Imaging}},\ }\href {https://doi.org/10.3847/1538-4365/aafbea} {\bibfield  {journal} {\bibinfo  {journal} {\apjs}\ }\textbf {\bibinfo {volume} {240}},\ \bibinfo {eid} {30} (\bibinfo {year} {2019})},\ \Eprint {https://arxiv.org/abs/1901.03337} {arXiv:1901.03337 [astro-ph.IM]} \BibitemShut {NoStop}%
\bibitem [{\citenamefont {{Gaia Collaboration}}\ \emph {et~al.}(2016)\citenamefont {{Gaia Collaboration}}, \citenamefont {{Prusti}}, \citenamefont {{de Bruijne}}, \citenamefont {{Brown}}, \citenamefont {{Vallenari}}, \citenamefont {{Babusiaux}}, \citenamefont {{Bailer-Jones}}, \citenamefont {{Bastian}}, \citenamefont {{Biermann}}, \citenamefont {{Evans}},\ and\ \citenamefont {et~al.}}]{GaiaCollaboration2016}%
  \BibitemOpen
  \bibfield  {author} {\bibinfo {author} {\bibnamefont {{Gaia Collaboration}}}, \bibinfo {author} {\bibfnamefont {T.}~\bibnamefont {{Prusti}}}, \bibinfo {author} {\bibfnamefont {J.~H.~J.}\ \bibnamefont {{de Bruijne}}}, \bibinfo {author} {\bibfnamefont {A.~G.~A.}\ \bibnamefont {{Brown}}}, \bibinfo {author} {\bibfnamefont {A.}~\bibnamefont {{Vallenari}}}, \bibinfo {author} {\bibfnamefont {C.}~\bibnamefont {{Babusiaux}}}, \bibinfo {author} {\bibfnamefont {C.~A.~L.}\ \bibnamefont {{Bailer-Jones}}}, \bibinfo {author} {\bibfnamefont {U.}~\bibnamefont {{Bastian}}}, \bibinfo {author} {\bibfnamefont {M.}~\bibnamefont {{Biermann}}}, \bibinfo {author} {\bibfnamefont {D.~W.}\ \bibnamefont {{Evans}}},\ and\ \bibinfo {author} {\bibnamefont {et~al.}},\ }\bibfield  {title} {\bibinfo {title} {{The Gaia mission}},\ }\href {https://doi.org/10.1051/0004-6361/201629272} {\bibfield  {journal} {\bibinfo  {journal} {\aap}\ }\textbf {\bibinfo {volume} {595}},\ \bibinfo {eid} {A1} (\bibinfo {year} {2016})},\ \Eprint
  {https://arxiv.org/abs/1609.04153} {arXiv:1609.04153 [astro-ph.IM]} \BibitemShut {NoStop}%
\bibitem [{\citenamefont {{Storey-Fisher}}\ \emph {et~al.}(2024)\citenamefont {{Storey-Fisher}}, \citenamefont {{Hogg}}, \citenamefont {{Rix}}, \citenamefont {{Eilers}}, \citenamefont {{Fabbian}}, \citenamefont {{Blanton}},\ and\ \citenamefont {{Alonso}}}]{Storey-Fisher2023}%
  \BibitemOpen
  \bibfield  {author} {\bibinfo {author} {\bibfnamefont {K.}~\bibnamefont {{Storey-Fisher}}}, \bibinfo {author} {\bibfnamefont {D.~W.}\ \bibnamefont {{Hogg}}}, \bibinfo {author} {\bibfnamefont {H.-W.}\ \bibnamefont {{Rix}}}, \bibinfo {author} {\bibfnamefont {A.-C.}\ \bibnamefont {{Eilers}}}, \bibinfo {author} {\bibfnamefont {G.}~\bibnamefont {{Fabbian}}}, \bibinfo {author} {\bibfnamefont {M.~R.}\ \bibnamefont {{Blanton}}},\ and\ \bibinfo {author} {\bibfnamefont {D.}~\bibnamefont {{Alonso}}},\ }\bibfield  {title} {\bibinfo {title} {{Quaia, the Gaia-unWISE Quasar Catalog: An All-sky Spectroscopic Quasar Sample}},\ }\href {https://doi.org/10.3847/1538-4357/ad1328} {\bibfield  {journal} {\bibinfo  {journal} {\apj}\ }\textbf {\bibinfo {volume} {964}},\ \bibinfo {eid} {69} (\bibinfo {year} {2024})},\ \Eprint {https://arxiv.org/abs/2306.17749} {arXiv:2306.17749 [astro-ph.GA]} \BibitemShut {NoStop}%
\bibitem [{\citenamefont {{Lyke}}\ \emph {et~al.}(2020)\citenamefont {{Lyke}}, \citenamefont {{Higley}}, \citenamefont {{McLane}}, \citenamefont {{Schurhammer}}, \citenamefont {{Myers}}, \citenamefont {{Ross}}, \citenamefont {{Dawson}}, \citenamefont {{Chabanier}}, \citenamefont {{Martini}}, \citenamefont {{Busca}}, \citenamefont {{Mas des Bourboux}}, \citenamefont {{Salvato}}, \citenamefont {{Streblyanska}}, \citenamefont {{Zarrouk}}, \citenamefont {{Burtin}}, \citenamefont {{Anderson}}, \citenamefont {{Bautista}}, \citenamefont {{Bizyaev}}, \citenamefont {{Brandt}}, \citenamefont {{Brinkmann}}, \citenamefont {{Brownstein}}, \citenamefont {{Comparat}}, \citenamefont {{Green}}, \citenamefont {{de la Macorra}}, \citenamefont {{Mu{\~n}oz Guti{\'e}rrez}}, \citenamefont {{Hou}}, \citenamefont {{Newman}}, \citenamefont {{Palanque-Delabrouille}}, \citenamefont {{P{\^a}ris}}, \citenamefont {{Percival}}, \citenamefont {{Petitjean}}, \citenamefont {{Rich}}, \citenamefont {{Rossi}}, \citenamefont {{Schneider}},
  \citenamefont {{Smith}}, \citenamefont {{Vivek}},\ and\ \citenamefont {{Weaver}}}]{Lyke2020}%
  \BibitemOpen
  \bibfield  {author} {\bibinfo {author} {\bibfnamefont {B.~W.}\ \bibnamefont {{Lyke}}}, \bibinfo {author} {\bibfnamefont {A.~N.}\ \bibnamefont {{Higley}}}, \bibinfo {author} {\bibfnamefont {J.~N.}\ \bibnamefont {{McLane}}}, \bibinfo {author} {\bibfnamefont {D.~P.}\ \bibnamefont {{Schurhammer}}}, \bibinfo {author} {\bibfnamefont {A.~D.}\ \bibnamefont {{Myers}}}, \bibinfo {author} {\bibfnamefont {A.~J.}\ \bibnamefont {{Ross}}}, \bibinfo {author} {\bibfnamefont {K.}~\bibnamefont {{Dawson}}}, \bibinfo {author} {\bibfnamefont {S.}~\bibnamefont {{Chabanier}}}, \bibinfo {author} {\bibfnamefont {P.}~\bibnamefont {{Martini}}}, \bibinfo {author} {\bibfnamefont {N.~G.}\ \bibnamefont {{Busca}}}, \bibinfo {author} {\bibfnamefont {H.~d.}\ \bibnamefont {{Mas des Bourboux}}}, \bibinfo {author} {\bibfnamefont {M.}~\bibnamefont {{Salvato}}}, \bibinfo {author} {\bibfnamefont {A.}~\bibnamefont {{Streblyanska}}}, \bibinfo {author} {\bibfnamefont {P.}~\bibnamefont {{Zarrouk}}}, \bibinfo {author} {\bibfnamefont {E.}~\bibnamefont
  {{Burtin}}}, \bibinfo {author} {\bibfnamefont {S.~F.}\ \bibnamefont {{Anderson}}}, \bibinfo {author} {\bibfnamefont {J.}~\bibnamefont {{Bautista}}}, \bibinfo {author} {\bibfnamefont {D.}~\bibnamefont {{Bizyaev}}}, \bibinfo {author} {\bibfnamefont {W.~N.}\ \bibnamefont {{Brandt}}}, \bibinfo {author} {\bibfnamefont {J.}~\bibnamefont {{Brinkmann}}}, \bibinfo {author} {\bibfnamefont {J.~R.}\ \bibnamefont {{Brownstein}}}, \bibinfo {author} {\bibfnamefont {J.}~\bibnamefont {{Comparat}}}, \bibinfo {author} {\bibfnamefont {P.}~\bibnamefont {{Green}}}, \bibinfo {author} {\bibfnamefont {A.}~\bibnamefont {{de la Macorra}}}, \bibinfo {author} {\bibfnamefont {A.}~\bibnamefont {{Mu{\~n}oz Guti{\'e}rrez}}}, \bibinfo {author} {\bibfnamefont {J.}~\bibnamefont {{Hou}}}, \bibinfo {author} {\bibfnamefont {J.~A.}\ \bibnamefont {{Newman}}}, \bibinfo {author} {\bibfnamefont {N.}~\bibnamefont {{Palanque-Delabrouille}}}, \bibinfo {author} {\bibfnamefont {I.}~\bibnamefont {{P{\^a}ris}}}, \bibinfo {author} {\bibfnamefont {W.~J.}\
  \bibnamefont {{Percival}}}, \bibinfo {author} {\bibfnamefont {P.}~\bibnamefont {{Petitjean}}}, \bibinfo {author} {\bibfnamefont {J.}~\bibnamefont {{Rich}}}, \bibinfo {author} {\bibfnamefont {G.}~\bibnamefont {{Rossi}}}, \bibinfo {author} {\bibfnamefont {D.~P.}\ \bibnamefont {{Schneider}}}, \bibinfo {author} {\bibfnamefont {A.}~\bibnamefont {{Smith}}}, \bibinfo {author} {\bibfnamefont {M.}~\bibnamefont {{Vivek}}},\ and\ \bibinfo {author} {\bibfnamefont {B.~A.}\ \bibnamefont {{Weaver}}},\ }\bibfield  {title} {\bibinfo {title} {{The Sloan Digital Sky Survey Quasar Catalog: Sixteenth Data Release}},\ }\href {https://doi.org/10.3847/1538-4365/aba623} {\bibfield  {journal} {\bibinfo  {journal} {\apjs}\ }\textbf {\bibinfo {volume} {250}},\ \bibinfo {eid} {8} (\bibinfo {year} {2020})},\ \Eprint {https://arxiv.org/abs/2007.09001} {arXiv:2007.09001 [astro-ph.GA]} \BibitemShut {NoStop}%
\bibitem [{\citenamefont {{Alonso}}\ \emph {et~al.}(2023)\citenamefont {{Alonso}}, \citenamefont {{Fabbian}}, \citenamefont {{Storey-Fisher}}, \citenamefont {{Eilers}}, \citenamefont {{Garc{\'\i}a-Garc{\'\i}a}}, \citenamefont {{Hogg}},\ and\ \citenamefont {{Rix}}}]{Alonso2023}%
  \BibitemOpen
  \bibfield  {author} {\bibinfo {author} {\bibfnamefont {D.}~\bibnamefont {{Alonso}}}, \bibinfo {author} {\bibfnamefont {G.}~\bibnamefont {{Fabbian}}}, \bibinfo {author} {\bibfnamefont {K.}~\bibnamefont {{Storey-Fisher}}}, \bibinfo {author} {\bibfnamefont {A.-C.}\ \bibnamefont {{Eilers}}}, \bibinfo {author} {\bibfnamefont {C.}~\bibnamefont {{Garc{\'\i}a-Garc{\'\i}a}}}, \bibinfo {author} {\bibfnamefont {D.~W.}\ \bibnamefont {{Hogg}}},\ and\ \bibinfo {author} {\bibfnamefont {H.-W.}\ \bibnamefont {{Rix}}},\ }\bibfield  {title} {\bibinfo {title} {{Constraining cosmology with the Gaia-unWISE Quasar Catalog and CMB lensing: structure growth}},\ }\href {https://doi.org/10.1088/1475-7516/2023/11/043} {\bibfield  {journal} {\bibinfo  {journal} {\jcap}\ }\textbf {\bibinfo {volume} {2023}},\ \bibinfo {eid} {043} (\bibinfo {year} {2023})},\ \Eprint {https://arxiv.org/abs/2306.17748} {arXiv:2306.17748 [astro-ph.CO]} \BibitemShut {NoStop}%
\bibitem [{\citenamefont {{Lenz}}\ \emph {et~al.}(2019)\citenamefont {{Lenz}}, \citenamefont {{Dor{\'e}}},\ and\ \citenamefont {{Lagache}}}]{Lenz2019}%
  \BibitemOpen
  \bibfield  {author} {\bibinfo {author} {\bibfnamefont {D.}~\bibnamefont {{Lenz}}}, \bibinfo {author} {\bibfnamefont {O.}~\bibnamefont {{Dor{\'e}}}},\ and\ \bibinfo {author} {\bibfnamefont {G.}~\bibnamefont {{Lagache}}},\ }\bibfield  {title} {\bibinfo {title} {{Large-scale Maps of the Cosmic Infrared Background from Planck}},\ }\href {https://doi.org/10.3847/1538-4357/ab3c2b} {\bibfield  {journal} {\bibinfo  {journal} {\apj}\ }\textbf {\bibinfo {volume} {883}},\ \bibinfo {eid} {75} (\bibinfo {year} {2019})},\ \Eprint {https://arxiv.org/abs/1905.00426} {arXiv:1905.00426 [astro-ph.CO]} \BibitemShut {NoStop}%
\bibitem [{\citenamefont {{Maniyar}}\ \emph {et~al.}(2018)\citenamefont {{Maniyar}}, \citenamefont {{B{\'e}thermin}},\ and\ \citenamefont {{Lagache}}}]{Maniyar2018}%
  \BibitemOpen
  \bibfield  {author} {\bibinfo {author} {\bibfnamefont {A.~S.}\ \bibnamefont {{Maniyar}}}, \bibinfo {author} {\bibfnamefont {M.}~\bibnamefont {{B{\'e}thermin}}},\ and\ \bibinfo {author} {\bibfnamefont {G.}~\bibnamefont {{Lagache}}},\ }\bibfield  {title} {\bibinfo {title} {{Star formation history from the cosmic infrared background anisotropies}},\ }\href {https://doi.org/10.1051/0004-6361/201732499} {\bibfield  {journal} {\bibinfo  {journal} {\aap}\ }\textbf {\bibinfo {volume} {614}},\ \bibinfo {eid} {A39} (\bibinfo {year} {2018})},\ \Eprint {https://arxiv.org/abs/1801.10146} {arXiv:1801.10146 [astro-ph.CO]} \BibitemShut {NoStop}%
\bibitem [{\citenamefont {{Limber}}(1953)}]{Limber1953}%
  \BibitemOpen
  \bibfield  {author} {\bibinfo {author} {\bibfnamefont {D.~N.}\ \bibnamefont {{Limber}}},\ }\bibfield  {title} {\bibinfo {title} {{The Analysis of Counts of the Extragalactic Nebulae in Terms of a Fluctuating Density Field.}},\ }\href {https://doi.org/10.1086/145672} {\bibfield  {journal} {\bibinfo  {journal} {\apj}\ }\textbf {\bibinfo {volume} {117}},\ \bibinfo {pages} {134} (\bibinfo {year} {1953})}\BibitemShut {NoStop}%
\bibitem [{\citenamefont {{LoVerde}}\ and\ \citenamefont {{Afshordi}}(2008)}]{Loverde2008}%
  \BibitemOpen
  \bibfield  {author} {\bibinfo {author} {\bibfnamefont {M.}~\bibnamefont {{LoVerde}}}\ and\ \bibinfo {author} {\bibfnamefont {N.}~\bibnamefont {{Afshordi}}},\ }\bibfield  {title} {\bibinfo {title} {{Extended Limber approximation}},\ }\href {https://doi.org/10.1103/PhysRevD.78.123506} {\bibfield  {journal} {\bibinfo  {journal} {\prd}\ }\textbf {\bibinfo {volume} {78}},\ \bibinfo {eid} {123506} (\bibinfo {year} {2008})},\ \Eprint {https://arxiv.org/abs/0809.5112} {arXiv:0809.5112 [astro-ph]} \BibitemShut {NoStop}%
\bibitem [{\citenamefont {{Chisari}}\ \emph {et~al.}(2019)\citenamefont {{Chisari}}, \citenamefont {{Alonso}}, \citenamefont {{Krause}}, \citenamefont {{Leonard}}, \citenamefont {{Bull}}, \citenamefont {{Neveu}}, \citenamefont {{Villarreal}}, \citenamefont {{Singh}}, \citenamefont {{McClintock}}, \citenamefont {{Ellison}}, \citenamefont {{Du}}, \citenamefont {{Zuntz}}, \citenamefont {{Mead}}, \citenamefont {{Joudaki}}, \citenamefont {{Lorenz}}, \citenamefont {{Tr{\"o}ster}}, \citenamefont {{Sanchez}}, \citenamefont {{Lanusse}}, \citenamefont {{Ishak}}, \citenamefont {{Hlozek}}, \citenamefont {{Blazek}}, \citenamefont {{Campagne}}, \citenamefont {{Almoubayyed}}, \citenamefont {{Eifler}}, \citenamefont {{Kirby}}, \citenamefont {{Kirkby}}, \citenamefont {{Plaszczynski}}, \citenamefont {{Slosar}}, \citenamefont {{Vrastil}}, \citenamefont {{Wagoner}},\ and\ \citenamefont {{LSST Dark Energy Science Collaboration}}}]{Chisari2019}%
  \BibitemOpen
  \bibfield  {author} {\bibinfo {author} {\bibfnamefont {N.~E.}\ \bibnamefont {{Chisari}}}, \bibinfo {author} {\bibfnamefont {D.}~\bibnamefont {{Alonso}}}, \bibinfo {author} {\bibfnamefont {E.}~\bibnamefont {{Krause}}}, \bibinfo {author} {\bibfnamefont {C.~D.}\ \bibnamefont {{Leonard}}}, \bibinfo {author} {\bibfnamefont {P.}~\bibnamefont {{Bull}}}, \bibinfo {author} {\bibfnamefont {J.}~\bibnamefont {{Neveu}}}, \bibinfo {author} {\bibfnamefont {A.~S.}\ \bibnamefont {{Villarreal}}}, \bibinfo {author} {\bibfnamefont {S.}~\bibnamefont {{Singh}}}, \bibinfo {author} {\bibfnamefont {T.}~\bibnamefont {{McClintock}}}, \bibinfo {author} {\bibfnamefont {J.}~\bibnamefont {{Ellison}}}, \bibinfo {author} {\bibfnamefont {Z.}~\bibnamefont {{Du}}}, \bibinfo {author} {\bibfnamefont {J.}~\bibnamefont {{Zuntz}}}, \bibinfo {author} {\bibfnamefont {A.}~\bibnamefont {{Mead}}}, \bibinfo {author} {\bibfnamefont {S.}~\bibnamefont {{Joudaki}}}, \bibinfo {author} {\bibfnamefont {C.~S.}\ \bibnamefont {{Lorenz}}}, \bibinfo {author}
  {\bibfnamefont {T.}~\bibnamefont {{Tr{\"o}ster}}}, \bibinfo {author} {\bibfnamefont {J.}~\bibnamefont {{Sanchez}}}, \bibinfo {author} {\bibfnamefont {F.}~\bibnamefont {{Lanusse}}}, \bibinfo {author} {\bibfnamefont {M.}~\bibnamefont {{Ishak}}}, \bibinfo {author} {\bibfnamefont {R.}~\bibnamefont {{Hlozek}}}, \bibinfo {author} {\bibfnamefont {J.}~\bibnamefont {{Blazek}}}, \bibinfo {author} {\bibfnamefont {J.-E.}\ \bibnamefont {{Campagne}}}, \bibinfo {author} {\bibfnamefont {H.}~\bibnamefont {{Almoubayyed}}}, \bibinfo {author} {\bibfnamefont {T.}~\bibnamefont {{Eifler}}}, \bibinfo {author} {\bibfnamefont {M.}~\bibnamefont {{Kirby}}}, \bibinfo {author} {\bibfnamefont {D.}~\bibnamefont {{Kirkby}}}, \bibinfo {author} {\bibfnamefont {S.}~\bibnamefont {{Plaszczynski}}}, \bibinfo {author} {\bibfnamefont {A.}~\bibnamefont {{Slosar}}}, \bibinfo {author} {\bibfnamefont {M.}~\bibnamefont {{Vrastil}}}, \bibinfo {author} {\bibfnamefont {E.~L.}\ \bibnamefont {{Wagoner}}},\ and\ \bibinfo {author} {\bibnamefont {{LSST Dark
  Energy Science Collaboration}}},\ }\bibfield  {title} {\bibinfo {title} {{Core Cosmology Library: Precision Cosmological Predictions for LSST}},\ }\href {https://doi.org/10.3847/1538-4365/ab1658} {\bibfield  {journal} {\bibinfo  {journal} {\apjs}\ }\textbf {\bibinfo {volume} {242}},\ \bibinfo {eid} {2} (\bibinfo {year} {2019})},\ \Eprint {https://arxiv.org/abs/1812.05995} {arXiv:1812.05995 [astro-ph.CO]} \BibitemShut {NoStop}%
\bibitem [{\citenamefont {{Takahashi}}\ \emph {et~al.}(2012)\citenamefont {{Takahashi}}, \citenamefont {{Sato}}, \citenamefont {{Nishimichi}}, \citenamefont {{Taruya}},\ and\ \citenamefont {{Oguri}}}]{Takahashi2012}%
  \BibitemOpen
  \bibfield  {author} {\bibinfo {author} {\bibfnamefont {R.}~\bibnamefont {{Takahashi}}}, \bibinfo {author} {\bibfnamefont {M.}~\bibnamefont {{Sato}}}, \bibinfo {author} {\bibfnamefont {T.}~\bibnamefont {{Nishimichi}}}, \bibinfo {author} {\bibfnamefont {A.}~\bibnamefont {{Taruya}}},\ and\ \bibinfo {author} {\bibfnamefont {M.}~\bibnamefont {{Oguri}}},\ }\bibfield  {title} {\bibinfo {title} {{Revising the Halofit Model for the Nonlinear Matter Power Spectrum}},\ }\href {https://doi.org/10.1088/0004-637X/761/2/152} {\bibfield  {journal} {\bibinfo  {journal} {\apj}\ }\textbf {\bibinfo {volume} {761}},\ \bibinfo {eid} {152} (\bibinfo {year} {2012})},\ \Eprint {https://arxiv.org/abs/1208.2701} {arXiv:1208.2701 [astro-ph.CO]} \BibitemShut {NoStop}%
\bibitem [{\citenamefont {{Laurent}}\ \emph {et~al.}(2017)\citenamefont {{Laurent}}, \citenamefont {{Eftekharzadeh}}, \citenamefont {{Le Goff}}, \citenamefont {{Myers}}, \citenamefont {{Burtin}}, \citenamefont {{White}}, \citenamefont {{Ross}}, \citenamefont {{Tinker}}, \citenamefont {{Tojeiro}}, \citenamefont {{Bautista}}, \citenamefont {{Brinkmann}}, \citenamefont {{Comparat}}, \citenamefont {{Dawson}}, \citenamefont {{du Mas des Bourboux}}, \citenamefont {{Kneib}}, \citenamefont {{McGreer}}, \citenamefont {{Palanque-Delabrouille}}, \citenamefont {{Percival}}, \citenamefont {{Prada}}, \citenamefont {{Rossi}}, \citenamefont {{Schneider}}, \citenamefont {{Weinberg}}, \citenamefont {{Y{\`e}che}}, \citenamefont {{Zarrouk}},\ and\ \citenamefont {{Zhao}}}]{Laurent2017}%
  \BibitemOpen
  \bibfield  {author} {\bibinfo {author} {\bibfnamefont {P.}~\bibnamefont {{Laurent}}}, \bibinfo {author} {\bibfnamefont {S.}~\bibnamefont {{Eftekharzadeh}}}, \bibinfo {author} {\bibfnamefont {J.-M.}\ \bibnamefont {{Le Goff}}}, \bibinfo {author} {\bibfnamefont {A.}~\bibnamefont {{Myers}}}, \bibinfo {author} {\bibfnamefont {E.}~\bibnamefont {{Burtin}}}, \bibinfo {author} {\bibfnamefont {M.}~\bibnamefont {{White}}}, \bibinfo {author} {\bibfnamefont {A.~J.}\ \bibnamefont {{Ross}}}, \bibinfo {author} {\bibfnamefont {J.}~\bibnamefont {{Tinker}}}, \bibinfo {author} {\bibfnamefont {R.}~\bibnamefont {{Tojeiro}}}, \bibinfo {author} {\bibfnamefont {J.}~\bibnamefont {{Bautista}}}, \bibinfo {author} {\bibfnamefont {J.}~\bibnamefont {{Brinkmann}}}, \bibinfo {author} {\bibfnamefont {J.}~\bibnamefont {{Comparat}}}, \bibinfo {author} {\bibfnamefont {K.}~\bibnamefont {{Dawson}}}, \bibinfo {author} {\bibfnamefont {H.}~\bibnamefont {{du Mas des Bourboux}}}, \bibinfo {author} {\bibfnamefont {J.-P.}\ \bibnamefont {{Kneib}}},
  \bibinfo {author} {\bibfnamefont {I.~D.}\ \bibnamefont {{McGreer}}}, \bibinfo {author} {\bibfnamefont {N.}~\bibnamefont {{Palanque-Delabrouille}}}, \bibinfo {author} {\bibfnamefont {W.~J.}\ \bibnamefont {{Percival}}}, \bibinfo {author} {\bibfnamefont {F.}~\bibnamefont {{Prada}}}, \bibinfo {author} {\bibfnamefont {G.}~\bibnamefont {{Rossi}}}, \bibinfo {author} {\bibfnamefont {D.~P.}\ \bibnamefont {{Schneider}}}, \bibinfo {author} {\bibfnamefont {D.}~\bibnamefont {{Weinberg}}}, \bibinfo {author} {\bibfnamefont {C.}~\bibnamefont {{Y{\`e}che}}}, \bibinfo {author} {\bibfnamefont {P.}~\bibnamefont {{Zarrouk}}},\ and\ \bibinfo {author} {\bibfnamefont {G.-B.}\ \bibnamefont {{Zhao}}},\ }\bibfield  {title} {\bibinfo {title} {{Clustering of quasars in SDSS-IV eBOSS: study of potential systematics and bias determination}},\ }\href {https://doi.org/10.1088/1475-7516/2017/07/017} {\bibfield  {journal} {\bibinfo  {journal} {\jcap}\ }\textbf {\bibinfo {volume} {2017}},\ \bibinfo {eid} {017} (\bibinfo {year} {2017})},\
  \Eprint {https://arxiv.org/abs/1705.04718} {arXiv:1705.04718 [astro-ph.CO]} \BibitemShut {NoStop}%
\bibitem [{\citenamefont {{G{\'o}rski}}\ \emph {et~al.}(2005)\citenamefont {{G{\'o}rski}}, \citenamefont {{Hivon}}, \citenamefont {{Banday}}, \citenamefont {{Wandelt}}, \citenamefont {{Hansen}}, \citenamefont {{Reinecke}},\ and\ \citenamefont {{Bartelmann}}}]{Gorski2005}%
  \BibitemOpen
  \bibfield  {author} {\bibinfo {author} {\bibfnamefont {K.~M.}\ \bibnamefont {{G{\'o}rski}}}, \bibinfo {author} {\bibfnamefont {E.}~\bibnamefont {{Hivon}}}, \bibinfo {author} {\bibfnamefont {A.~J.}\ \bibnamefont {{Banday}}}, \bibinfo {author} {\bibfnamefont {B.~D.}\ \bibnamefont {{Wandelt}}}, \bibinfo {author} {\bibfnamefont {F.~K.}\ \bibnamefont {{Hansen}}}, \bibinfo {author} {\bibfnamefont {M.}~\bibnamefont {{Reinecke}}},\ and\ \bibinfo {author} {\bibfnamefont {M.}~\bibnamefont {{Bartelmann}}},\ }\bibfield  {title} {\bibinfo {title} {{HEALPix: A Framework for High-Resolution Discretization and Fast Analysis of Data Distributed on the Sphere}},\ }\href {https://doi.org/10.1086/427976} {\bibfield  {journal} {\bibinfo  {journal} {\apj}\ }\textbf {\bibinfo {volume} {622}},\ \bibinfo {pages} {759} (\bibinfo {year} {2005})},\ \Eprint {https://arxiv.org/abs/astro-ph/0409513} {arXiv:astro-ph/0409513 [astro-ph]} \BibitemShut {NoStop}%
\bibitem [{\citenamefont {{Zonca}}\ \emph {et~al.}(2019)\citenamefont {{Zonca}}, \citenamefont {{Singer}}, \citenamefont {{Lenz}}, \citenamefont {{Reinecke}}, \citenamefont {{Rosset}}, \citenamefont {{Hivon}},\ and\ \citenamefont {{Gorski}}}]{Zonca2019}%
  \BibitemOpen
  \bibfield  {author} {\bibinfo {author} {\bibfnamefont {A.}~\bibnamefont {{Zonca}}}, \bibinfo {author} {\bibfnamefont {L.}~\bibnamefont {{Singer}}}, \bibinfo {author} {\bibfnamefont {D.}~\bibnamefont {{Lenz}}}, \bibinfo {author} {\bibfnamefont {M.}~\bibnamefont {{Reinecke}}}, \bibinfo {author} {\bibfnamefont {C.}~\bibnamefont {{Rosset}}}, \bibinfo {author} {\bibfnamefont {E.}~\bibnamefont {{Hivon}}},\ and\ \bibinfo {author} {\bibfnamefont {K.}~\bibnamefont {{Gorski}}},\ }\bibfield  {title} {\bibinfo {title} {{healpy: equal area pixelization and spherical harmonics transforms for data on the sphere in Python}},\ }\href {https://doi.org/10.21105/joss.01298} {\bibfield  {journal} {\bibinfo  {journal} {The Journal of Open Source Software}\ }\textbf {\bibinfo {volume} {4}},\ \bibinfo {eid} {1298} (\bibinfo {year} {2019})}\BibitemShut {NoStop}%
\bibitem [{\citenamefont {{Planck Collaboration}}\ \emph {et~al.}(2020{\natexlab{b}})\citenamefont {{Planck Collaboration}}, \citenamefont {{Aghanim}}, \citenamefont {{Akrami}}, \citenamefont {{Ashdown}}, \citenamefont {{Aumont}}, \citenamefont {{Baccigalupi}}, \citenamefont {{Ballardini}}, \citenamefont {{Banday}}, \citenamefont {{Barreiro}}, \citenamefont {{Bartolo}}, \citenamefont {{Basak}}, \citenamefont {{Benabed}}, \citenamefont {{Bernard}}, \citenamefont {{Bersanelli}}, \citenamefont {{Bielewicz}}, \citenamefont {{Bond}}, \citenamefont {{Borrill}}, \citenamefont {{Bouchet}}, \citenamefont {{Boulanger}}, \citenamefont {{Bucher}}, \citenamefont {{Burigana}}, \citenamefont {{Calabrese}}, \citenamefont {{Cardoso}}, \citenamefont {{Carron}}, \citenamefont {{Challinor}}, \citenamefont {{Chiang}}, \citenamefont {{Colombo}}, \citenamefont {{Combet}}, \citenamefont {{Couchot}}, \citenamefont {{Crill}}, \citenamefont {{Cuttaia}}, \citenamefont {{de Bernardis}}, \citenamefont {{de Rosa}}, \citenamefont {{de
  Zotti}}, \citenamefont {{Delabrouille}}, \citenamefont {{Delouis}}, \citenamefont {{Di Valentino}}, \citenamefont {{Diego}}, \citenamefont {{Dor{\'e}}}, \citenamefont {{Douspis}}, \citenamefont {{Ducout}}, \citenamefont {{Dupac}}, \citenamefont {{Efstathiou}}, \citenamefont {{Elsner}}, \citenamefont {{En{\ss}lin}}, \citenamefont {{Eriksen}}, \citenamefont {{Falgarone}}, \citenamefont {{Fantaye}}, \citenamefont {{Finelli}}, \citenamefont {{Frailis}}, \citenamefont {{Fraisse}}, \citenamefont {{Franceschi}}, \citenamefont {{Frolov}}, \citenamefont {{Galeotta}}, \citenamefont {{Galli}}, \citenamefont {{Ganga}}, \citenamefont {{G{\'e}nova-Santos}}, \citenamefont {{Gerbino}}, \citenamefont {{Ghosh}}, \citenamefont {{Gonz{\'a}lez-Nuevo}}, \citenamefont {{G{\'o}rski}}, \citenamefont {{Gratton}}, \citenamefont {{Gruppuso}}, \citenamefont {{Gudmundsson}}, \citenamefont {{Handley}}, \citenamefont {{Hansen}}, \citenamefont {{Henrot-Versill{\'e}}}, \citenamefont {{Herranz}}, \citenamefont {{Hivon}}, \citenamefont
  {{Huang}}, \citenamefont {{Jaffe}}, \citenamefont {{Jones}}, \citenamefont {{Karakci}}, \citenamefont {{Keih{\"a}nen}}, \citenamefont {{Keskitalo}}, \citenamefont {{Kiiveri}}, \citenamefont {{Kim}}, \citenamefont {{Kisner}}, \citenamefont {{Krachmalnicoff}}, \citenamefont {{Kunz}}, \citenamefont {{Kurki-Suonio}}, \citenamefont {{Lagache}}, \citenamefont {{Lamarre}}, \citenamefont {{Lasenby}}, \citenamefont {{Lattanzi}}, \citenamefont {{Lawrence}}, \citenamefont {{Levrier}}, \citenamefont {{Liguori}}, \citenamefont {{Lilje}}, \citenamefont {{Lindholm}}, \citenamefont {{L{\'o}pez-Caniego}}, \citenamefont {{Ma}}, \citenamefont {{Mac{\'\i}as-P{\'e}rez}}, \citenamefont {{Maggio}}, \citenamefont {{Maino}}, \citenamefont {{Mandolesi}}, \citenamefont {{Mangilli}}, \citenamefont {{Martin}}, \citenamefont {{Mart{\'\i}nez-Gonz{\'a}lez}}, \citenamefont {{Matarrese}}, \citenamefont {{Mauri}}, \citenamefont {{McEwen}}, \citenamefont {{Melchiorri}}, \citenamefont {{Mennella}}, \citenamefont {{Migliaccio}}, \citenamefont
  {{Miville-Desch{\^e}nes}}, \citenamefont {{Molinari}}, \citenamefont {{Moneti}}, \citenamefont {{Montier}}, \citenamefont {{Morgante}}, \citenamefont {{Moss}}, \citenamefont {{Mottet}}, \citenamefont {{Natoli}}, \citenamefont {{Pagano}}, \citenamefont {{Paoletti}}, \citenamefont {{Partridge}}, \citenamefont {{Patanchon}}, \citenamefont {{Patrizii}}, \citenamefont {{Perdereau}}, \citenamefont {{Perrotta}}, \citenamefont {{Pettorino}}, \citenamefont {{Piacentini}}, \citenamefont {{Puget}}, \citenamefont {{Rachen}}, \citenamefont {{Reinecke}}, \citenamefont {{Remazeilles}}, \citenamefont {{Renzi}}, \citenamefont {{Rocha}}, \citenamefont {{Roudier}}, \citenamefont {{Salvati}}, \citenamefont {{Sandri}}, \citenamefont {{Savelainen}}, \citenamefont {{Scott}}, \citenamefont {{Sirignano}}, \citenamefont {{Sirri}}, \citenamefont {{Spencer}}, \citenamefont {{Sunyaev}}, \citenamefont {{Suur-Uski}}, \citenamefont {{Tauber}}, \citenamefont {{Tavagnacco}}, \citenamefont {{Tenti}}, \citenamefont {{Toffolatti}},
  \citenamefont {{Tomasi}}, \citenamefont {{Tristram}}, \citenamefont {{Trombetti}}, \citenamefont {{Valiviita}}, \citenamefont {{Vansyngel}}, \citenamefont {{Van Tent}}, \citenamefont {{Vibert}}, \citenamefont {{Vielva}}, \citenamefont {{Villa}}, \citenamefont {{Vittorio}}, \citenamefont {{Wandelt}}, \citenamefont {{Wehus}},\ and\ \citenamefont {{Zonca}}}]{Planck2018}%
  \BibitemOpen
  \bibfield  {author} {\bibinfo {author} {\bibnamefont {{Planck Collaboration}}}, \bibinfo {author} {\bibfnamefont {N.}~\bibnamefont {{Aghanim}}}, \bibinfo {author} {\bibfnamefont {Y.}~\bibnamefont {{Akrami}}}, \bibinfo {author} {\bibfnamefont {M.}~\bibnamefont {{Ashdown}}}, \bibinfo {author} {\bibfnamefont {J.}~\bibnamefont {{Aumont}}}, \bibinfo {author} {\bibfnamefont {C.}~\bibnamefont {{Baccigalupi}}}, \bibinfo {author} {\bibfnamefont {M.}~\bibnamefont {{Ballardini}}}, \bibinfo {author} {\bibfnamefont {A.~J.}\ \bibnamefont {{Banday}}}, \bibinfo {author} {\bibfnamefont {R.~B.}\ \bibnamefont {{Barreiro}}}, \bibinfo {author} {\bibfnamefont {N.}~\bibnamefont {{Bartolo}}}, \bibinfo {author} {\bibfnamefont {S.}~\bibnamefont {{Basak}}}, \bibinfo {author} {\bibfnamefont {K.}~\bibnamefont {{Benabed}}}, \bibinfo {author} {\bibfnamefont {J.~P.}\ \bibnamefont {{Bernard}}}, \bibinfo {author} {\bibfnamefont {M.}~\bibnamefont {{Bersanelli}}}, \bibinfo {author} {\bibfnamefont {P.}~\bibnamefont {{Bielewicz}}}, \bibinfo
  {author} {\bibfnamefont {J.~R.}\ \bibnamefont {{Bond}}}, \bibinfo {author} {\bibfnamefont {J.}~\bibnamefont {{Borrill}}}, \bibinfo {author} {\bibfnamefont {F.~R.}\ \bibnamefont {{Bouchet}}}, \bibinfo {author} {\bibfnamefont {F.}~\bibnamefont {{Boulanger}}}, \bibinfo {author} {\bibfnamefont {M.}~\bibnamefont {{Bucher}}}, \bibinfo {author} {\bibfnamefont {C.}~\bibnamefont {{Burigana}}}, \bibinfo {author} {\bibfnamefont {E.}~\bibnamefont {{Calabrese}}}, \bibinfo {author} {\bibfnamefont {J.~F.}\ \bibnamefont {{Cardoso}}}, \bibinfo {author} {\bibfnamefont {J.}~\bibnamefont {{Carron}}}, \bibinfo {author} {\bibfnamefont {A.}~\bibnamefont {{Challinor}}}, \bibinfo {author} {\bibfnamefont {H.~C.}\ \bibnamefont {{Chiang}}}, \bibinfo {author} {\bibfnamefont {L.~P.~L.}\ \bibnamefont {{Colombo}}}, \bibinfo {author} {\bibfnamefont {C.}~\bibnamefont {{Combet}}}, \bibinfo {author} {\bibfnamefont {F.}~\bibnamefont {{Couchot}}}, \bibinfo {author} {\bibfnamefont {B.~P.}\ \bibnamefont {{Crill}}}, \bibinfo {author}
  {\bibfnamefont {F.}~\bibnamefont {{Cuttaia}}}, \bibinfo {author} {\bibfnamefont {P.}~\bibnamefont {{de Bernardis}}}, \bibinfo {author} {\bibfnamefont {A.}~\bibnamefont {{de Rosa}}}, \bibinfo {author} {\bibfnamefont {G.}~\bibnamefont {{de Zotti}}}, \bibinfo {author} {\bibfnamefont {J.}~\bibnamefont {{Delabrouille}}}, \bibinfo {author} {\bibfnamefont {J.~M.}\ \bibnamefont {{Delouis}}}, \bibinfo {author} {\bibfnamefont {E.}~\bibnamefont {{Di Valentino}}}, \bibinfo {author} {\bibfnamefont {J.~M.}\ \bibnamefont {{Diego}}}, \bibinfo {author} {\bibfnamefont {O.}~\bibnamefont {{Dor{\'e}}}}, \bibinfo {author} {\bibfnamefont {M.}~\bibnamefont {{Douspis}}}, \bibinfo {author} {\bibfnamefont {A.}~\bibnamefont {{Ducout}}}, \bibinfo {author} {\bibfnamefont {X.}~\bibnamefont {{Dupac}}}, \bibinfo {author} {\bibfnamefont {G.}~\bibnamefont {{Efstathiou}}}, \bibinfo {author} {\bibfnamefont {F.}~\bibnamefont {{Elsner}}}, \bibinfo {author} {\bibfnamefont {T.~A.}\ \bibnamefont {{En{\ss}lin}}}, \bibinfo {author} {\bibfnamefont
  {H.~K.}\ \bibnamefont {{Eriksen}}}, \bibinfo {author} {\bibfnamefont {E.}~\bibnamefont {{Falgarone}}}, \bibinfo {author} {\bibfnamefont {Y.}~\bibnamefont {{Fantaye}}}, \bibinfo {author} {\bibfnamefont {F.}~\bibnamefont {{Finelli}}}, \bibinfo {author} {\bibfnamefont {M.}~\bibnamefont {{Frailis}}}, \bibinfo {author} {\bibfnamefont {A.~A.}\ \bibnamefont {{Fraisse}}}, \bibinfo {author} {\bibfnamefont {E.}~\bibnamefont {{Franceschi}}}, \bibinfo {author} {\bibfnamefont {A.}~\bibnamefont {{Frolov}}}, \bibinfo {author} {\bibfnamefont {S.}~\bibnamefont {{Galeotta}}}, \bibinfo {author} {\bibfnamefont {S.}~\bibnamefont {{Galli}}}, \bibinfo {author} {\bibfnamefont {K.}~\bibnamefont {{Ganga}}}, \bibinfo {author} {\bibfnamefont {R.~T.}\ \bibnamefont {{G{\'e}nova-Santos}}}, \bibinfo {author} {\bibfnamefont {M.}~\bibnamefont {{Gerbino}}}, \bibinfo {author} {\bibfnamefont {T.}~\bibnamefont {{Ghosh}}}, \bibinfo {author} {\bibfnamefont {J.}~\bibnamefont {{Gonz{\'a}lez-Nuevo}}}, \bibinfo {author} {\bibfnamefont {K.~M.}\
  \bibnamefont {{G{\'o}rski}}}, \bibinfo {author} {\bibfnamefont {S.}~\bibnamefont {{Gratton}}}, \bibinfo {author} {\bibfnamefont {A.}~\bibnamefont {{Gruppuso}}}, \bibinfo {author} {\bibfnamefont {J.~E.}\ \bibnamefont {{Gudmundsson}}}, \bibinfo {author} {\bibfnamefont {W.}~\bibnamefont {{Handley}}}, \bibinfo {author} {\bibfnamefont {F.~K.}\ \bibnamefont {{Hansen}}}, \bibinfo {author} {\bibfnamefont {S.}~\bibnamefont {{Henrot-Versill{\'e}}}}, \bibinfo {author} {\bibfnamefont {D.}~\bibnamefont {{Herranz}}}, \bibinfo {author} {\bibfnamefont {E.}~\bibnamefont {{Hivon}}}, \bibinfo {author} {\bibfnamefont {Z.}~\bibnamefont {{Huang}}}, \bibinfo {author} {\bibfnamefont {A.~H.}\ \bibnamefont {{Jaffe}}}, \bibinfo {author} {\bibfnamefont {W.~C.}\ \bibnamefont {{Jones}}}, \bibinfo {author} {\bibfnamefont {A.}~\bibnamefont {{Karakci}}}, \bibinfo {author} {\bibfnamefont {E.}~\bibnamefont {{Keih{\"a}nen}}}, \bibinfo {author} {\bibfnamefont {R.}~\bibnamefont {{Keskitalo}}}, \bibinfo {author} {\bibfnamefont {K.}~\bibnamefont
  {{Kiiveri}}}, \bibinfo {author} {\bibfnamefont {J.}~\bibnamefont {{Kim}}}, \bibinfo {author} {\bibfnamefont {T.~S.}\ \bibnamefont {{Kisner}}}, \bibinfo {author} {\bibfnamefont {N.}~\bibnamefont {{Krachmalnicoff}}}, \bibinfo {author} {\bibfnamefont {M.}~\bibnamefont {{Kunz}}}, \bibinfo {author} {\bibfnamefont {H.}~\bibnamefont {{Kurki-Suonio}}}, \bibinfo {author} {\bibfnamefont {G.}~\bibnamefont {{Lagache}}}, \bibinfo {author} {\bibfnamefont {J.~M.}\ \bibnamefont {{Lamarre}}}, \bibinfo {author} {\bibfnamefont {A.}~\bibnamefont {{Lasenby}}}, \bibinfo {author} {\bibfnamefont {M.}~\bibnamefont {{Lattanzi}}}, \bibinfo {author} {\bibfnamefont {C.~R.}\ \bibnamefont {{Lawrence}}}, \bibinfo {author} {\bibfnamefont {F.}~\bibnamefont {{Levrier}}}, \bibinfo {author} {\bibfnamefont {M.}~\bibnamefont {{Liguori}}}, \bibinfo {author} {\bibfnamefont {P.~B.}\ \bibnamefont {{Lilje}}}, \bibinfo {author} {\bibfnamefont {V.}~\bibnamefont {{Lindholm}}}, \bibinfo {author} {\bibfnamefont {M.}~\bibnamefont {{L{\'o}pez-Caniego}}},
  \bibinfo {author} {\bibfnamefont {Y.~Z.}\ \bibnamefont {{Ma}}}, \bibinfo {author} {\bibfnamefont {J.~F.}\ \bibnamefont {{Mac{\'\i}as-P{\'e}rez}}}, \bibinfo {author} {\bibfnamefont {G.}~\bibnamefont {{Maggio}}}, \bibinfo {author} {\bibfnamefont {D.}~\bibnamefont {{Maino}}}, \bibinfo {author} {\bibfnamefont {N.}~\bibnamefont {{Mandolesi}}}, \bibinfo {author} {\bibfnamefont {A.}~\bibnamefont {{Mangilli}}}, \bibinfo {author} {\bibfnamefont {P.~G.}\ \bibnamefont {{Martin}}}, \bibinfo {author} {\bibfnamefont {E.}~\bibnamefont {{Mart{\'\i}nez-Gonz{\'a}lez}}}, \bibinfo {author} {\bibfnamefont {S.}~\bibnamefont {{Matarrese}}}, \bibinfo {author} {\bibfnamefont {N.}~\bibnamefont {{Mauri}}}, \bibinfo {author} {\bibfnamefont {J.~D.}\ \bibnamefont {{McEwen}}}, \bibinfo {author} {\bibfnamefont {A.}~\bibnamefont {{Melchiorri}}}, \bibinfo {author} {\bibfnamefont {A.}~\bibnamefont {{Mennella}}}, \bibinfo {author} {\bibfnamefont {M.}~\bibnamefont {{Migliaccio}}}, \bibinfo {author} {\bibfnamefont {M.~A.}\ \bibnamefont
  {{Miville-Desch{\^e}nes}}}, \bibinfo {author} {\bibfnamefont {D.}~\bibnamefont {{Molinari}}}, \bibinfo {author} {\bibfnamefont {A.}~\bibnamefont {{Moneti}}}, \bibinfo {author} {\bibfnamefont {L.}~\bibnamefont {{Montier}}}, \bibinfo {author} {\bibfnamefont {G.}~\bibnamefont {{Morgante}}}, \bibinfo {author} {\bibfnamefont {A.}~\bibnamefont {{Moss}}}, \bibinfo {author} {\bibfnamefont {S.}~\bibnamefont {{Mottet}}}, \bibinfo {author} {\bibfnamefont {P.}~\bibnamefont {{Natoli}}}, \bibinfo {author} {\bibfnamefont {L.}~\bibnamefont {{Pagano}}}, \bibinfo {author} {\bibfnamefont {D.}~\bibnamefont {{Paoletti}}}, \bibinfo {author} {\bibfnamefont {B.}~\bibnamefont {{Partridge}}}, \bibinfo {author} {\bibfnamefont {G.}~\bibnamefont {{Patanchon}}}, \bibinfo {author} {\bibfnamefont {L.}~\bibnamefont {{Patrizii}}}, \bibinfo {author} {\bibfnamefont {O.}~\bibnamefont {{Perdereau}}}, \bibinfo {author} {\bibfnamefont {F.}~\bibnamefont {{Perrotta}}}, \bibinfo {author} {\bibfnamefont {V.}~\bibnamefont {{Pettorino}}}, \bibinfo
  {author} {\bibfnamefont {F.}~\bibnamefont {{Piacentini}}}, \bibinfo {author} {\bibfnamefont {J.~L.}\ \bibnamefont {{Puget}}}, \bibinfo {author} {\bibfnamefont {J.~P.}\ \bibnamefont {{Rachen}}}, \bibinfo {author} {\bibfnamefont {M.}~\bibnamefont {{Reinecke}}}, \bibinfo {author} {\bibfnamefont {M.}~\bibnamefont {{Remazeilles}}}, \bibinfo {author} {\bibfnamefont {A.}~\bibnamefont {{Renzi}}}, \bibinfo {author} {\bibfnamefont {G.}~\bibnamefont {{Rocha}}}, \bibinfo {author} {\bibfnamefont {G.}~\bibnamefont {{Roudier}}}, \bibinfo {author} {\bibfnamefont {L.}~\bibnamefont {{Salvati}}}, \bibinfo {author} {\bibfnamefont {M.}~\bibnamefont {{Sandri}}}, \bibinfo {author} {\bibfnamefont {M.}~\bibnamefont {{Savelainen}}}, \bibinfo {author} {\bibfnamefont {D.}~\bibnamefont {{Scott}}}, \bibinfo {author} {\bibfnamefont {C.}~\bibnamefont {{Sirignano}}}, \bibinfo {author} {\bibfnamefont {G.}~\bibnamefont {{Sirri}}}, \bibinfo {author} {\bibfnamefont {L.~D.}\ \bibnamefont {{Spencer}}}, \bibinfo {author} {\bibfnamefont
  {R.}~\bibnamefont {{Sunyaev}}}, \bibinfo {author} {\bibfnamefont {A.~S.}\ \bibnamefont {{Suur-Uski}}}, \bibinfo {author} {\bibfnamefont {J.~A.}\ \bibnamefont {{Tauber}}}, \bibinfo {author} {\bibfnamefont {D.}~\bibnamefont {{Tavagnacco}}}, \bibinfo {author} {\bibfnamefont {M.}~\bibnamefont {{Tenti}}}, \bibinfo {author} {\bibfnamefont {L.}~\bibnamefont {{Toffolatti}}}, \bibinfo {author} {\bibfnamefont {M.}~\bibnamefont {{Tomasi}}}, \bibinfo {author} {\bibfnamefont {M.}~\bibnamefont {{Tristram}}}, \bibinfo {author} {\bibfnamefont {T.}~\bibnamefont {{Trombetti}}}, \bibinfo {author} {\bibfnamefont {J.}~\bibnamefont {{Valiviita}}}, \bibinfo {author} {\bibfnamefont {F.}~\bibnamefont {{Vansyngel}}}, \bibinfo {author} {\bibfnamefont {B.}~\bibnamefont {{Van Tent}}}, \bibinfo {author} {\bibfnamefont {L.}~\bibnamefont {{Vibert}}}, \bibinfo {author} {\bibfnamefont {P.}~\bibnamefont {{Vielva}}}, \bibinfo {author} {\bibfnamefont {F.}~\bibnamefont {{Villa}}}, \bibinfo {author} {\bibfnamefont {N.}~\bibnamefont
  {{Vittorio}}}, \bibinfo {author} {\bibfnamefont {B.~D.}\ \bibnamefont {{Wandelt}}}, \bibinfo {author} {\bibfnamefont {I.~K.}\ \bibnamefont {{Wehus}}},\ and\ \bibinfo {author} {\bibfnamefont {A.}~\bibnamefont {{Zonca}}},\ }\bibfield  {title} {\bibinfo {title} {{Planck 2018 results. III. High Frequency Instrument data processing and frequency maps}},\ }\href {https://doi.org/10.1051/0004-6361/201832909} {\bibfield  {journal} {\bibinfo  {journal} {\aap}\ }\textbf {\bibinfo {volume} {641}},\ \bibinfo {eid} {A3} (\bibinfo {year} {2020}{\natexlab{b}})},\ \Eprint {https://arxiv.org/abs/1807.06207} {arXiv:1807.06207 [astro-ph.CO]} \BibitemShut {NoStop}%
\bibitem [{\citenamefont {{Braun}}\ \emph {et~al.}(2008)\citenamefont {{Braun}}, \citenamefont {{Dumm}}, \citenamefont {{De Palma}}, \citenamefont {{Finley}}, \citenamefont {{Karle}},\ and\ \citenamefont {{Montaruli}}}]{Braun2008}%
  \BibitemOpen
  \bibfield  {author} {\bibinfo {author} {\bibfnamefont {J.}~\bibnamefont {{Braun}}}, \bibinfo {author} {\bibfnamefont {J.}~\bibnamefont {{Dumm}}}, \bibinfo {author} {\bibfnamefont {F.}~\bibnamefont {{De Palma}}}, \bibinfo {author} {\bibfnamefont {C.}~\bibnamefont {{Finley}}}, \bibinfo {author} {\bibfnamefont {A.}~\bibnamefont {{Karle}}},\ and\ \bibinfo {author} {\bibfnamefont {T.}~\bibnamefont {{Montaruli}}},\ }\bibfield  {title} {\bibinfo {title} {{Methods for point source analysis in high energy neutrino telescopes}},\ }\href {https://doi.org/10.1016/j.astropartphys.2008.02.007} {\bibfield  {journal} {\bibinfo  {journal} {Astroparticle Physics}\ }\textbf {\bibinfo {volume} {29}},\ \bibinfo {pages} {299} (\bibinfo {year} {2008})},\ \Eprint {https://arxiv.org/abs/0801.1604} {arXiv:0801.1604 [astro-ph]} \BibitemShut {NoStop}%
\bibitem [{\citenamefont {{Abbasi}}\ \emph {et~al.}(2011)\citenamefont {{Abbasi}}, \citenamefont {{Abdou}}, \citenamefont {{Abu-Zayyad}}, \citenamefont {{Adams}}, \citenamefont {{Aguilar}}, \citenamefont {{Ahlers}}, \citenamefont {{Andeen}}, \citenamefont {{Auffenberg}}, \citenamefont {{Bai}}, \citenamefont {{Baker}},\ and\ \citenamefont {et~al.}}]{Abbasi2011}%
  \BibitemOpen
  \bibfield  {author} {\bibinfo {author} {\bibfnamefont {R.}~\bibnamefont {{Abbasi}}}, \bibinfo {author} {\bibfnamefont {Y.}~\bibnamefont {{Abdou}}}, \bibinfo {author} {\bibfnamefont {T.}~\bibnamefont {{Abu-Zayyad}}}, \bibinfo {author} {\bibfnamefont {J.}~\bibnamefont {{Adams}}}, \bibinfo {author} {\bibfnamefont {J.~A.}\ \bibnamefont {{Aguilar}}}, \bibinfo {author} {\bibfnamefont {M.}~\bibnamefont {{Ahlers}}}, \bibinfo {author} {\bibfnamefont {K.}~\bibnamefont {{Andeen}}}, \bibinfo {author} {\bibfnamefont {J.}~\bibnamefont {{Auffenberg}}}, \bibinfo {author} {\bibfnamefont {X.}~\bibnamefont {{Bai}}}, \bibinfo {author} {\bibfnamefont {M.}~\bibnamefont {{Baker}}},\ and\ \bibinfo {author} {\bibnamefont {et~al.}},\ }\bibfield  {title} {\bibinfo {title} {{Time-integrated Searches for Point-like Sources of Neutrinos with the 40-string IceCube Detector}},\ }\href {https://doi.org/10.1088/0004-637X/732/1/18} {\bibfield  {journal} {\bibinfo  {journal} {\apj}\ }\textbf {\bibinfo {volume} {732}},\ \bibinfo {eid} {18}
  (\bibinfo {year} {2011})},\ \Eprint {https://arxiv.org/abs/1012.2137} {arXiv:1012.2137 [astro-ph.HE]} \BibitemShut {NoStop}%
\bibitem [{\citenamefont {{Abbasi}}\ \emph {et~al.}(2023)\citenamefont {{Abbasi}}, \citenamefont {{Ackermann}}, \citenamefont {{Adams}}, \citenamefont {{Agarwalla}}, \citenamefont {{Aguilar}}, \citenamefont {{Ahlers}}, \citenamefont {{Alameddine}}, \citenamefont {{Amin}}, \citenamefont {{Andeen}}, \citenamefont {{Anton}},\ and\ \citenamefont {et~al.}}]{2304.01174}%
  \BibitemOpen
  \bibfield  {author} {\bibinfo {author} {\bibfnamefont {R.}~\bibnamefont {{Abbasi}}}, \bibinfo {author} {\bibfnamefont {M.}~\bibnamefont {{Ackermann}}}, \bibinfo {author} {\bibfnamefont {J.}~\bibnamefont {{Adams}}}, \bibinfo {author} {\bibfnamefont {S.~K.}\ \bibnamefont {{Agarwalla}}}, \bibinfo {author} {\bibfnamefont {J.~A.}\ \bibnamefont {{Aguilar}}}, \bibinfo {author} {\bibfnamefont {M.}~\bibnamefont {{Ahlers}}}, \bibinfo {author} {\bibfnamefont {J.~M.}\ \bibnamefont {{Alameddine}}}, \bibinfo {author} {\bibfnamefont {N.~M.}\ \bibnamefont {{Amin}}}, \bibinfo {author} {\bibfnamefont {K.}~\bibnamefont {{Andeen}}}, \bibinfo {author} {\bibfnamefont {G.}~\bibnamefont {{Anton}}},\ and\ \bibinfo {author} {\bibnamefont {et~al.}},\ }\bibfield  {title} {\bibinfo {title} {{IceCat-1: The IceCube Event Catalog of Alert Tracks}},\ }\href {https://doi.org/10.3847/1538-4365/acfa95} {\bibfield  {journal} {\bibinfo  {journal} {\apjs}\ }\textbf {\bibinfo {volume} {269}},\ \bibinfo {eid} {25} (\bibinfo {year} {2023})},\
  \Eprint {https://arxiv.org/abs/2304.01174} {arXiv:2304.01174 [astro-ph.HE]} \BibitemShut {NoStop}%
\bibitem [{\citenamefont {{Alonso}}\ \emph {et~al.}(2019)\citenamefont {{Alonso}}, \citenamefont {{Sanchez}}, \citenamefont {{Slosar}},\ and\ \citenamefont {{LSST Dark Energy Science Collaboration}}}]{Alonso2019}%
  \BibitemOpen
  \bibfield  {author} {\bibinfo {author} {\bibfnamefont {D.}~\bibnamefont {{Alonso}}}, \bibinfo {author} {\bibfnamefont {J.}~\bibnamefont {{Sanchez}}}, \bibinfo {author} {\bibfnamefont {A.}~\bibnamefont {{Slosar}}},\ and\ \bibinfo {author} {\bibnamefont {{LSST Dark Energy Science Collaboration}}},\ }\bibfield  {title} {\bibinfo {title} {{A unified pseudo-C$_{{\ensuremath{\ell}}}$ framework}},\ }\href {https://doi.org/10.1093/mnras/stz093} {\bibfield  {journal} {\bibinfo  {journal} {\mnras}\ }\textbf {\bibinfo {volume} {484}},\ \bibinfo {pages} {4127} (\bibinfo {year} {2019})},\ \Eprint {https://arxiv.org/abs/1809.09603} {arXiv:1809.09603 [astro-ph.CO]} \BibitemShut {NoStop}%
\bibitem [{\citenamefont {{Hivon}}\ \emph {et~al.}(2002)\citenamefont {{Hivon}}, \citenamefont {{G{\'o}rski}}, \citenamefont {{Netterfield}}, \citenamefont {{Crill}}, \citenamefont {{Prunet}},\ and\ \citenamefont {{Hansen}}}]{Hivon2002}%
  \BibitemOpen
  \bibfield  {author} {\bibinfo {author} {\bibfnamefont {E.}~\bibnamefont {{Hivon}}}, \bibinfo {author} {\bibfnamefont {K.~M.}\ \bibnamefont {{G{\'o}rski}}}, \bibinfo {author} {\bibfnamefont {C.~B.}\ \bibnamefont {{Netterfield}}}, \bibinfo {author} {\bibfnamefont {B.~P.}\ \bibnamefont {{Crill}}}, \bibinfo {author} {\bibfnamefont {S.}~\bibnamefont {{Prunet}}},\ and\ \bibinfo {author} {\bibfnamefont {F.}~\bibnamefont {{Hansen}}},\ }\bibfield  {title} {\bibinfo {title} {{MASTER of the Cosmic Microwave Background Anisotropy Power Spectrum: A Fast Method for Statistical Analysis of Large and Complex Cosmic Microwave Background Data Sets}},\ }\href {https://doi.org/10.1086/338126} {\bibfield  {journal} {\bibinfo  {journal} {\apj}\ }\textbf {\bibinfo {volume} {567}},\ \bibinfo {pages} {2} (\bibinfo {year} {2002})},\ \Eprint {https://arxiv.org/abs/astro-ph/0105302} {arXiv:astro-ph/0105302 [astro-ph]} \BibitemShut {NoStop}%
\bibitem [{\citenamefont {{Garc{\'\i}a-Garc{\'\i}a}}\ \emph {et~al.}(2019)\citenamefont {{Garc{\'\i}a-Garc{\'\i}a}}, \citenamefont {{Alonso}},\ and\ \citenamefont {{Bellini}}}]{Garcia-Garcia2019}%
  \BibitemOpen
  \bibfield  {author} {\bibinfo {author} {\bibfnamefont {C.}~\bibnamefont {{Garc{\'\i}a-Garc{\'\i}a}}}, \bibinfo {author} {\bibfnamefont {D.}~\bibnamefont {{Alonso}}},\ and\ \bibinfo {author} {\bibfnamefont {E.}~\bibnamefont {{Bellini}}},\ }\bibfield  {title} {\bibinfo {title} {{Disconnected pseudo-C$_{l}$ covariances for projected large-scale structure data}},\ }\href {https://doi.org/10.1088/1475-7516/2019/11/043} {\bibfield  {journal} {\bibinfo  {journal} {\jcap}\ }\textbf {\bibinfo {volume} {2019}},\ \bibinfo {eid} {043} (\bibinfo {year} {2019})},\ \Eprint {https://arxiv.org/abs/1906.11765} {arXiv:1906.11765 [astro-ph.CO]} \BibitemShut {NoStop}%
\bibitem [{\citenamefont {{LSST Science Collaboration}}\ \emph {et~al.}(2009)\citenamefont {{LSST Science Collaboration}}, \citenamefont {{Abell}}, \citenamefont {{Allison}}, \citenamefont {{Anderson}}, \citenamefont {{Andrew}}, \citenamefont {{Angel}}, \citenamefont {{Armus}}, \citenamefont {{Arnett}}, \citenamefont {{Asztalos}}, \citenamefont {{Axelrod}},\ and\ \citenamefont {et~al.}}]{0912.0201}%
  \BibitemOpen
  \bibfield  {author} {\bibinfo {author} {\bibnamefont {{LSST Science Collaboration}}}, \bibinfo {author} {\bibfnamefont {P.~A.}\ \bibnamefont {{Abell}}}, \bibinfo {author} {\bibfnamefont {J.}~\bibnamefont {{Allison}}}, \bibinfo {author} {\bibfnamefont {S.~F.}\ \bibnamefont {{Anderson}}}, \bibinfo {author} {\bibfnamefont {J.~R.}\ \bibnamefont {{Andrew}}}, \bibinfo {author} {\bibfnamefont {J.~R.~P.}\ \bibnamefont {{Angel}}}, \bibinfo {author} {\bibfnamefont {L.}~\bibnamefont {{Armus}}}, \bibinfo {author} {\bibfnamefont {D.}~\bibnamefont {{Arnett}}}, \bibinfo {author} {\bibfnamefont {S.~J.}\ \bibnamefont {{Asztalos}}}, \bibinfo {author} {\bibfnamefont {T.~S.}\ \bibnamefont {{Axelrod}}},\ and\ \bibinfo {author} {\bibnamefont {et~al.}},\ }\bibfield  {title} {\bibinfo {title} {{LSST Science Book, Version 2.0}},\ }\href {https://doi.org/10.48550/arXiv.0912.0201} {\bibfield  {journal} {\bibinfo  {journal} {arXiv e-prints}\ ,\ \bibinfo {eid} {arXiv:0912.0201}} (\bibinfo {year} {2009})},\ \Eprint
  {https://arxiv.org/abs/0912.0201} {arXiv:0912.0201 [astro-ph.IM]} \BibitemShut {NoStop}%
\bibitem [{\citenamefont {{Tung}}\ \emph {et~al.}(2021)\citenamefont {{Tung}}, \citenamefont {{Glauch}}, \citenamefont {{Larson}}, \citenamefont {{Pizzuto}}, \citenamefont {{Reimann}},\ and\ \citenamefont {{Taboada}}}]{Tung2021}%
  \BibitemOpen
  \bibfield  {author} {\bibinfo {author} {\bibfnamefont {C.}~\bibnamefont {{Tung}}}, \bibinfo {author} {\bibfnamefont {T.}~\bibnamefont {{Glauch}}}, \bibinfo {author} {\bibfnamefont {M.}~\bibnamefont {{Larson}}}, \bibinfo {author} {\bibfnamefont {A.}~\bibnamefont {{Pizzuto}}}, \bibinfo {author} {\bibfnamefont {R.}~\bibnamefont {{Reimann}}},\ and\ \bibinfo {author} {\bibfnamefont {I.}~\bibnamefont {{Taboada}}},\ }\bibfield  {title} {\bibinfo {title} {{FIRESONG: A python package to simulate populations of extragalactic neutrino sources}},\ }\href {https://doi.org/10.21105/joss.03194} {\bibfield  {journal} {\bibinfo  {journal} {The Journal of Open Source Software}\ }\textbf {\bibinfo {volume} {6}},\ \bibinfo {pages} {3194} (\bibinfo {year} {2021})}\BibitemShut {NoStop}%
\bibitem [{\citenamefont {{Aartsen}}\ \emph {et~al.}(2021)\citenamefont {{Aartsen}}, \citenamefont {{Abbasi}}, \citenamefont {{Ackermann}}, \citenamefont {{Adams}}, \citenamefont {{Aguilar}}, \citenamefont {{Ahlers}}, \citenamefont {{Ahrens}}, \citenamefont {{Alispach}}, \citenamefont {{Allison}}, \citenamefont {{Amin}},\ and\ \citenamefont {et~al.}}]{2008.04323}%
  \BibitemOpen
  \bibfield  {author} {\bibinfo {author} {\bibfnamefont {M.~G.}\ \bibnamefont {{Aartsen}}}, \bibinfo {author} {\bibfnamefont {R.}~\bibnamefont {{Abbasi}}}, \bibinfo {author} {\bibfnamefont {M.}~\bibnamefont {{Ackermann}}}, \bibinfo {author} {\bibfnamefont {J.}~\bibnamefont {{Adams}}}, \bibinfo {author} {\bibfnamefont {J.~A.}\ \bibnamefont {{Aguilar}}}, \bibinfo {author} {\bibfnamefont {M.}~\bibnamefont {{Ahlers}}}, \bibinfo {author} {\bibfnamefont {M.}~\bibnamefont {{Ahrens}}}, \bibinfo {author} {\bibfnamefont {C.}~\bibnamefont {{Alispach}}}, \bibinfo {author} {\bibfnamefont {P.}~\bibnamefont {{Allison}}}, \bibinfo {author} {\bibfnamefont {N.~M.}\ \bibnamefont {{Amin}}},\ and\ \bibinfo {author} {\bibnamefont {et~al.}},\ }\bibfield  {title} {\bibinfo {title} {{IceCube-Gen2: the window to the extreme Universe}},\ }\href {https://doi.org/10.1088/1361-6471/abbd48} {\bibfield  {journal} {\bibinfo  {journal} {Journal of Physics G Nuclear Physics}\ }\textbf {\bibinfo {volume} {48}},\ \bibinfo {eid} {060501}
  (\bibinfo {year} {2021})},\ \Eprint {https://arxiv.org/abs/2008.04323} {arXiv:2008.04323 [astro-ph.HE]} \BibitemShut {NoStop}%
\bibitem [{\citenamefont {{Stein Muzio}}\ and\ \citenamefont {{Globus}}(2023)}]{Muzio2023}%
  \BibitemOpen
  \bibfield  {author} {\bibinfo {author} {\bibfnamefont {M.}~\bibnamefont {{Stein Muzio}}}\ and\ \bibinfo {author} {\bibfnamefont {N.}~\bibnamefont {{Globus}}},\ }\bibfield  {title} {\bibinfo {title} {{Neutrino anisotropy as a probe of extreme astrophysical accelerators}},\ }\href {https://doi.org/10.48550/arXiv.2308.16225} {\bibfield  {journal} {\bibinfo  {journal} {arXiv e-prints}\ ,\ \bibinfo {eid} {arXiv:2308.16225}} (\bibinfo {year} {2023})},\ \Eprint {https://arxiv.org/abs/2308.16225} {arXiv:2308.16225 [astro-ph.HE]} \BibitemShut {NoStop}%
\bibitem [{\citenamefont {{Hahn}}\ \emph {et~al.}(2023)\citenamefont {{Hahn}}, \citenamefont {{Wilson}}, \citenamefont {{Ruiz-Macias}}, \citenamefont {{Cole}}, \citenamefont {{Weinberg}}, \citenamefont {{Moustakas}}, \citenamefont {{Kremin}}, \citenamefont {{Tinker}}, \citenamefont {{Smith}}, \citenamefont {{Wechsler}}, \citenamefont {{Ahlen}}, \citenamefont {{Alam}}, \citenamefont {{Bailey}}, \citenamefont {{Brooks}}, \citenamefont {{Cooper}}, \citenamefont {{Davis}}, \citenamefont {{Dawson}}, \citenamefont {{Dey}}, \citenamefont {{Dey}}, \citenamefont {{Eftekharzadeh}}, \citenamefont {{Eisenstein}}, \citenamefont {{Fanning}}, \citenamefont {{Forero-Romero}}, \citenamefont {{Frenk}}, \citenamefont {{Gazta{\~n}aga}}, \citenamefont {{Gontcho A Gontcho}}, \citenamefont {{Guy}}, \citenamefont {{Honscheid}}, \citenamefont {{Ishak}}, \citenamefont {{Juneau}}, \citenamefont {{Kehoe}}, \citenamefont {{Kisner}}, \citenamefont {{Lan}}, \citenamefont {{Landriau}}, \citenamefont {{Le Guillou}}, \citenamefont {{Levi}},
  \citenamefont {{Magneville}}, \citenamefont {{Martini}}, \citenamefont {{Meisner}}, \citenamefont {{Myers}}, \citenamefont {{Nie}}, \citenamefont {{Norberg}}, \citenamefont {{Palanque-Delabrouille}}, \citenamefont {{Percival}}, \citenamefont {{Poppett}}, \citenamefont {{Prada}}, \citenamefont {{Raichoor}}, \citenamefont {{Ross}}, \citenamefont {{Safonova}}, \citenamefont {{Saulder}}, \citenamefont {{Schlafly}}, \citenamefont {{Schlegel}}, \citenamefont {{Sierra-Porta}}, \citenamefont {{Tarle}}, \citenamefont {{Weaver}}, \citenamefont {{Y{\`e}che}}, \citenamefont {{Zarrouk}}, \citenamefont {{Zhou}}, \citenamefont {{Zhou}},\ and\ \citenamefont {{Zou}}}]{Hahn2023}%
  \BibitemOpen
  \bibfield  {author} {\bibinfo {author} {\bibfnamefont {C.}~\bibnamefont {{Hahn}}}, \bibinfo {author} {\bibfnamefont {M.~J.}\ \bibnamefont {{Wilson}}}, \bibinfo {author} {\bibfnamefont {O.}~\bibnamefont {{Ruiz-Macias}}}, \bibinfo {author} {\bibfnamefont {S.}~\bibnamefont {{Cole}}}, \bibinfo {author} {\bibfnamefont {D.~H.}\ \bibnamefont {{Weinberg}}}, \bibinfo {author} {\bibfnamefont {J.}~\bibnamefont {{Moustakas}}}, \bibinfo {author} {\bibfnamefont {A.}~\bibnamefont {{Kremin}}}, \bibinfo {author} {\bibfnamefont {J.~L.}\ \bibnamefont {{Tinker}}}, \bibinfo {author} {\bibfnamefont {A.}~\bibnamefont {{Smith}}}, \bibinfo {author} {\bibfnamefont {R.~H.}\ \bibnamefont {{Wechsler}}}, \bibinfo {author} {\bibfnamefont {S.}~\bibnamefont {{Ahlen}}}, \bibinfo {author} {\bibfnamefont {S.}~\bibnamefont {{Alam}}}, \bibinfo {author} {\bibfnamefont {S.}~\bibnamefont {{Bailey}}}, \bibinfo {author} {\bibfnamefont {D.}~\bibnamefont {{Brooks}}}, \bibinfo {author} {\bibfnamefont {A.~P.}\ \bibnamefont {{Cooper}}}, \bibinfo {author}
  {\bibfnamefont {T.~M.}\ \bibnamefont {{Davis}}}, \bibinfo {author} {\bibfnamefont {K.}~\bibnamefont {{Dawson}}}, \bibinfo {author} {\bibfnamefont {A.}~\bibnamefont {{Dey}}}, \bibinfo {author} {\bibfnamefont {B.}~\bibnamefont {{Dey}}}, \bibinfo {author} {\bibfnamefont {S.}~\bibnamefont {{Eftekharzadeh}}}, \bibinfo {author} {\bibfnamefont {D.~J.}\ \bibnamefont {{Eisenstein}}}, \bibinfo {author} {\bibfnamefont {K.}~\bibnamefont {{Fanning}}}, \bibinfo {author} {\bibfnamefont {J.~E.}\ \bibnamefont {{Forero-Romero}}}, \bibinfo {author} {\bibfnamefont {C.~S.}\ \bibnamefont {{Frenk}}}, \bibinfo {author} {\bibfnamefont {E.}~\bibnamefont {{Gazta{\~n}aga}}}, \bibinfo {author} {\bibfnamefont {S.}~\bibnamefont {{Gontcho A Gontcho}}}, \bibinfo {author} {\bibfnamefont {J.}~\bibnamefont {{Guy}}}, \bibinfo {author} {\bibfnamefont {K.}~\bibnamefont {{Honscheid}}}, \bibinfo {author} {\bibfnamefont {M.}~\bibnamefont {{Ishak}}}, \bibinfo {author} {\bibfnamefont {S.}~\bibnamefont {{Juneau}}}, \bibinfo {author} {\bibfnamefont
  {R.}~\bibnamefont {{Kehoe}}}, \bibinfo {author} {\bibfnamefont {T.}~\bibnamefont {{Kisner}}}, \bibinfo {author} {\bibfnamefont {T.-W.}\ \bibnamefont {{Lan}}}, \bibinfo {author} {\bibfnamefont {M.}~\bibnamefont {{Landriau}}}, \bibinfo {author} {\bibfnamefont {L.}~\bibnamefont {{Le Guillou}}}, \bibinfo {author} {\bibfnamefont {M.~E.}\ \bibnamefont {{Levi}}}, \bibinfo {author} {\bibfnamefont {C.}~\bibnamefont {{Magneville}}}, \bibinfo {author} {\bibfnamefont {P.}~\bibnamefont {{Martini}}}, \bibinfo {author} {\bibfnamefont {A.}~\bibnamefont {{Meisner}}}, \bibinfo {author} {\bibfnamefont {A.~D.}\ \bibnamefont {{Myers}}}, \bibinfo {author} {\bibfnamefont {J.}~\bibnamefont {{Nie}}}, \bibinfo {author} {\bibfnamefont {P.}~\bibnamefont {{Norberg}}}, \bibinfo {author} {\bibfnamefont {N.}~\bibnamefont {{Palanque-Delabrouille}}}, \bibinfo {author} {\bibfnamefont {W.~J.}\ \bibnamefont {{Percival}}}, \bibinfo {author} {\bibfnamefont {C.}~\bibnamefont {{Poppett}}}, \bibinfo {author} {\bibfnamefont {F.}~\bibnamefont
  {{Prada}}}, \bibinfo {author} {\bibfnamefont {A.}~\bibnamefont {{Raichoor}}}, \bibinfo {author} {\bibfnamefont {A.~J.}\ \bibnamefont {{Ross}}}, \bibinfo {author} {\bibfnamefont {S.}~\bibnamefont {{Safonova}}}, \bibinfo {author} {\bibfnamefont {C.}~\bibnamefont {{Saulder}}}, \bibinfo {author} {\bibfnamefont {E.}~\bibnamefont {{Schlafly}}}, \bibinfo {author} {\bibfnamefont {D.}~\bibnamefont {{Schlegel}}}, \bibinfo {author} {\bibfnamefont {D.}~\bibnamefont {{Sierra-Porta}}}, \bibinfo {author} {\bibfnamefont {G.}~\bibnamefont {{Tarle}}}, \bibinfo {author} {\bibfnamefont {B.~A.}\ \bibnamefont {{Weaver}}}, \bibinfo {author} {\bibfnamefont {C.}~\bibnamefont {{Y{\`e}che}}}, \bibinfo {author} {\bibfnamefont {P.}~\bibnamefont {{Zarrouk}}}, \bibinfo {author} {\bibfnamefont {R.}~\bibnamefont {{Zhou}}}, \bibinfo {author} {\bibfnamefont {Z.}~\bibnamefont {{Zhou}}},\ and\ \bibinfo {author} {\bibfnamefont {H.}~\bibnamefont {{Zou}}},\ }\bibfield  {title} {\bibinfo {title} {{The DESI Bright Galaxy Survey: Final Target
  Selection, Design, and Validation}},\ }\href {https://doi.org/10.3847/1538-3881/accff8} {\bibfield  {journal} {\bibinfo  {journal} {\aj}\ }\textbf {\bibinfo {volume} {165}},\ \bibinfo {eid} {253} (\bibinfo {year} {2023})},\ \Eprint {https://arxiv.org/abs/2208.08512} {arXiv:2208.08512 [astro-ph.CO]} \BibitemShut {NoStop}%
\bibitem [{\citenamefont {{Harris}}\ \emph {et~al.}(2020)\citenamefont {{Harris}}, \citenamefont {{Millman}}, \citenamefont {{van der Walt}}, \citenamefont {{Gommers}}, \citenamefont {{Virtanen}}, \citenamefont {{Cournapeau}}, \citenamefont {{Wieser}}, \citenamefont {{Taylor}}, \citenamefont {{Berg}}, \citenamefont {{Smith}}, \citenamefont {{Kern}}, \citenamefont {{Picus}}, \citenamefont {{Hoyer}}, \citenamefont {{van Kerkwijk}}, \citenamefont {{Brett}}, \citenamefont {{Haldane}}, \citenamefont {{del R{\'\i}o}}, \citenamefont {{Wiebe}}, \citenamefont {{Peterson}}, \citenamefont {{G{\'e}rard-Marchant}}, \citenamefont {{Sheppard}}, \citenamefont {{Reddy}}, \citenamefont {{Weckesser}}, \citenamefont {{Abbasi}}, \citenamefont {{Gohlke}},\ and\ \citenamefont {{Oliphant}}}]{Harris2020}%
  \BibitemOpen
  \bibfield  {author} {\bibinfo {author} {\bibfnamefont {C.~R.}\ \bibnamefont {{Harris}}}, \bibinfo {author} {\bibfnamefont {K.~J.}\ \bibnamefont {{Millman}}}, \bibinfo {author} {\bibfnamefont {S.~J.}\ \bibnamefont {{van der Walt}}}, \bibinfo {author} {\bibfnamefont {R.}~\bibnamefont {{Gommers}}}, \bibinfo {author} {\bibfnamefont {P.}~\bibnamefont {{Virtanen}}}, \bibinfo {author} {\bibfnamefont {D.}~\bibnamefont {{Cournapeau}}}, \bibinfo {author} {\bibfnamefont {E.}~\bibnamefont {{Wieser}}}, \bibinfo {author} {\bibfnamefont {J.}~\bibnamefont {{Taylor}}}, \bibinfo {author} {\bibfnamefont {S.}~\bibnamefont {{Berg}}}, \bibinfo {author} {\bibfnamefont {N.~J.}\ \bibnamefont {{Smith}}}, \bibinfo {author} {\bibfnamefont {R.}~\bibnamefont {{Kern}}}, \bibinfo {author} {\bibfnamefont {M.}~\bibnamefont {{Picus}}}, \bibinfo {author} {\bibfnamefont {S.}~\bibnamefont {{Hoyer}}}, \bibinfo {author} {\bibfnamefont {M.~H.}\ \bibnamefont {{van Kerkwijk}}}, \bibinfo {author} {\bibfnamefont {M.}~\bibnamefont {{Brett}}}, \bibinfo
  {author} {\bibfnamefont {A.}~\bibnamefont {{Haldane}}}, \bibinfo {author} {\bibfnamefont {J.~F.}\ \bibnamefont {{del R{\'\i}o}}}, \bibinfo {author} {\bibfnamefont {M.}~\bibnamefont {{Wiebe}}}, \bibinfo {author} {\bibfnamefont {P.}~\bibnamefont {{Peterson}}}, \bibinfo {author} {\bibfnamefont {P.}~\bibnamefont {{G{\'e}rard-Marchant}}}, \bibinfo {author} {\bibfnamefont {K.}~\bibnamefont {{Sheppard}}}, \bibinfo {author} {\bibfnamefont {T.}~\bibnamefont {{Reddy}}}, \bibinfo {author} {\bibfnamefont {W.}~\bibnamefont {{Weckesser}}}, \bibinfo {author} {\bibfnamefont {H.}~\bibnamefont {{Abbasi}}}, \bibinfo {author} {\bibfnamefont {C.}~\bibnamefont {{Gohlke}}},\ and\ \bibinfo {author} {\bibfnamefont {T.~E.}\ \bibnamefont {{Oliphant}}},\ }\bibfield  {title} {\bibinfo {title} {{Array programming with NumPy}},\ }\href {https://doi.org/10.1038/s41586-020-2649-2} {\bibfield  {journal} {\bibinfo  {journal} {\nat}\ }\textbf {\bibinfo {volume} {585}},\ \bibinfo {pages} {357} (\bibinfo {year} {2020})},\ \Eprint
  {https://arxiv.org/abs/2006.10256} {arXiv:2006.10256 [cs.MS]} \BibitemShut {NoStop}%
\bibitem [{\citenamefont {{Virtanen}}\ \emph {et~al.}(2020)\citenamefont {{Virtanen}}, \citenamefont {{Gommers}}, \citenamefont {{Oliphant}}, \citenamefont {{Haberland}}, \citenamefont {{Reddy}}, \citenamefont {{Cournapeau}}, \citenamefont {{Burovski}}, \citenamefont {{Peterson}}, \citenamefont {{Weckesser}}, \citenamefont {{Bright}}, \citenamefont {{van der Walt}}, \citenamefont {{Brett}}, \citenamefont {{Wilson}}, \citenamefont {{Millman}}, \citenamefont {{Mayorov}}, \citenamefont {{Nelson}}, \citenamefont {{Jones}}, \citenamefont {{Kern}}, \citenamefont {{Larson}}, \citenamefont {{Carey}}, \citenamefont {{Polat}}, \citenamefont {{Feng}}, \citenamefont {{Moore}}, \citenamefont {{VanderPlas}}, \citenamefont {{Laxalde}}, \citenamefont {{Perktold}}, \citenamefont {{Cimrman}}, \citenamefont {{Henriksen}}, \citenamefont {{Quintero}}, \citenamefont {{Harris}}, \citenamefont {{Archibald}}, \citenamefont {{Ribeiro}}, \citenamefont {{Pedregosa}}, \citenamefont {{van Mulbregt}},\ and\ \citenamefont {{SciPy 1. 0
  Contributors}}}]{Virtanen2020}%
  \BibitemOpen
  \bibfield  {author} {\bibinfo {author} {\bibfnamefont {P.}~\bibnamefont {{Virtanen}}}, \bibinfo {author} {\bibfnamefont {R.}~\bibnamefont {{Gommers}}}, \bibinfo {author} {\bibfnamefont {T.~E.}\ \bibnamefont {{Oliphant}}}, \bibinfo {author} {\bibfnamefont {M.}~\bibnamefont {{Haberland}}}, \bibinfo {author} {\bibfnamefont {T.}~\bibnamefont {{Reddy}}}, \bibinfo {author} {\bibfnamefont {D.}~\bibnamefont {{Cournapeau}}}, \bibinfo {author} {\bibfnamefont {E.}~\bibnamefont {{Burovski}}}, \bibinfo {author} {\bibfnamefont {P.}~\bibnamefont {{Peterson}}}, \bibinfo {author} {\bibfnamefont {W.}~\bibnamefont {{Weckesser}}}, \bibinfo {author} {\bibfnamefont {J.}~\bibnamefont {{Bright}}}, \bibinfo {author} {\bibfnamefont {S.~J.}\ \bibnamefont {{van der Walt}}}, \bibinfo {author} {\bibfnamefont {M.}~\bibnamefont {{Brett}}}, \bibinfo {author} {\bibfnamefont {J.}~\bibnamefont {{Wilson}}}, \bibinfo {author} {\bibfnamefont {K.~J.}\ \bibnamefont {{Millman}}}, \bibinfo {author} {\bibfnamefont {N.}~\bibnamefont {{Mayorov}}},
  \bibinfo {author} {\bibfnamefont {A.~R.~J.}\ \bibnamefont {{Nelson}}}, \bibinfo {author} {\bibfnamefont {E.}~\bibnamefont {{Jones}}}, \bibinfo {author} {\bibfnamefont {R.}~\bibnamefont {{Kern}}}, \bibinfo {author} {\bibfnamefont {E.}~\bibnamefont {{Larson}}}, \bibinfo {author} {\bibfnamefont {C.~J.}\ \bibnamefont {{Carey}}}, \bibinfo {author} {\bibfnamefont {{\.I}.}~\bibnamefont {{Polat}}}, \bibinfo {author} {\bibfnamefont {Y.}~\bibnamefont {{Feng}}}, \bibinfo {author} {\bibfnamefont {E.~W.}\ \bibnamefont {{Moore}}}, \bibinfo {author} {\bibfnamefont {J.}~\bibnamefont {{VanderPlas}}}, \bibinfo {author} {\bibfnamefont {D.}~\bibnamefont {{Laxalde}}}, \bibinfo {author} {\bibfnamefont {J.}~\bibnamefont {{Perktold}}}, \bibinfo {author} {\bibfnamefont {R.}~\bibnamefont {{Cimrman}}}, \bibinfo {author} {\bibfnamefont {I.}~\bibnamefont {{Henriksen}}}, \bibinfo {author} {\bibfnamefont {E.~A.}\ \bibnamefont {{Quintero}}}, \bibinfo {author} {\bibfnamefont {C.~R.}\ \bibnamefont {{Harris}}}, \bibinfo {author}
  {\bibfnamefont {A.~M.}\ \bibnamefont {{Archibald}}}, \bibinfo {author} {\bibfnamefont {A.~H.}\ \bibnamefont {{Ribeiro}}}, \bibinfo {author} {\bibfnamefont {F.}~\bibnamefont {{Pedregosa}}}, \bibinfo {author} {\bibfnamefont {P.}~\bibnamefont {{van Mulbregt}}},\ and\ \bibinfo {author} {\bibnamefont {{SciPy 1. 0 Contributors}}},\ }\bibfield  {title} {\bibinfo {title} {{SciPy 1.0: fundamental algorithms for scientific computing in Python}},\ }\href {https://doi.org/10.1038/s41592-019-0686-2} {\bibfield  {journal} {\bibinfo  {journal} {Nature Methods}\ }\textbf {\bibinfo {volume} {17}},\ \bibinfo {pages} {261} (\bibinfo {year} {2020})},\ \Eprint {https://arxiv.org/abs/1907.10121} {arXiv:1907.10121 [cs.MS]} \BibitemShut {NoStop}%
\bibitem [{\citenamefont {{Hunter}}(2007)}]{Hunter2007}%
  \BibitemOpen
  \bibfield  {author} {\bibinfo {author} {\bibfnamefont {J.~D.}\ \bibnamefont {{Hunter}}},\ }\bibfield  {title} {\bibinfo {title} {{Matplotlib: A 2D Graphics Environment}},\ }\href {https://doi.org/10.1109/MCSE.2007.55} {\bibfield  {journal} {\bibinfo  {journal} {Computing in Science and Engineering}\ }\textbf {\bibinfo {volume} {9}},\ \bibinfo {pages} {90} (\bibinfo {year} {2007})}\BibitemShut {NoStop}%
\end{thebibliography}%

\end{document}